\documentclass[twocolumn,aps,prd,showpacs]{revtex4}
\usepackage{graphicx}
\usepackage{natbib}
\usepackage{amsbsy,amssymb,amsmath}
\usepackage[usenames]{color}
\usepackage{psfrag}
\usepackage[ps2pdf,colorlinks,bookmarks]{hyperref}
\definecolor{linkblue}{rgb}{0,0,0.8}
\definecolor{linkgreen}{rgb}{0,0.5,0}
\hypersetup{pdfpagemode=None, pdfstartview=FitH, linkcolor=linkblue, %
            citecolor=linkgreen, urlcolor=linkblue}

\newcommand{\beq}{\begin{equation}}
\newcommand{\eeq}{\end{equation}}

\def\Teds{T_0^{\rm EdS}}
\def\Heds{H^{\rm EdS}}

\def\zeds{z^{\rm EdS}}
\def\zreeds{z_{\rm re}^{\rm EdS}}
\def\ld{\left}
\def\rd{\right}
\def\fr{\frac}
\def\oo{\frac{1}}
\def\half{\frac{1}{2}}
\def\n{{\bf{\hat{n}}}}
\def\np{{\bf{\hat{n}}^{\prime}}}
\def\lcdm{$\Lambda$CDM}

\begin{document}

\title{Precision cosmology defeats void models for acceleration}

\author{Adam Moss} 
\email{adammoss@phas.ubc.ca}
\affiliation{Department of Physics and Astronomy, %
University of British Columbia, %
Vancouver, BC, V6T 1Z1  Canada}

\author{James P. Zibin} 
\email{zibin@phas.ubc.ca}
\affiliation{Department of Physics and Astronomy, %
University of British Columbia, %
Vancouver, BC, V6T 1Z1  Canada}

\author{Douglas Scott} 
\email{dscott@phas.ubc.ca}
\affiliation{Department of Physics and Astronomy, %
University of British Columbia, %
Vancouver, BC, V6T 1Z1  Canada}

\date{\today}

\begin{abstract}
The suggestion that we occupy a privileged position near the centre of a large, nonlinear, and nearly spherical void has recently attracted much attention as an alternative to dark energy. Putting aside the philosophical problems with this scenario,  we perform the most complete and up-to-date comparison with cosmological data. We use supernovae and the {\em full} cosmic microwave background spectrum as the basis of our analysis. We also include constraints from radial baryonic acoustic oscillations, the local Hubble rate, age, big bang nucleosynthesis, the Compton $y$-distortion, and for the first time include the local amplitude of matter fluctuations, $\sigma_8$. These all paint a consistent picture in which voids are in severe tension with the data.  In particular, void models predict a very low local Hubble rate, suffer from an ``old age problem'', and predict much less local structure than is observed.
\end{abstract}


\maketitle

\section{Introduction}

Over the last decade the foundations of the standard model of cosmology (SMC; see, e.g., \cite{scott06}) have been laid. The dominant components of the SMC at this current epoch are baryonic and cold dark matter, together with dark energy, whose equation of state is close to that of a cosmological constant. The dark energy is driving the accelerated expansion of a flat, homogenous and isotropic Friedmann-Lema\^itre-Robertson-Walker (FLRW) background. This standard model is also referred to as the (flat) $\Lambda$ cold dark matter (\lcdm) model.

Because of a perceived lack of elegance or naturalness in the SMC, several alternatives have been pursued.  In recent years there has been a resurgence of interest in inhomogeneous cosmological models as an alternative to dark energy~\cite{Caldwell:2007yu,GarciaBellido:2008nz,Zibin:2008vk,February:2009pv} (see~\cite{Tomita:2009ar} for a recent review).  This has been partly motivated by the fact that the {\em local} expansion rate is greater in underdense regions of the Universe (for growing modes). If an underdensity or {\em void} is large enough in amplitude (density fluctuation $\delta \rho /\rho \sim 1$) and physical size ($\sim$$1 \, {\rm Gpc}$), one can mimic the acceleration due to dark energy if one occupies a privileged position near the centre of the void.  This is essentially because a greater expansion rate near the observer {\em spatially} can be hard to distinguish from an expansion rate increasing {\em in time,} for observations confined to the past light cone.  These models break with the standard assumption of cosmological-scale homogeneity, and retain isotropy at the cost of a violation of the Copernican principle.

There are several serious physical and philosophical objections to this scenario. First, the probability of producing a void of the required amplitude and size in standard structure formation models is extremely small~\cite{2010MNRAS.401..547H}. Second, the probability of a random observer lying close enough to the centre of the void to avoid a large cosmic microwave background (CMB) dipole is also very low: these models appear to strongly violate the Copernican principle~\cite{Alnes:2006pf,Blomqvist:2009ps,Kodama:2010gr,Foreman:2010uj}. Finally, the void would need to be close to spherical to match the observed isotropy in the Universe. These features make the model somewhat unappealing from the point of view of Occam's Razor.

However, \lcdm\ appears to suffer from severe fine-tuning problems of its own and it is conceivable that unknown physics could produce the conditions necessary for such a void. Therefore, it is prudent to ignore our philosophical prejudices and use the available cosmological observations to decide the issue.  Given the increasing scope and precision of those observations, which largely support the SMC, demonstrating the viability of models that depart so dramatically from \lcdm\ would certainly appear unlikely.  Nevertheless, the relevance of this line of inquiry extends beyond particular models of dark energy, and refuting these inhomogeneous models would, importantly, strengthen our confidence in the fundamental assumptions of our standard cosmological models.  There is presently little {\em direct} evidence for cosmological-scale homogeneity, in particular.

The notion that a local underdensity might be an explanation for the observed redshift-luminosity distance relation of Type Ia supernovae (SNe) appears to have been first proposed in Refs.~\cite{Tomita:1999qn,Goodwin:1999ej,Celerier:1999hp}.   To describe such a large void, the spherically symmetric Lema\^itre-Tolman-Bondi (LTB) spacetime~\cite{lemaitre33,tolman34,bondi47} is usually used.  This is an exact solution of Einstein's equations, with a realistic pressureless matter source.  Most early studies focused on fitting the SN data to LTB models of voids.  This is not a good way of discriminating a void from \lcdm, since it is possible to construct a radial density profile of a void which {\em exactly} matches the redshift-luminosity distance relation of \lcdm\ (see, e.g.,~\cite{Yoo:2008su}).  This highlights the need to consider other observations.  More recently, studies have included baryon acoustic oscillations (BAO)~\cite{GarciaBellido:2008nz}, spectral distortions in the CMB~\cite{Caldwell:2007yu}, the kinematic Sunyaev-Zeldovich effect~\cite{GarciaBellido:2008gd}, and estimates of the Hubble rate~\cite{February:2009pv}.

Crucial evidence for the SMC is provided by measurements of the CMB, which give us a precise picture of the Universe at early times. The CMB also constrains the distance to the last scattering surface (LSS) via the angular scale of the acoustic peaks in the observed power spectrum. Other authors have used this angular scale {\em by itself} in their analysis of voids, but the radial degree of freedom can again be tuned to match that of \lcdm~\cite{Alnes:2005rw,Bolejko:2008cm,GarciaBellido:2008yq,Yoo:2010qy}. 

In Zibin, Moss, and Scott~\cite{Zibin:2008vk} (hereafter ZMS), we stressed the importance of using the {\em full} CMB anisotropy power spectrum (see also~\cite{Alexander:2007xx}) and introduced the {\em effective Einstein-de Sitter} (EdS) approach to calculating the anisotropies in arbitrary LTB models.  This allowed us to place much tighter constraints on void models than previous studies.  We also explored a very wide region of void radial profile space using a spline parameterization, which allowed us to identify models which fitted the CMB + SN data.  However, we found that models which matched the CMB power spectrum had a local Hubble rate, $H_0$, so low as to rule them out.  We argued that it would require significant fine tuning of the primordial perturbation spectrum in order to circumvent this conclusion, but we did not investigate this in detail.  More recently, Ref.~\cite{Vonlanthen:2010cd} stressed the importance of obtaining model-independent constraints from the CMB in a broader context.

Another class of observations that could potentially put strong constraints on void models of acceleration is that of structure in the late Universe.  The evolution of matter perturbations on LTB backgrounds is expected to differ significantly from that in the SMC, so observations of the late-time matter power spectrum shape or amplitude, weak lensing, or integrated Sachs-Wolfe (ISW) effect would likely help to rule out void models.  Unfortunately, the evolution of perturbations on LTB spacetimes is an extremely difficult problem~\cite{ccf09}, and no exact calculations have yet been carried out.  An approximation method which drastically simplifies the calculations was proposed in Ref.~\cite{z08} (see also~\cite{Dunsby:2010ts}); however, again, no confrontation with data has yet been performed.  A large part of the difficulty is that observations, e.g.\ of the matter power spectrum, are often presented in a model-dependent manner, in that an underlying FLRW background is assumed.

In this work we undertake an extended update of ZMS, using the latest (and additional) cosmological data. We improve our method in several respects, and also consider extensions to the ``basic'' void model.   We begin in Sec.~\ref{sec:overview} with a summary of the exact general relativistic framework within which we perform our analysis.  In Sec.~\ref{sec:basicconstr} we describe our effective EdS approach to calculating CMB spectra, and describe the other data we consider, as well as our statistical approach.  We also enumerate our improvements to the approach of ZMS, which include updated CMB and supernova data and advances in our numerical and fitting procedures.  Our analysis also includes new types of observations, such as the $y$-distortion of the CMB frequency spectrum, as well as the first constraints on void models using the amplitude of local structure.  We avoid the problem of model dependence in examining structure data by considering only the most local observations.  Section~\ref{sec:basicconstr} closes with the constraints which form our main results.  We then consider whether modifications to the primordial spectrum can alleviate the problems facing voids in Sec.~\ref{sec:extensions}, before examining an interesting multi-valued region of parameter space.  We close with conclusions in Sec.~\ref{sec:concl}.  Throughout this paper we set $c = 1$.

\section{Overview of LTB framework}
\label{sec:overview}

\subsection{Exact solution}

We consider an observer in an isotropic, inhomogeneous Universe which at late times is dominated by pressureless matter. Such a model is described by the LTB spacetime, in which the Einstein equations can be solved exactly. The line element of the LTB metric is given by \beq ds^2 = -dt^2 + \fr{Y'^2}{1 - K}dr^2 + Y^2d\Omega^2\,, \eeq    where a prime denotes the derivative with respect to comoving radial coordinate $r$, and $t$ is the proper time along the comoving worldlines. The curvature function $K = K(r)$ is a free radial function (bounded by $K < 1$), and the areal radius $Y = Y(t,r)$ is given parametrically by the exact solution \beq \label{YsolnKneg} Y = \frac{M}{K}(1 - \cosh\eta)\,, \quad t - t_{\rm B} = \fr{M}{(-K)^{3/2}}(\sinh \eta - \eta)\,, \eeq for $ K < 0$; \beq \label{YsolnKpos} Y = \frac{M}{K}(1 - \cos \eta)\,, \quad t - t_{\rm B} = \fr{M}{K^{3/2}}(\eta - \sin \eta )\,, \eeq for $ 0 < K < 1$; and \beq \label{YsolnK0} Y = \left( \frac{9M}{2} \right)^{1/3} \left(  t - t_{\rm B}\right)^{2/3} \,, \eeq for $K=0$. Here, the free radial function $t_{\rm B} = t_{\rm B} (r)$ is known as the ``bang time'', since the cosmological singularity occurs at $t = t_{\rm B}(r)$. In this work we assume a homogeneous big bang and set $t_{\rm B}=0$.  In this case the void only has a growing mode~\cite{silk77,z08}, and hence we can use standard cosmological perturbation theory up to the epoch where the void becomes nonlinear.  The function $M=M(r)$ is a further arbitrary radial function.  However, one of these free radial functions is a gauge mode corresponding to a rescaling of the radial coordinate.  We set $M=r^3$, which imposes no restriction on the solutions except that they will not be valid past the ``equator'' of a spatially closed model.  The LTB spacetime is then completely specified once the single free function $K(r)$ is fixed.

Covariant physical quantities within the LTB spacetime are given by (see, e.g.,~\cite{z08})
\begin{eqnarray}
4 \pi G \rho &=& \frac{M'}{ Y^2Y'}\,,\label{eqn:rhodef} \\
\theta &=& H_{\rm R} + 2 H_{\rm T}\,, \\
\Sigma &=& \frac{2}{3}\left(H_{\rm R} - H_{\rm T} \right)\,, \\
^{(3)}R &=& \fr{2(KY)'}{Y^2Y'},\label{eqn:RK}
\end{eqnarray}
where $\rho$, $\theta$, and $\Sigma$ are the comoving matter density, expansion, and shear, respectively, and $^{(3)}R$ is the Ricci curvature of the spatial comoving-orthogonal hypersurfaces. The radial and transverse expansion rates are given by $H_{\rm R} = {\dot Y}'/Y'$ and $H_{\rm T} = {\dot Y}/Y$, where the overdot denotes the derivative with respect to $t$. We define the local density parameter by $\Omega^{\rm loc}_{\rm m} \equiv 24\pi G\rho/\theta^2$. These quantities reduce to the standard FLRW expressions ($\theta = 3 H$, $\Sigma=0$, $\Omega^{\rm loc}_{\rm m} = \Omega_{\rm m}$) when $Y(t,r)=a(t) r$, where $a(t)$ is the FLRW scale factor and $H=\dot{a}/a$ the Hubble rate.  Along the centre of symmetry in an arbitrary LTB spacetime, we have $\Sigma = 0$ and $H_{\rm R} = H_{\rm T} \equiv H$.

The null radial incoming geodesics ($ds^2 = d\Omega^2 = 0$) are described by the equations \beq \label{eqn:geo} \fr{dt}{dr} = \fr{-Y'}{\sqrt{1 - K}}\,, \qquad \fr{dz}{dr} = \fr{(1 + z){\dot Y}'}{\sqrt{1 - K}}\,, \eeq where the redshift $z$ is measured along the past light cone, increasing from $z = 0$ at the origin. The luminosity and angular diameter distances from the centre to redshift $z$ are then \beq d_{\rm L} = \left(1+z\right)^2 Y\,, \quad d_{\rm A} = Y\,. \eeq

\subsection{Numerical implementation}

In our numerical calculations, instead of fixing $K(r)$ we found it more convenient to express the single free radial function by the early-time comoving density perturbation, $\delta (t_i,r) \equiv \left[\rho(t_i,r) - \rho_{\rm FLRW}(t_i) \right] / \rho_{\rm FLRW}(t_i) $, where $t_i$ is a time near last scattering.  Because we only consider the growing mode of the void, $\delta (t_i,r)$ can be considered a linear fluctuation from FLRW. In this case the curvature function $K(r)$ can be determined from the density perturbation using the linear growing mode relation between the comoving density perturbation and curvature (see, e.g.,~\cite{z08}) and Eq.~(\ref{eqn:RK}).  The result is \beq \label{eqn:curv} (r K)' = \frac{40 \pi G}{3} \rho_{\rm FLRW}(t_i) Y^2 \delta (t_i,r)\,, \eeq where in the early linear regime $Y = a_i r$ and $a_i = a(t_i)$ is the initial scale factor, which we arbitrarily set to $a_i = 10^{-3}$.

The LTB equations were coded in a Fortran module to interface with the CMB code \textsc{camb}~\cite{Lewis:1999bs}. We first relate the curvature function to the early density profile by Eq.~(\ref{eqn:curv}), and discard models with $1 - K(r)<10^{-3}$ to ensure numerical stability. Given the Hubble rate $H_0 \equiv 100\, h_0 \, {\rm km} \, {\rm s^{-1}} \, {\rm Mpc^{-1}}$ observed today ($t = t_0$) at the void centre, we evaluate the time $t_0$ using the exact solutions.

We then integrate Eq.~(\ref{eqn:geo}) along the past light cone from the centre today to a redshift far outside the void. During integration, we store quantities such as $t$, $z$, and $d_{\rm L}$ with the integration variable $r$ in a finely-spaced array. These are then interpolated using cubic splines when intermediate values are required (such as fitting redshift-luminosity distances of SNe). 

In some models (typically very deep voids) the radial expansion rate can become negative, when overdense regions break from the background expansion and begin to contract.  As can be seen from Eq.~(\ref{eqn:geo}), negative $H_{\rm R} = {\dot Y}'/Y'$ implies that the redshift will {\em decrease} with increasing radial coordinate down the light cone.  This leads to an interesting class of models with {\em multi-valued} distance-redshift relations~\cite{Mustapha:1997xb} (see also~\cite{Biswas:2006ub}). In extreme cases the redshift can even become negative. We flag these multi-valued models and do not use them in our basic constraints, due to uncertainties in interpreting fits to SNe. The multi-valued cases will be discussed further in Sec.~\ref{sec:deep}.

After overdense regions contract sufficiently, they can experience a {\em shell-crossing singularity,} when $Y'$ crosses zero and hence the density diverges according to Eq.~(\ref{eqn:rhodef}).  These singularities are a symptom of the pressureless assumption of LTB: in reality, interactions on small scales would modify the dynamics and the LTB solution would become invalid.  Therefore, we perform checks in our code and discard models in which $Y' = 0$ anywhere on the past light cone.  (Of course, shell crossings may still occur in the future, but we do not need the solution in such unobservable regions.)


\section{Basic constraints}
\label{sec:basicconstr}

In this section we detail the basic cosmological constraints on LTB models. We have made several changes and additions to our analysis in ZMS, most of which are discussed in more detail later. In summary, these are

\begin{itemize}
\item Cosmological data have moved on.  We use CMB data from the Wilkinson Microwave Anisotropy Probe (WMAP) 7-year release~\cite{Jarosik:2010iu,Komatsu:2010fb} and the latest Union2 SN compilation~\cite{Amanullah:2010vv}, which has nearly double the number of SNe and improved outlier rejection compared with our previous analysis.
\item We change the integration variable of the LTB equations to use the radial coordinate rather than redshift. This allows us to find models  with multi-valued distance-redshift relations and to more easily reject models with shell crossings.
\item A more efficient method is used to search through the void parameter space when fitting to CMB observations. This involves sampling from a set of effective parameters derived from CMB data.
\item We investigate an alternative parameterization of the void profile, which allows us to separate the physics of fitting voids to SN and CMB data. 
\item We investigate voids embedded in a spatially curved FLRW background.
\item We discuss additional cosmological constraints, such as those from the Compton $y$-distortion of the  CMB blackbody spectrum. We also present the first estimates of the local amplitude of matter fluctuations smoothed over $8\, h_0^{-1} \, {\rm Mpc}$ spheres, $\sigma_8$.
\end{itemize}


\subsection{Formalism} \label{sec:formalism}

\subsubsection{CMB spectra}
\label{sec:formalism_CMBspectra}

The effective EdS method introduced in ZMS forms the basis for our computations of CMB power spectra in LTB models. The method can also be applied to {\em any} model with an expansion history different from EdS (FLRW with pure dust source and vanishing spatial curvature) from recombination to today. The basic idea is as follows: we find the parameters of an effective EdS model which has the {\em same} physics at recombination as the void model, as well as the {\em same} angular diameter distance to the LSS, in proper units at last scattering. This ensures that the $C_{\ell}$'s will be identical between the effective EdS and LTB models, apart from any sources of secondary anisotropy between today and the LSS.  The spectra can then be readily calculated by feeding the effective parameters into public CMB codes. Using EdS for the effective model is convenient since it ensures that no secondary ISW component will be present.

To generate the effective model, we must match the angular diameter distance to the LSS between the EdS model and the void model, and also match the Hubble rate and density of relativistic species and matter components at the LSS (i.e.\ the physics at last scattering). The extra degree of freedom required in the effective EdS model comes from our ability to specify a CMB temperature for the effective model, different from the actual $T_0 = 2.726 \pm 0.001\, {\rm K}$~\cite{Fixsen:2009ug} which we observe today.  (Note that the effective model will include substantial radiation at early times, and hence should not properly be called ``EdS''.  However, for the purposes of feeding effective late-time parameters into CMB codes, it is essentially EdS.)

We follow the procedure given in ZMS, generalized slightly to accomodate void models that asymptote to spatially curved FLRW, rather than just to flat FLRW.  To specify an LTB model we must fix the free radial profile and the Hubble rate at the observation point today, $H_0$.  Once this is done, we first compute the coordinates $(t_{\rm m},r_{\rm m})$ at a ``midpoint'' redshift $z_{\rm m}$ down the light cone, far outside the void. This redshift is chosen such that the background LTB shear is negligible, and the radiation density (treating radiation as a test field on the LTB background) is small compared to the matter density. In our numerical calculations we use $z_{\rm m} = 100$, and have checked that our results are not affected significantly within the range $50<z_{\rm m}<200$.  We cannot choose $z_{\rm m}$ to coincide with the LSS, since radiation is important at background level there and the LTB solution does not include radiation.  However, as long as the background shear is negligible beyond $z_{\rm m}$, matching at $z_{\rm m}$ will be essentially equivalent to matching at the LSS.  The basic idea is that where radiation is unimportant (for $z < z_{\rm m}$), the LTB model can be used, while where background shear is unimportant (for $z_{\rm m} < z < z_{\rm LSS}$), the {\em true} inhomogeneous spacetime (i.e.\ including radiation as a source) is very closely approximated by an FLRW matter plus radiation model, so that matching at $z_{\rm m}$ is essentially equivalent to matching at $z_{\rm LSS}$.

We then compute the Hubble rate, $H_{\rm m}$, at $z_{\rm m}$ in the LTB model and integrate back up the light cone into the effective EdS model to comoving coordinate $r^{\rm EdS} = 0$.  In the case that the void becomes asymptotically spatially flat, we ensure that the proper distance to the LSS is correct by setting $a(r^{\rm EdS}_{\rm m})r^{\rm EdS}_{\rm m} = Y(t_{\rm m},r_{\rm m})$, for EdS scale factor $a$. This allows us to calculate the effective EdS mean temperature and Hubble rate via \beq \Teds = T_0(1 + \zeds_0)\,, \quad \Heds_0 = H_{\rm m}\ld(\fr{1 + \zeds_0}{1 + z_{\rm m}}\rd)^{3/2}\,, \label{eqn:CMBeffTH}\eeq where \beq 1 + \zeds_0 = \fr{1 + z_{\rm m}}{\ld(1 + a^{\rm EdS}_{\rm m}r^{\rm EdS}_{\rm m}H_{\rm m}/2\rd)^2}. \eeq (Note that the quantity $\zeds_0$ was called $\zeds_{\rm m}$ in ZMS, and that in general $\zeds_0 \ne 0$, since we choose the values of the redshifts $z_{\rm m}$ at the ``midpoint'' to be identical in both void and effective EdS models.)  The effective parameters $\Teds$ and $\Heds_0$ define the point in EdS which observes the same primary CMB as the specified LTB model.

   In the case that the LTB model asymptotes to spatially curved FLRW, we still choose our effective model to be EdS, but we must modify the above calculation.  In particular, we modify the matching condition to $a(r^{\rm EdS}_{\rm m})r^{\rm EdS}_{\rm m} = Y(t_{\rm m},r_{\rm m}) + \Delta d_{\rm A}$, where
\begin{eqnarray}
\Delta d_{\rm A}
   &\equiv& \ld[d_{\rm A}(z_{\rm LSS}) - d_{\rm A}(z_{\rm m})\rd]_{\substack{\rm curved \\ \rm FLRW}}\nonumber\\
        &-& \ld[d_{\rm A}(z_{\rm LSS}) - d_{\rm A}(z_{\rm m})\rd]_{\substack{\rm flat \\ \rm FLRW}}
\end{eqnarray}
is the difference in the angular diameter distance increment from $z_{\rm m}$ to $z_{\rm LSS}$ between the actual curved FLRW model and flat FLRW.  In this expression, the curved FLRW values are calculated in the FLRW model which has the same matter and radiation densities and curvature at $z_{\rm m}$ as does the actual void model.  For flat FLRW, the increment from $z_{\rm m}$ to $z_{\rm LSS}$ is calculated directly by integrating the relation \beq a_{\rm m}\fr{dr}{dz} = \fr{1}{(1 + z_{\rm m})H(z)}. \label{adrdz} \eeq For the curved FLRW case, Eq.~(\ref{adrdz}) only provides the {\em coordinate} increment $a_{\rm m}\Delta r$; this must be further translated into an increment in $d_{\rm A}$ using the relations \beq a_{\rm m} = \fr{1}{H_{\rm m}\sqrt{|1 - \Omega^{\rm loc}_{\rm m}(z_{\rm m})|}},\qquad d_{\rm A} = a\sin r \eeq (and similarly for the open FLRW case).  With this correction, the effective EdS model will produce essentially exactly the correct primary CMB as long as the LTB model becomes essentially homogeneous (shear-free) by $z_{\rm m}$.

\begin{figure}
\centering 
\includegraphics[width=0.9\columnwidth,angle=0]{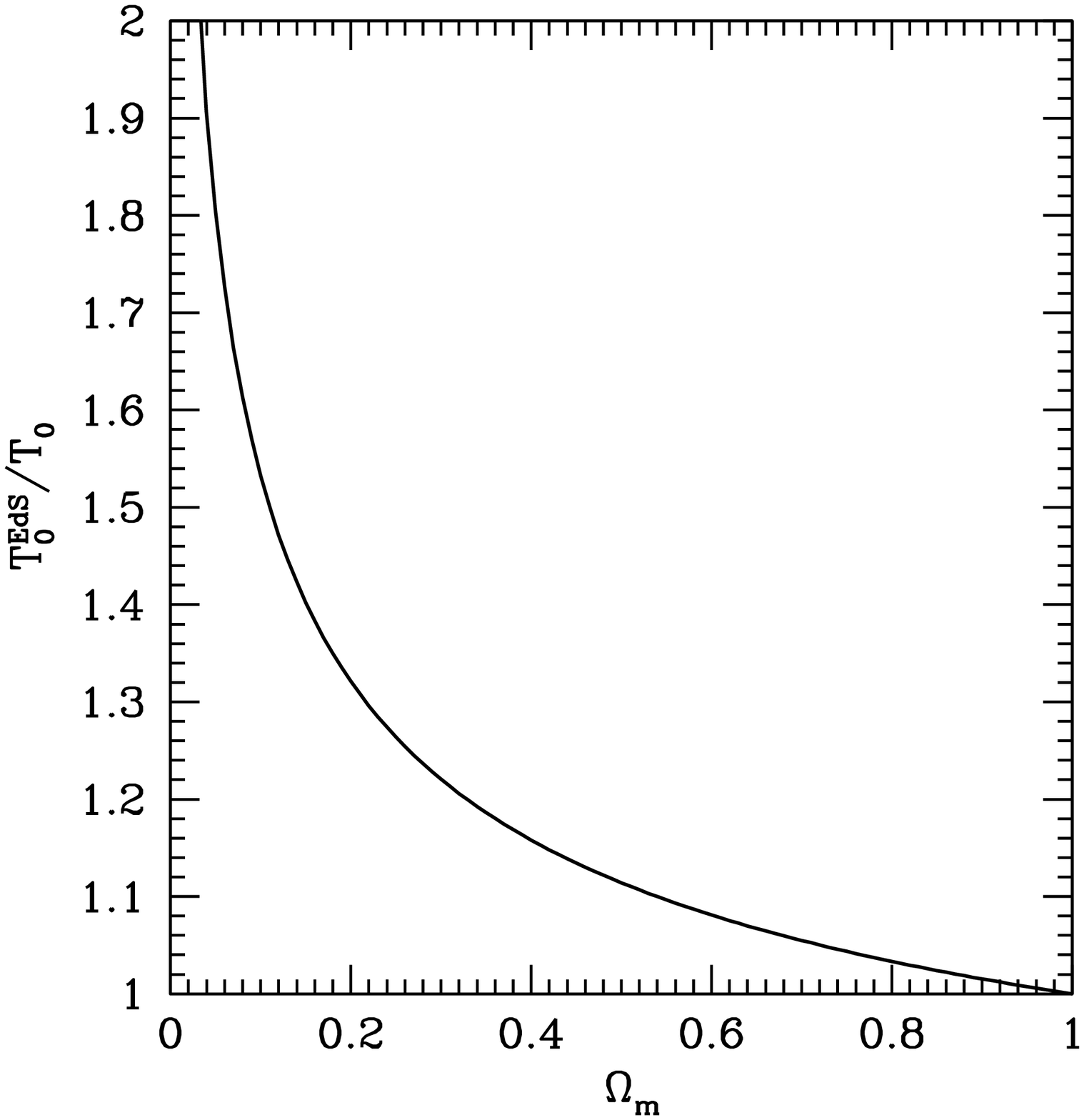}
\includegraphics[width=0.9\columnwidth,angle=0]{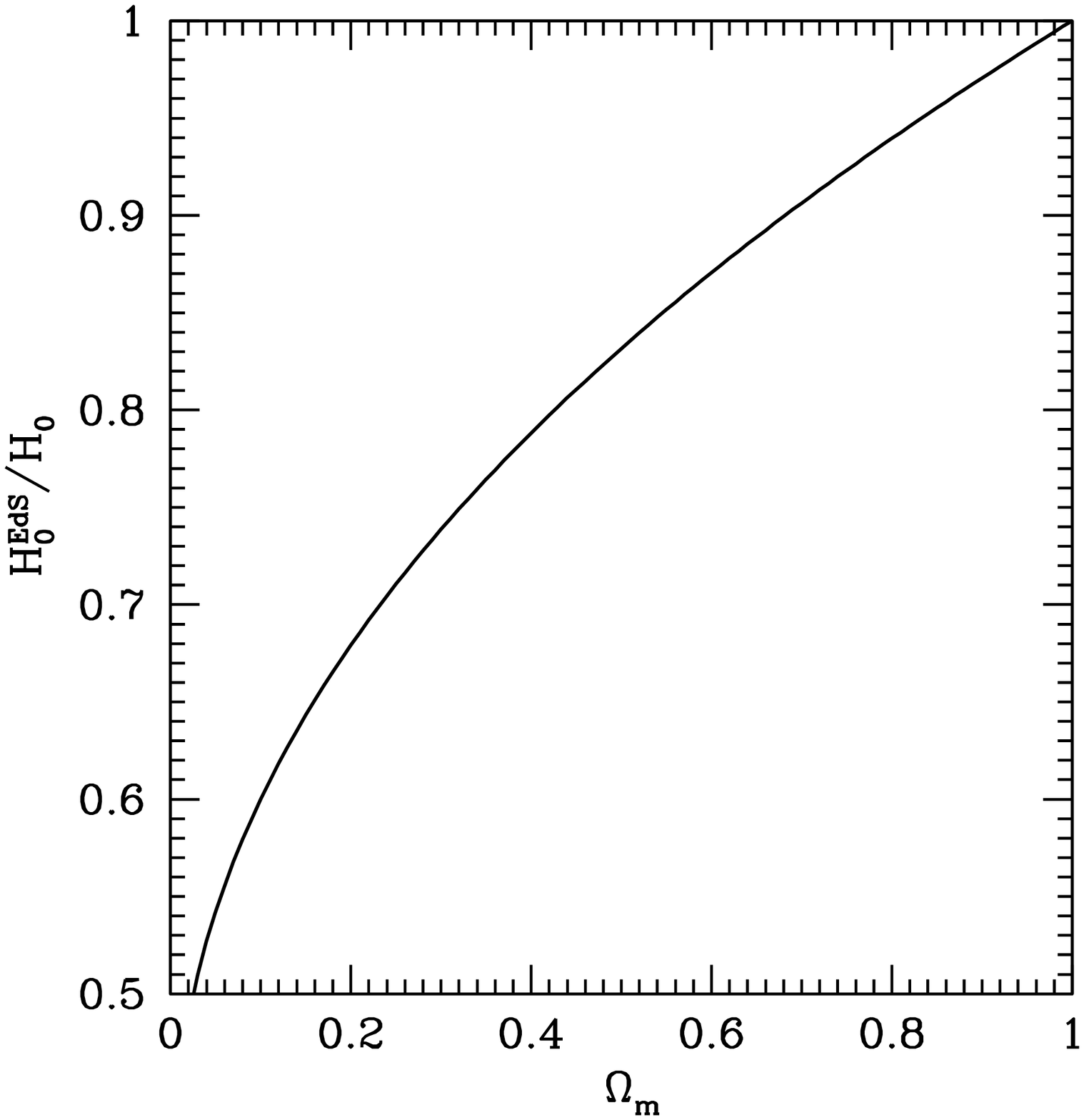}
\caption{\label{fig:effective_eds} Effective EdS parameters $\Teds$ and $\Heds_0$ as a function of the matter density $\Omega_{\rm m}$ in the corresponding \lcdm\ cosmology.}
\end{figure}

It is easy to evaluate the effective EdS parameters for a \lcdm\ cosmology instead of an LTB model. They can be found by substituting the quantities
\begin{widetext}
\beq \label{eqn:eff_lam} H_{\rm m} \approx H_0 \sqrt {\Omega_{\rm m} \left( 1+z_{\rm m} \right)^3}\,, \quad  1 + \zeds_0  = \frac{1+{z_{\rm m}}}{\left\{1+\half\sqrt{1+z_{\rm m}} \int_{0}^{z_{\rm m}} dz  \left[ \left(1+z \right)^3 + \Omega_{\rm m}^{-1}-1 \right]^{-1/2}  \right\}^2} \, \eeq
\end{widetext}
into Eq.~(\ref{eqn:CMBeffTH}), where $\Omega_{\rm m}$ is the matter density parameter today in \lcdm.  The integral in Eq.~(\ref{eqn:eff_lam}) can be computed numerically, and is insensitive to the choice of $z_{\rm m}$ at the level of $< 0.1\%$  for $z_{\rm m} > 5$. Plots of the effective parameters are shown in Fig.~\ref{fig:effective_eds}. As $\Omega_{\rm m} \rightarrow 1$ we find $\Teds \rightarrow T_0$ and $\Heds_0 \rightarrow H_0$ as expected, while for low $\Omega_{\rm m}$ the effective parameters are very different from the actual observed parameters. For the WMAP 7-year best-fit \lcdm\ model, given by $\Omega_{\rm m} = 0.262$ and $H_0 = 71.4 \, {\rm km} \, {\rm s^{-1}} \, {\rm Mpc^{-1}}$~\cite{Komatsu:2010fb}, we find $\Teds = 3.416 \, {\rm K}$ and $ \Heds_0 =  51.2 \, {\rm km} \, {\rm s^{-1}} \, {\rm Mpc^{-1}}$.  A void model must have effective EdS parameters close to these values if it is to produce CMB power spectra similar to those actually observed.

\subsubsection{Low $\ell$ CMB}

The $C_{\ell}$'s of the effective model are essentially identical to the actual model on intermediate and small angular scales. For temperature, however, the \lcdm\ model creates an ISW anisotropy on large scales, due to decaying gravitational potentials along the line of sight. A void model is also expected to create an ISW signal, but the complications of evolving perturbations on an LTB background have prohibited a rigourous calculation of the LTB ISW effect, so that the size and shape of the ISW signal is uncertain. 

In order for the effective EdS method to remain viable, however, we introduce a marginalization uncertainty over an ISW template; large angle data are necessary for constraining the amplitude and spectral index of initial fluctuations, so one would lose important predictive power by applying a simple cutoff in $\ell$. Specifically, the $C_{\ell}$ spectrum is given by \beq C_{\ell} = 4 \pi  \int d (\ln k) \mathcal{P}_{\rm S}(k) |\Delta_{\ell}(k,t_{0})|^{2}\,, \eeq where $\Delta_{\ell}(k,t_{0})$ is the associated multipole moment for the photon distribution and $\mathcal{P}_{\rm S}(k)$ is the initial comoving curvature power spectrum, parameterized by $\mathcal{P}_{\rm S}(k)=A_{\rm S} \left( k/k_0 \right)^{n_{\rm S}-1}$, where $A_{\rm S}$ is the initial scalar amplitude, $n_{\rm S}$ the scalar spectral index, and $k_0$ the pivot scale (which we fix to $ 0.05 \, {\rm Mpc}^{-1}$). We modify the transfer function to \beq \Delta_{\ell}(k, t_{0}) = \Delta^{\rm EdS}_{\ell}(k,t_{0}) + A_{\rm ISW} \Delta^{\rm ISW}_{\ell}(k,t_{0})\,, \eeq where $A_{\rm ISW}$ is the ISW amplitude. Since the data on large scales are limited by cosmic variance, the precise shape of $\Delta^{\rm ISW}_{\ell}(k,\tau_{0})$ is not very well constrained, and so we use a template  from the WMAP7 \lcdm\ model. This choice provides a conservative estimate of the expected ISW signal from a void.

\begin{figure}
\centering 
\mbox{\resizebox{\columnwidth}{!}{\includegraphics[angle=0]{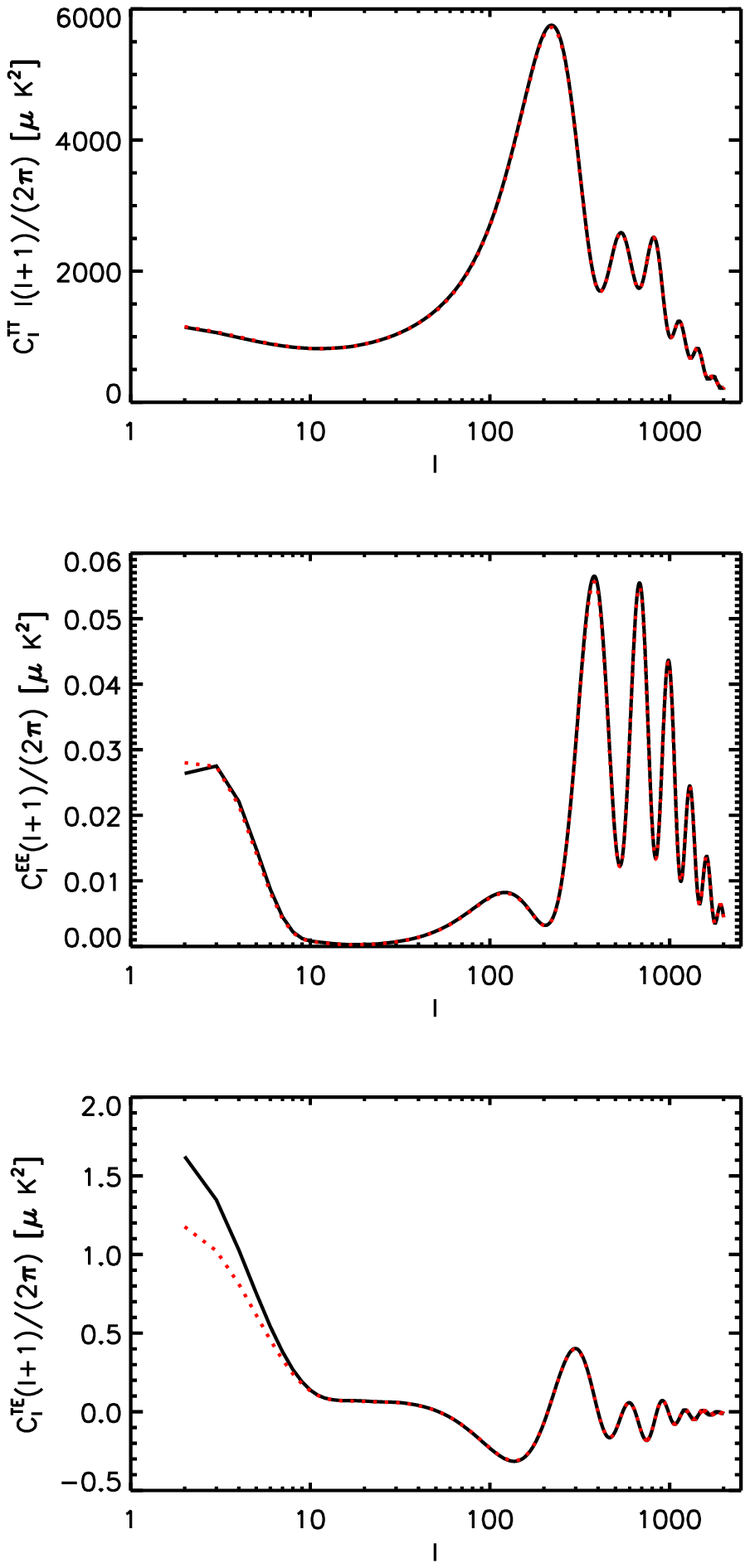}}} \caption{\label{fig:cls_eff} CMB power spectra for the WMAP7 \lcdm\ cosmology (solid curves) along with the corresponding effective EdS model (dotted curves). We show the TT (top panel), EE (middle panel), and TE (bottom panel) spectra.}
\end{figure}

The effective EdS parameters can now be used in any of various CMB anisotropy codes; we used \textsc{camb}~\cite{Lewis:1999bs}. The full set of parameters specifying the effective model are the baryonic matter fraction $f_{\rm b} \equiv \rho_{\rm b}/\rho_{\rm m}$ (which we take to be spatially constant throughout this work); $\Teds$; $\Heds_0 =  100\, h_0^{\rm EdS} \, {\rm km} \, {\rm s^{-1}} \, {\rm Mpc^{-1}}$; $A_{\rm S}$; $n_{\rm S}$; $A_{\rm ISW}$; and the redshift of reionization $z_{\rm re}^{\rm EdS}$. We show the temperature and polarization power spectra of the effective EdS model for  WMAP7 \lcdm\ in Fig.~\ref{fig:cls_eff}. Here, we have fixed the reionization optical depth to the WMAP7 value, so that $z_{\rm re}$ is lower in EdS compared to \lcdm\ ($\sim$8 as opposed to $\sim$10).

For the TT spectrum, the agreement between $C_{\rm \ell}$'s is excellent at all $\ell$. For polarization, particularly the TE cross spectrum, there is still a discrepancy on large scales. These modes are sourced at $z < z_{\rm re}$, so presumably the difference arises from the slightly different reionization history of EdS and (for TE) the cross-correlation with our simplified ISW template.

The generation of large angle polarization is an interesting question in LTB models. Electron scatterers  in the region of the void at $z < z_{\rm re}$ will see a temperature anisotropy dominated by a dipole (we discuss this further in the following section). Each scatterer will also see a smaller local quadrupole, with a  relative value larger in the void periphery. This quadrupole does {\em not} induce any polarization for an observer at the centre of the void though, since it is parallel to the radial direction.

The void will, however, have a different reionization history than for the effective EdS model. Our earlier work in ZMS showed that a large overdense outer shell at $z > 1$ is required to fit the CMB data (in the absence of background spatial curvature), and the precise details of this shell will modify the reionization history, and hence the low $\ell$ polarization.  For this reason, together with our simplified treatment of the ISW cross-correlation, we choose to ignore TE data on large scales, using a cutoff of $\ell = 50$. Since $z_{\rm re}$ is less constrained by the data without large scale TE, we apply a conservative prior on $z_{\rm re}$ in our analysis, the details of which are given in  Sec.~\ref{sec:method}.

\subsubsection{Dipole anisotropy}

Off-centre observers in an LTB model will see a CMB dipole, $\Delta T/T =\beta \, \cos \theta$, where $\theta$ is the angle with respect to the radial direction.  This dipole will be the dominant anisotropy sufficiently close to the centre.  As we will see, even for central observers this will have important consequences: the scatterers along the observer's light cone will produce spectral distortions in the CMB seen at the centre.  To compute $\beta(z)$, the dipole seen by scatterers at redshift $z$ down the light cone of the central observer, we can consider the propagation of incoming and outgoing radial geodesics towards the scatterer, and assume the anisotropy is a pure dipole.  (Incoming/outgoing refer to the direction {\em at the scatterer.})  More detailed calculations using non-radial geodesics show that this is a good approximation, apart from in the peripheral region of the void and further outwards, where the anisotropy subtends a smaller angle on the sky~\cite{Alnes:2006pf,Zibin_bang}.

To calculate $\beta(z)$, we follow a similar procedure to calculating the effective EdS CMB parameters. Concretely, we first evaluate the coordinates $(t_{\rm s}, r_{\rm s})$ of a scatterer at $z_{\rm s}$. We then continue the light cone integration to a position far from the void centre, choosing $z_{\rm in}=1000$ (i.e.\ approximately the redshift of the LSS), to find the originating coordinates $(t_{\rm in}, r_{\rm in})$ of the radial incoming geodesic.  To complete the calculation, we compute the redshift $z_{\rm out}$ of an {\em outgoing} geodesic, also originating at time $t_{\rm out } = t_{\rm in}$, which propagates back through the void centre and arrives at the same $(t_{\rm s}, r_{\rm s})$ as the incoming geodesic.  We use a value of $z_{\rm in} = 1000$, larger than $z_{\rm m} = 100$ used in the CMB calculations, since the outgoing ray generally passes through the void and the spatial curvature (and shear) is non-neglible for some models within the void at $t(z = 100)$. Moreover, the precise origin of the photons can affect the dipole at the observer, as shown in Sec.~\ref{sec:resultsdipole}, so we choose the LSS explicitly.  The condition $t_{\rm out } = t_{\rm in}$ ensures that both radial rays originated at essentially the same density, which defines the moment of last scattering in the absence of isocurvature modes. The dipole anisotropy observed at $z_{\rm s}$ is then \beq \beta = \frac{ z_{\rm out}- z_{\rm in}}{2+ z_{\rm out}+z_{\rm in}}\,. \eeq This dipole is indistinguishable from a peculiar velocity at the scatterer.

\subsubsection{Compton $y$-distortion}

Non-local processes can also provide complimentary constraints on LTB models~\cite{Goodman:1995dt,Caldwell:2007yu}. These arise from the re-scattering of photons in the region of the void after the Universe reionized at $z_{\rm re}$. One such process is the Compton $y$-distortion of the CMB blackbody spectrum. The $y$-distortion in the single-scattering and linear approximations is given by~\cite{Stebbins,Caldwell:2007yu}
\begin{widetext}
\beq \label{eqn:compton} y(\n) = \frac{3}{32 \pi} \int dz \frac{d \tau}{dz} \int d^2\np \left[ 1+ \left(\n \cdot \np \right)^2 \right] \left[ {\rm \Theta} \left(\np,\np, {\it z} \right) -  {\rm \Theta} \left(\n,\np,{\it z} \right)  \right]^2\,, \eeq
\end{widetext}
where $\Theta=\Delta T/T$ is the temperature anisotropy at the scatterer and $\tau$ is the optical depth. The vector $\n$ is the direction from observer to scatterer and $\np$ the direction of the photon arriving at the scatterer from the LSS. When the temperature anisotropy is dominated by the induced dipole, we have $\Theta \left(\n,\np, {\it z} \right) = \beta (z) \cos \theta$, with $\cos \theta= \n \cdot \np$. 

When the dipole dominates, we  can complete the angular part of Eq.~(\ref{eqn:compton}) analytically to leave \beq y= \frac{7}{10} \int_{0}^{z_{\rm re}} dz \frac{d \tau}{dz} \beta(z)^2 \,, \label{eqn:ydipapprox}\eeq where we integrate up to the redshift of reionization $z_{\rm re}$. The  redshift dependence of the optical depth is given by
\begin{eqnarray}
\frac{d \tau}{dz} &=& \sigma_{\rm T} \,n_e(z) \frac{dt}{dz}\nonumber\\
                  &=& \frac{ \sigma_{\rm T} \theta^2  f_{\rm b} \left(1-Y_{\rm He}/2 \right) \Omega_{\rm m}^{\rm loc} (z) }{24 \pi G m_p }\frac{dt}{dz},
\end{eqnarray}
with $\sigma_{\rm T}$ being the Thomson cross section, $n_e$ the electron number density, $Y_{\rm He}$ the helium mass fraction, and $m_p$ the proton mass. In our calculations we fix $Y_{\rm He}=0.24$.

\subsubsection{Baryon acoustic oscillation scale}
\label{sec:formalismbao}

LTB models generically exhibit shear at background level, so that $H_{\rm R} \ne H_{\rm T}$ for $r>0$, and this anisotropy in the expansion increases as the void grows and becomes nonlinear. Therefore, proper length scales, which are isotropic in the FLRW regime at early times, also become different in the radial and transverse directions at late times. One scale of particular cosmological interest is the sound horizon at the drag epoch, defined as the time when baryons are released from the Compton drag of photons. This can be approximated in an FLRW cosmology by~\cite{eisenstein} \beq s_{\rm p}(z) = \frac{44.5 \ln \left[9.83/(\Omega_{\rm m}  h_0^2) \right] }{(1 + z)\sqrt{1+ 10 (\Omega_{\rm b} h_0^2)^{3/4}}} \, {\rm Mpc}\,. \label{eqn:soundhor} \eeq Importantly, $s_{\rm p}(z)$ is the {\em proper} sound horizon at redshift $z$, which is a {\em model-independent} measure.  In Fourier space, this scale leads to the characteristic series of baryon acoustic oscillations observed in the matter power spectrum.

To calculate the observed BAO scales in the LTB model, we first generate a new effective EdS model, following the method introduced in ZMS.  This effective model will share the same early physics as the LTB model, but unlike the case of the CMB, here there is no need to match the angular diameter distance to the LSS.  In order to allow us to use Eq.~(\ref{eqn:soundhor}), we choose $\Teds = T_0$ for the effective model.  Substituting $\zeds_0 = 0$ in Eq.~(\ref{eqn:CMBeffTH}) then gives \beq \Heds_0 = \fr{H_{\rm m}}{(1 + z_{\rm m})^{3/2}}, \eeq where we again choose $z_{\rm m} = 100$.  To evaluate $s_{\rm p}(z_{\rm m})$, we then substitute this effective Hubble rate, together with $\Omega_{\rm m} = 1$ and $\Omega_{\rm b} = f_{\rm b}$, into Eq.~(\ref{eqn:soundhor}), checking that these effective parameters are within the regime of validity of that approximation.

Since the effective EdS and LTB models share the same early physics, we can conclude that the (essentially isotropic) sound horizon at coordinates $(t(z_{\rm m}),r(z_{\rm m})) \equiv (t_{\rm m},r_{\rm m})$ in the LTB model is given by $s_{\rm p}(z_{\rm m})$.  But, at the early time $t_{\rm m}$, the LTB spacetime is very close to FLRW.  Therefore, we can conclude that the sound horizon takes essentially the same value $s_{\rm p}(z_{\rm m})$ at $(t_{\rm m},r)$, for any $r$ in the LTB model (assuming no significant {\em isocurvature} mode associated with the void).  Thus, to evaluate the BAO scales at redshift $z$ down the observer's light cone, i.e.\ at coordinates $(t(z),r(z))$, we only need to evolve the scale $s_{\rm p}(z_{\rm m})$ according to the radial and transverse linear expansions in the sheared LTB background between the points $(t_{\rm m},r(z))$ and $(t(z),r(z))$.  The radial and transverse {\em proper} BAO scales at redshift $z$ are then, respectively,
\begin{eqnarray}
l^{\rm BAO}_{\rm R} (z) &=& s_{\rm p}(z_{\rm m})\frac{Y'(t(z),r(z))}{Y'(t_{\rm m}, r(z))}\,,\\
l^{\rm BAO}_{\rm T} (z) &=& s_{\rm p}(z_{\rm m})\frac{Y(t(z),r(z))}{Y(t_{\rm m}, r(z))}\,.
\end{eqnarray}
Finally, we can rewrite the radial and transverse proper BAO length scales as corresponding redshift and angular intervals at redshift $z$ in the LTB model.  The result is
\begin{eqnarray}
\Delta z (z) &=& (1+z)  l^{\rm BAO}_{\rm R} (z)  H_{\rm R} (t(z),r(z))\,,\label{eqn:Deltaz}\\
\Delta \theta (z) &=& \frac{ l^{\rm BAO}_{\rm T} (z)}{Y(t(z),r(z))}\,.\label{eqn:Deltath}
\end{eqnarray}

As stressed in ZMS, since these final quantities are directly measurable, they are model-independent, and hence can be unambiguously compared with data.  In addition, the radial scale $\Delta z(z)$ is expected to be a strong discriminator of void models.  This is because in the peripheral region of the void the radial expansion rate $H_{\rm R}(z)$ is generically suppressed, as dust from the void piles up.  This in turn leads to a suppression in the radial BAO length scale $l^{\rm BAO}_{\rm R}(z)$ in that region.  Therefore, according to Eq.~(\ref{eqn:Deltaz}), these two suppressions reinforce each other, resulting in a heavily suppressed radial redshift interval $\Delta z(z)$ in the void periphery.

In the literature on BAO observations, constraints are often expressed for the isotropized distance measure (see, e.g., \cite{Eisenstein:2005su}) \beq D_V(z) \equiv \ld(\fr{zd_{\rm A}^2(z)}{H(z)}\rd)^{1/3}. \eeq We do not attempt to estimate $D_V(z)$ for the LTB models for two reasons.  First, comparison with Eqs.~(\ref{eqn:Deltaz}) and (\ref{eqn:Deltath}) reveals that $D_V$ is related to a combination of radial and angular scales, but weighted more heavily to the angular scale.  The angular scale is determined in part by the angular diameter distance, $d_{\rm A}(z) = Y(z)$, which, for an LTB model that fits the SN data, must be close to that of the standard $\Lambda$ model.  Therefore we do not expect $D_V$ to be a strong discriminator of void models.  The second reason we avoid $D_V$ is that in its construction it assumes an FLRW background, and hence constraints on $D_V$ are not model-independent.

\subsubsection{Amplitude of local matter fluctuations, $\sigma_8$}
\label{sec:formalismsigma8}

   The evolution of structure on general LTB backgrounds is a very complex problem (see~\cite{ccf09}, and~\cite{z08} for an approximate approach).  Therefore, no predictions for the matter power spectrum in LTB models has yet been made.  This is unfortunate, since considerable information is available to us in the form of measurements of the power spectrum, and there are likely to be significant differences in the evolution of perturbations between LTB and FLRW models.

   However, in a certain region of the LTB spacetime we can in fact accurately calculate the evolution of perturbations, knowing only how the structures evolve on FLRW backgrounds.  In Ref.~\cite{z08} it was shown that the comoving matter density and expansion perturbations evolve according to
\begin{eqnarray}
\delta\dot\rho &=& -\theta\delta\rho - \rho\delta\theta,\label{eqn:denconslin}\\
\delta\dot\theta &=& -\fr{2}{3}\theta\delta\theta - 4\pi G\delta\rho - 3\Sigma\delta\Sigma,\label{eqn:dRaychaudlin}
\end{eqnarray}
where $\delta\Sigma$ is the shear perturbation associated with the structure.  These equations apply in general regions of the LTB spacetime.  However, they are not closed because of the $\delta\Sigma$ term, and that term will generally couple to perturbations in vector and tensor modes, leading to the immense complexity of the general problem.  But at the origin we have $\Sigma = 0$ by symmetry, and so the scalars $\delta\rho$ and $\delta\theta$ decouple from vectors and tensors.   Similarly, close enough to the origin that $\Sigma/\theta \ll 1$, we expect the decoupling to persist to good approximation.  Note that even though tensors generated where $\Sigma$ is large may propagate to the origin, near there they will not affect the evolution of the scalars at linear order because of the decoupling.

Therefore, where we can ignore the shear coupling term to a good approximation, Eqs.~(\ref{eqn:denconslin}) and (\ref{eqn:dRaychaudlin}) close, and their solutions are identical to those in the FLRW case.  Similarly, the {\em background} evolution ($\rho(t), \theta(t)$) of the LTB spacetime near the origin will match that of a spatially curved FLRW model.  The evolution of the scalar perturbations near the origin will then be the same as that in the FLRW model with the same background evolution as the origin of the LTB model.  It is simple to verify that for an LTB model with local Hubble rate $H_0$ and density parameter $\Omega_{\rm m}^{\rm loc}$ at proper time $t_0$ at the origin, the corresponding FLRW model would have the same Hubble rate and density parameter $\Omega_{\rm m} = \Omega_{\rm m}^{\rm loc}$ (and vanishing $\Lambda$) at the same $t_0$.  For the (underdense) void models considered here, this corresponds to an {\em open} FLRW model.

To calculate the matter power spectrum, we again use an effective model approach.  As before, the effective model will share the same early physics as the LTB model, but in this case the effective model will be the open FLRW model with parameters $H_0^{\rm open} = H_0$ and $\Omega_{\rm m}^{\rm open} = \Omega_{\rm m}^{\rm loc}(z = 0)$.  To determine the effective temperature, we first integrate down the LTB observer's light cone to redshift $z_{\rm m}$ to find $t_{\rm m} = t(z_{\rm m})$.  Then we evaluate the Hubble rate $H_{\rm c}$ and density parameter $\Omega_{\rm m,c}^{\rm loc}$ at coordinates $(t_{\rm m},0)$, i.e.\ at the centre at $t_{\rm m}$, using the exact LTB solution.  Then the temperature in the effective model can be calculated using the FLRW relation \beq T_0^{\rm open} = T_0(1 + z_{\rm m})\fr{H_0}{H_{\rm c}}\sqrt{\fr{1 - \Omega_{\rm m}^{\rm open}}{1 - \Omega_{\rm m,c}^{\rm loc}}}. \label{Topen} \eeq

To calculate the matter power spectrum (or its amplitude on scale $8/h_0$ Mpc, i.e.\ $\sigma_8$) near the origin in the LTB model, we then feed the effective parameters $H_0^{\rm open}$, $\Omega_{\rm m}^{\rm open}$, and $T_0^{\rm open}$ to \textsc{camb}.  Of course the result will depend on the {\em primordial} amplitude of perturbations, $A_{\rm S}$, which is an input parameter to \textsc{camb}, but this amplitude will be constrained when we fit to the CMB data.  Since we are assuming pure growing mode LTB models, the same primordial amplitude should apply to early times at the centre as at the LSS.

Note that in general we will have $T_0^{\rm open} \neq T_0$, i.e.\ the effective model will have a different radiation density today than the LTB model. However, the radiation density is so low today that it has negligible effect on the evolution of matter perturbations.  At early times, when radiation {\em is} important, our effective model shares the same physics as the void model, so the evolution of matter perturbations will be the same in the two models at all times.


\subsection{Method} \label{sec:method}

\subsubsection{$\Teds$, $\Heds_0$ covariance matrix}
\label{sec:methodcovarmat}

To calculate the CMB spectra for each void model we only need to know the effective parameters $\Teds$, $\Heds_0$ for the model (together with the primordial spectrum parameters, of course).  Since these effective parameters can be calculated very quickly, while the corresponding CMB spectra take considerably more time to generate, it is worthwhile to first investigate the posterior likelihood of these parameters (along with the other ``standard'' cosmological parameters) when applied to CMB data. If the posterior can accurately be characterized by a multi-variate Gaussian around the maximum likelihood, efficient searches of the void parameter space can be achieved by sampling from the covariance matrix of the effective parameters, rather than by re-fitting CMB data for each void model. 

We use \textsc{CosmoMC}~\cite{cosmomc} to generate Markov-Chain-Monte-Carlo (MCMC) chains to estimate confidence limits on the effective parameters. We use CMB data from WMAP7~\cite{Jarosik:2010iu,Komatsu:2010fb}, together with those from the ACBAR~\cite{acbar}, Boomerang~\cite{boom}, CBI~\cite{cbi}, and QUaD~\cite{quad} experiments, which observe to higher $\ell$. To compare the goodness-of-fit, we first fit a ``vanilla'' \lcdm\ model to our modified likelihood routine (ignoring polarization data at $\ell < 50$). The parameters in this model are: the baryon density, $\Omega_{\rm b}h_0^2$; cold dark matter density, $\Omega_{\rm c}h_0^2$; $h_0$; $z_{\rm re}$; $A_{\rm S}$; and $n_{\rm S}$.  We marginalize over the SZ amplitude $A_{\rm SZ}$, assuming the Komatsu and Seljak template~\cite{sztemplate}, and apply a prior of $z_{\rm re} > 8$. We do not include lensing of the CMB in our analysis, since at the moment we do not fully understand how LTB models modify the lensing signal.

For the effective EdS model, one degree of freedom is removed since $\Omega_{\rm m} = 1$, but we fit for $\Teds$ and marginalize over $A_{\rm ISW}$.  We apply a conservative prior of $  5 <  \zreeds < 15$, since the void model could have a lower effective $\zreeds$ than the {\em actual} $z_{\rm re}$.  In addition, we also investigate the case of spectral index running, such that the initial power spectrum is characterized by \beq \ln \mathcal{P}_{\rm S}(k)=\ln A_{\rm S} + \left(n_{\rm S}-1\right) \ln \left( k/k_0 \right) + \frac{n_{\rm run}}{2} \left[  \ln \left( k/k_0 \right) \right]^2\,. \label{eq:running}\eeq

\begin{figure*}
\centering \mbox{\resizebox{0.75\textwidth}{!}{\includegraphics[angle=0]{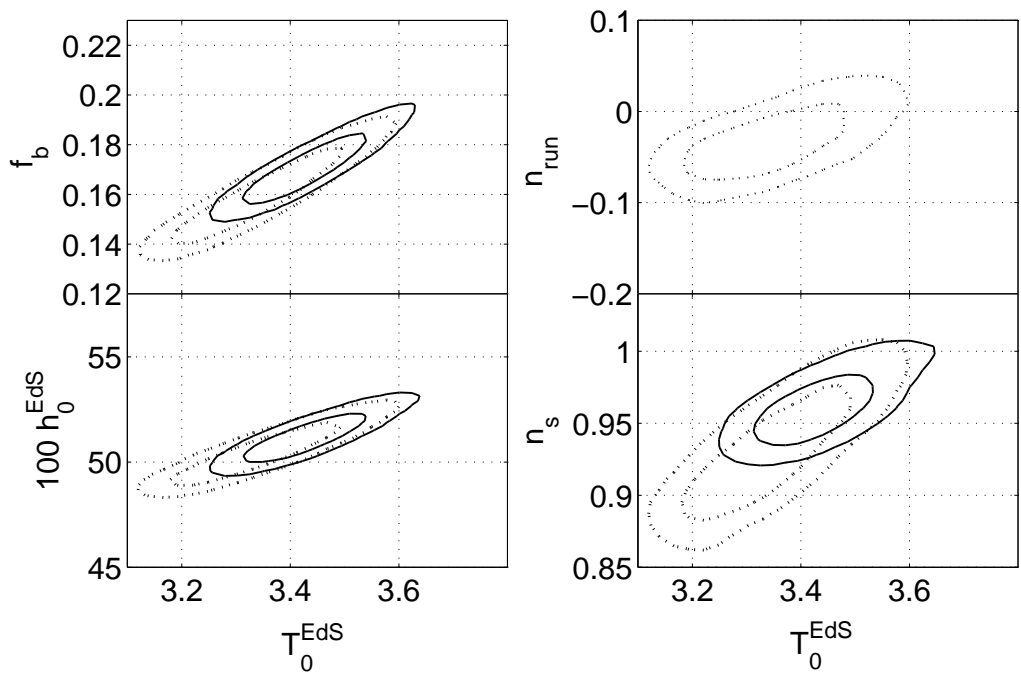}}} \caption{\label{fig:effective_eds_2d} Constraints on the effective EdS parameters from the CMB power spectrum data. Likelihood contours show the 68\% and 95\% confidence levels.  Dotted and solid contours are for models with and without spectral index running, respectively.}
\end{figure*}

In Fig.~\ref{fig:effective_eds_2d} we show a selection of 2D likelihoods from the MCMC chains for the effective EdS model parameter space. We define the best-fit likelihood relative to \lcdm\ as $\Delta\chi^2=-2\log({\mathcal L}_{\Lambda}/{\mathcal L}_{\rm EdS})$, such that positive values favour EdS. We find $\Delta\chi^2=0.6$ for the case of no running and 2.5 with running. The marginalized values of  $\Teds = \left( 3.43 \pm 0.08 \right) {\rm K}$ and $ h_0^{\rm EdS} =   0.512 \pm 0.008 $ (without running) are consistent with the best-fit WMAP7 \lcdm\ effective parameters calculated in Sec.~\ref{sec:formalism_CMBspectra}. We note that even with running, $\Teds > 3.17 \, {\rm K}$ at 2$\sigma$. This is important for void models, since an overdense outer shell (or positive spatial curvature) is required for $\Teds > T_0$, as highlighted in ZMS. Without such a feature, one would have to consider non-trivial modifications of the primordial spectrum to fit the CMB if $\Teds \approx T_0$. We return to this issue in Sec.~\ref{sec:power}.

\subsubsection{Void profile}

We define the initial radial density profile $\delta (t_i,r)$ of the LTB model in three ways. In the first, we repeat the analysis carried out in ZMS with updated cosmological data. Here, we fit a three-point cubic spline to the initial density fluctuation $\delta_j \equiv \delta(t_i,r_j)$, where  $j = 1$, $2$, $3$. We fix $r_1 = 0$ and enforce the void to be smooth at the origin and to smoothly approach EdS at large $r$ by setting $\delta'(r_1) = \delta'(r_3) = \delta_3 = 0$.  There are thus a total of four free profile parameters, and we find that additional spline points do not significantly improve the fit to the data.

In ZMS it was noted that a large overdense shell outside of the central void is required to fit the CMB power spectrum in asymptotically flat models. For the three-point spline parameterization, this shell will smoothly transition from the underdensity near the void centre. For this reason, we also find it convenient to decompose the initial density perturbation into the linear combination of profiles\beq
 \delta (t_i,r) = \delta_{\rm U} (r)  +  \delta_{\rm O} (r)\,,
\eeq where $\delta_{\rm U}$ represents the initial central underdensity and $\delta_{\rm O}$ the initial overdense shell. These profiles are given by the polynomial functions
\begin{widetext}
\beq \delta_{\rm U} (r) = \ld\{ \begin{array}{ll} \displaystyle \delta_0 \ld[1 - 3\ld(\fr{r}{r_0}\rd)^2
      + 2 \ld(\fr{r}{r_0}\rd)^3\rd], & \quad r \le r_0\,,\\
\displaystyle 0, & \quad r > r_0\,, \label{eqn:polyvoid}
\end{array} \rd.
\eeq \beq \delta_{\rm O} (r) = \ld\{ \begin{array}{ll}
\displaystyle 0, & \quad r \le r_1 -\Delta r \,, \\
\displaystyle \delta_1 \ld[ 3\ld(\fr{r+\Delta r  - r_1}{\Delta r }\rd)^2
      - 2 \ld(\fr{r+\Delta r - r_1}{\Delta r}\rd)^3\rd], & \quad r_1 - \Delta r < r \le r_1\,,\\
\displaystyle \delta_1 \ld[ 1- 3\ld(\fr{r  - r_1}{\Delta r }\rd)^2
      + 2 \ld(\fr{r - r_1}{\Delta r}\rd)^3\rd], & \quad r_1 < r \le r_1 + \Delta r \,,\\
\displaystyle 0, & \quad r > r_1 + \Delta r \,,
\end{array} \rd.
\eeq
\end{widetext}
such that $\delta  (t_i,r) $ smoothly matches to an EdS region between the void and shell. There are a total of five free parameters: the width and depth of the void, $r_0$ and $\delta_0$; and the width, height, and position of the shell, $\Delta r$, $\delta_1$, and $r_1$. However, the fit to the CMB power spectrum, requiring $\Teds  \approx 3.4  \,{\rm K}$, is primarily a function of the integrated shell density, so in our analysis we arbitrarily fix the shell position and width and vary only the height. We choose the position so that the shell lies beyond the SN data ($z>1.5$).

This ``polynomial + shell'' decomposition is useful as it separates the two main physical effects: the fit of the SN data to the central void; and the fit of the CMB to the outer shell. We find that some additional constraints (such as the Compton $y$-distortion and radial BAO scale) are sensitive to the details of the shell, which can more easily be investigated without the constraints the spline imposes on the shell position. 
 
In ZMS, we made the distinction between ``constrained'' profiles, for which $\mathcal{P} \equiv \int\delta\rho(t_i,r) r^2 dr \leq 0$, and ``unconstrained'', which are free. For the polynomial + shell model, a perfectly compensated void with $\mathcal{P} = 0$ can be obtained by setting \beq
 \delta_1 =  \frac{-\delta_0 r_0^3}{\Delta r \ld( 15 r_1^2 +2 \Delta r^2 \rd)}\,.
\eeq We do not enforce this condition.  Therefore, although our early-time profiles always satisfy $\delta\rho(t_i,r) = 0$ for $r$ lying between void and shell or exceeding some maximum radius, for the general uncompensated case the spacetime will depart from EdS at late times in these regions, due to the gravitational effect of the inhomogeneities.

Finally, we also consider voids embedded in a spatially curved FLRW background. Here, the curvature required to fit the CMB is distributed homogeneously rather than confined to a large overdense shell. In this case we set \beq
 \delta (t_i,r) = \delta_{\rm U} (r)  +  \delta_{1} \,,
\eeq where the central underdensity $\delta_{\rm U} (r)$ is given by the polynomial function~(\ref{eqn:polyvoid}) and $\delta_1$ is a constant.  For this ``polynomial + curvature'' parameterization, the curvature function at large $r$ then takes the FLRW form $K(r) \propto r^2$.  This corresponds to a spatially constant density parameter $\Omega_{\rm m}^{\rm loc}$ in this asymptotic region.  Note, however, that there is ambiguity in choosing the {\em time} at which to specify such a parameter in a globally non-FLRW background, and observations do not directly constrain the asymptotic curvature.


\subsubsection{Data fitting}

Measurements of the apparent magnitudes of Type Ia supernovae are important for constraining the void profile. In this work we use the recent Union2 compilation~\cite{Amanullah:2010vv} of 557 SNe in the range $z = 0.015$--$1.4$, which is nearly double the number of SNe used in ZMS, with improved rejection of outliers. The residual $\mu$ between the apparent and absolute magnitudes, $m$ and $M$, is defined as \beq \mu \equiv m - M = 5 \log_{10} \left[ \frac{d_{\rm L}}{\rm Mpc} \right] + 25\,. \eeq In our fitting we adopt the standard procedure of marginalizing over the unknown  absolute magnitude.


For the CMB we tested the spline profile both by sampling from the covariance matrix of the effective parameters and by re-fitting the CMB data each time.  The results in both cases were similar, but the former method was much faster, with combined CMB + SN constraints taking only a few hours to obtain. In our subsequent analysis we therefore use the covariance sampling.

As in ZMS, we applied a conservative prior of $\Omega_{\rm m}^{\rm loc} > 0.1$ at the void centre. This is consistent with estimates of the minimum matter density in the local Universe~\cite{Carlberg:1997zp,Fukugita:1997bi}. We investigate models with lower $\Omega_{\rm m}^{\rm loc} (z=0)$ in Sec.~\ref{sec:deep}.


\subsection{Results} \label{sec:results}
\subsubsection{Basic parameters} \label{sec:basicparams}

\begin{figure}
[!t] \centering \mbox{\resizebox{0.48\textwidth}{!}{\includegraphics[angle=0]{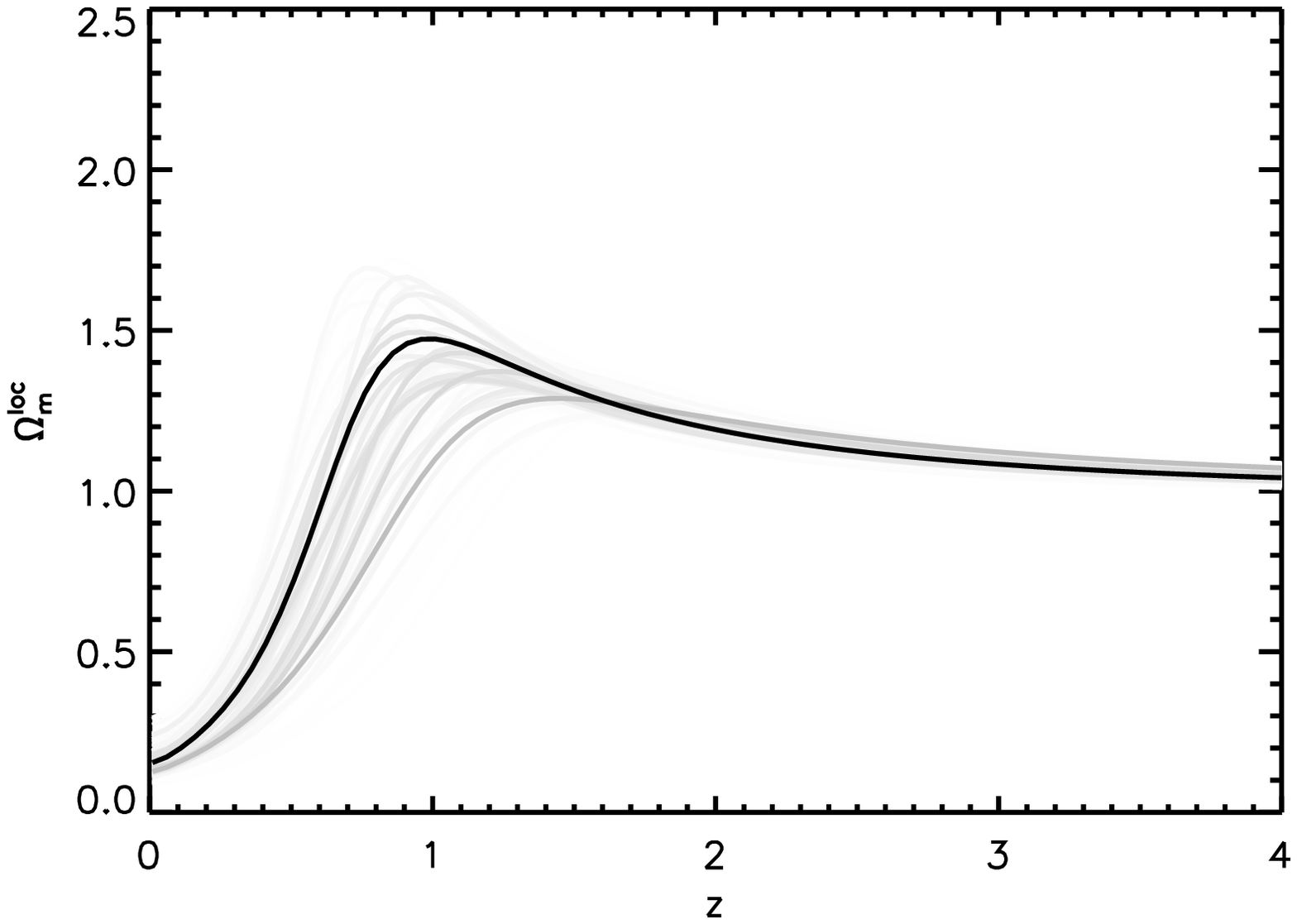}}} \mbox{\resizebox{0.48\textwidth}{!}{\includegraphics[angle=0]{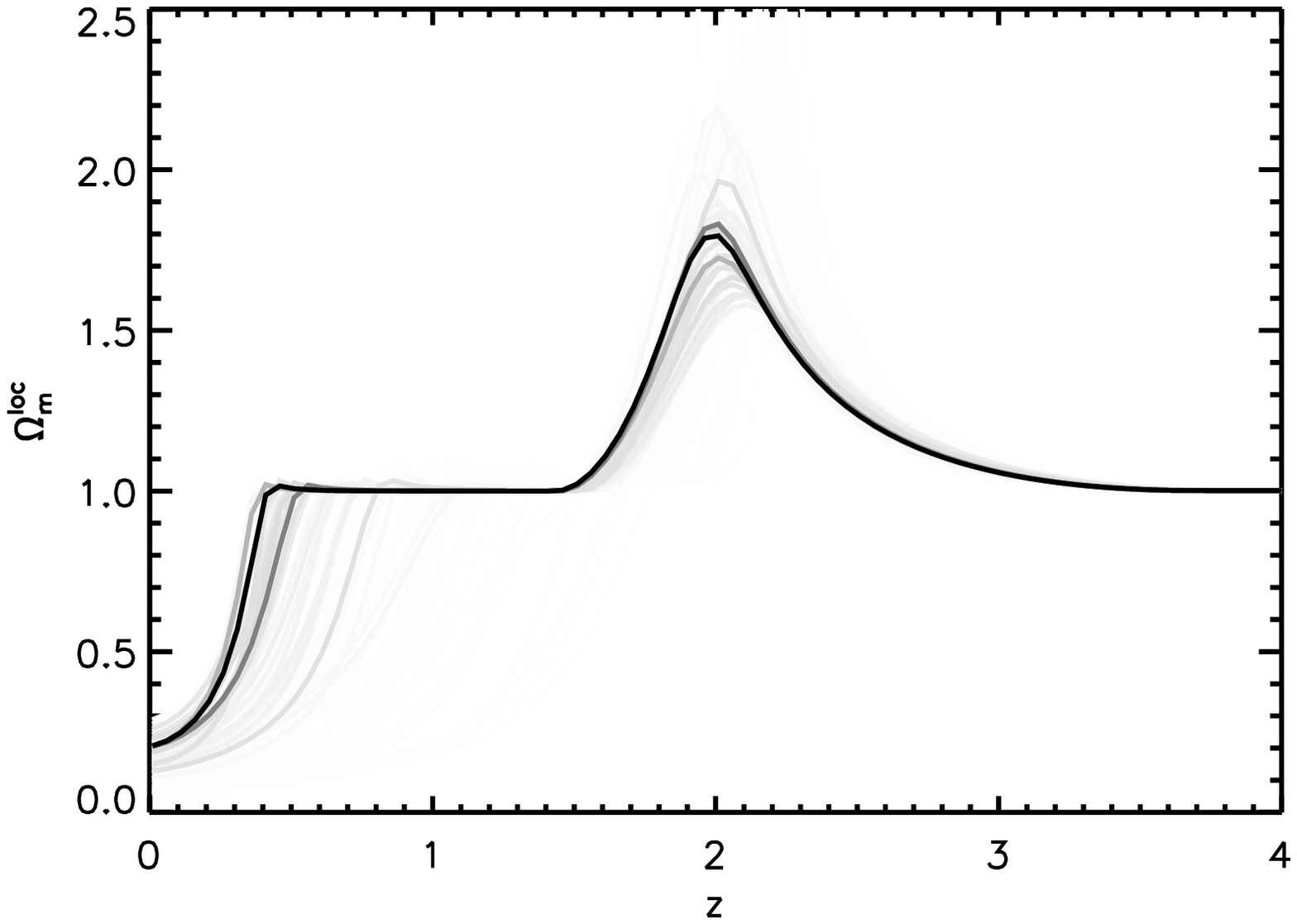}}}  \mbox{\resizebox{0.48\textwidth}{!}{\includegraphics[angle=0]{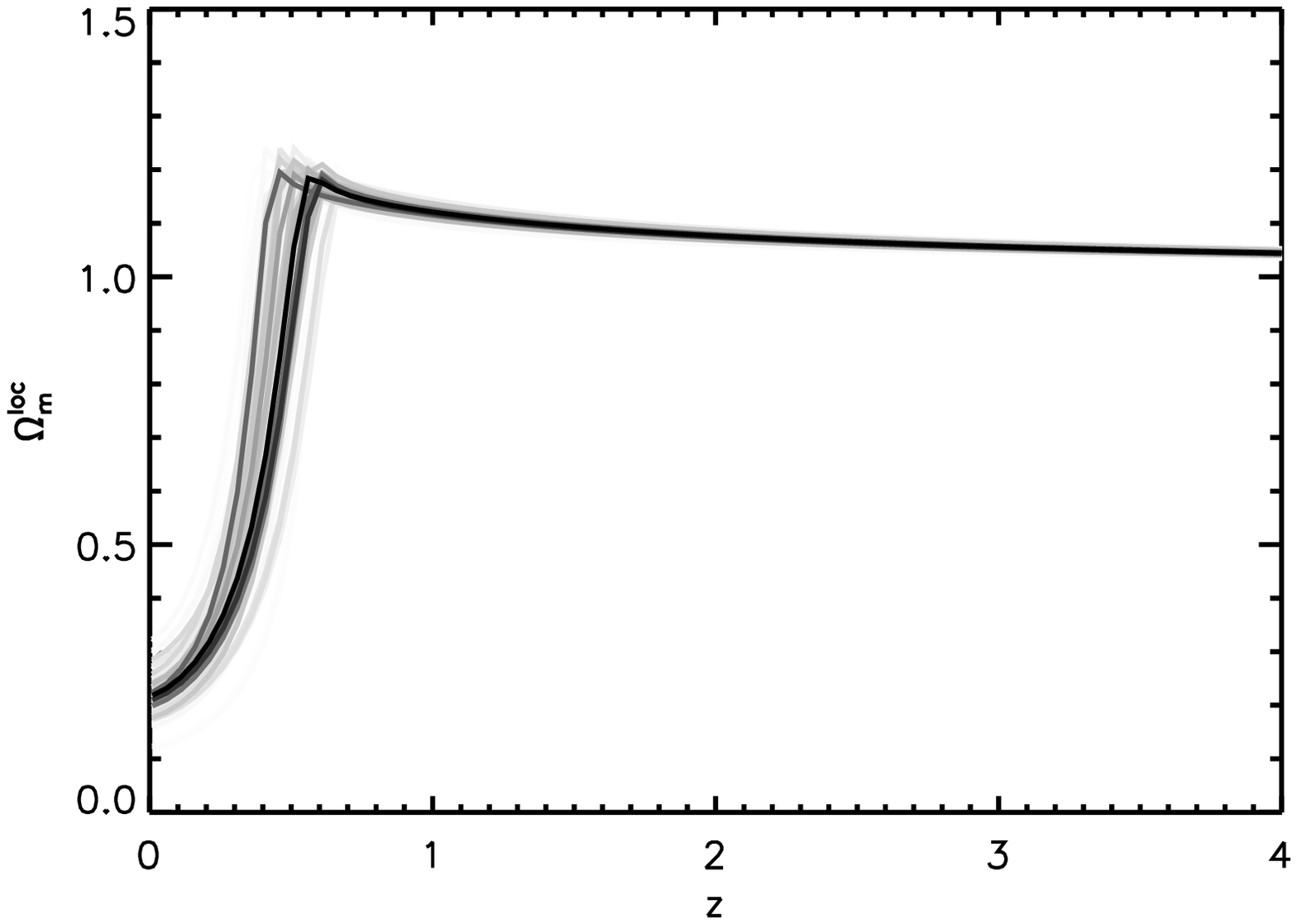}}} \caption{\label{fig:omega} Local density parameter $\Omega^{\rm loc}_{\rm m}$ versus redshift $z$ for the spline (top panel), polynomial + shell (middle panel), and polynomial + curvature (bottom panel) profiles.  The grayscale level indicates the relative likelihood in the fit to the CMB + SN data.}
\end{figure}

In Fig.~\ref{fig:omega} we plot the local density parameter $\Omega_{\rm m}^{\rm loc}  (z)$ along the observer's past light cone for several spline, polynomial + shell, and polynomial + curvature profiles. These are sampled from the MCMC chains, using CMB + SN data, with the grayscale level indicating the relative likelihood. Each curve has a similar shape for $z \alt 0.5$, since the profile in this region is constrained by the SN data, although the underdensity of the spline extends to slightly higher redshifts.  In the polynomial + shell model, the shell, located at $z \sim 2$, is quite separate from the void, but for the spline model the shell smoothly matches onto the void. The presence of positive curvature is highly significant, with shell amplitude $\delta_1 > 0$ at $8\sigma$ for polynomial + shell. Similarly, the polynomial + curvature model has $\delta_1 > 0$ at $7\sigma$, and density parameter $\Omega_{\rm m}^{\rm loc} = 1.146 \pm 0.013$ evaluated at $(t_0,r_{\rm LSS})$.  Each model fits the 557 SNe quite well, the best-fit $\chi^2_{\rm SNe }$ being $534$ for the spline, $539$ for the polynomial + shell, and $541$ for the polynomial + curvature. 

A more detailed model comparison between the three cases is somewhat hard to quantify, since error bars of the SNe are tuned to give a $\chi^2$ per degree of freedom of order unity for \lcdm. We find a corresponding $\chi^2_{\rm SNe } = 531$ for \lcdm\ using joint CMB + SN data, with the fit to the CMB component similar in the void and \lcdm\ models. These results indicate that, although the extra freedom of the spline improves the SN fit over the polynomial,  there is still a slight preference for \lcdm. This could be somewhat alleviated by introducing more control points to the spline, but the improvement in $\Delta \chi^2$ is not significant for the additional degrees of freedom. Even at this stage it is unlikely that a model comparison would significantly favour \lcdm\ over a three-point spline.

\begin{figure*}
\centering 
\mbox{\resizebox{\textwidth}{!}{\includegraphics[angle=0]{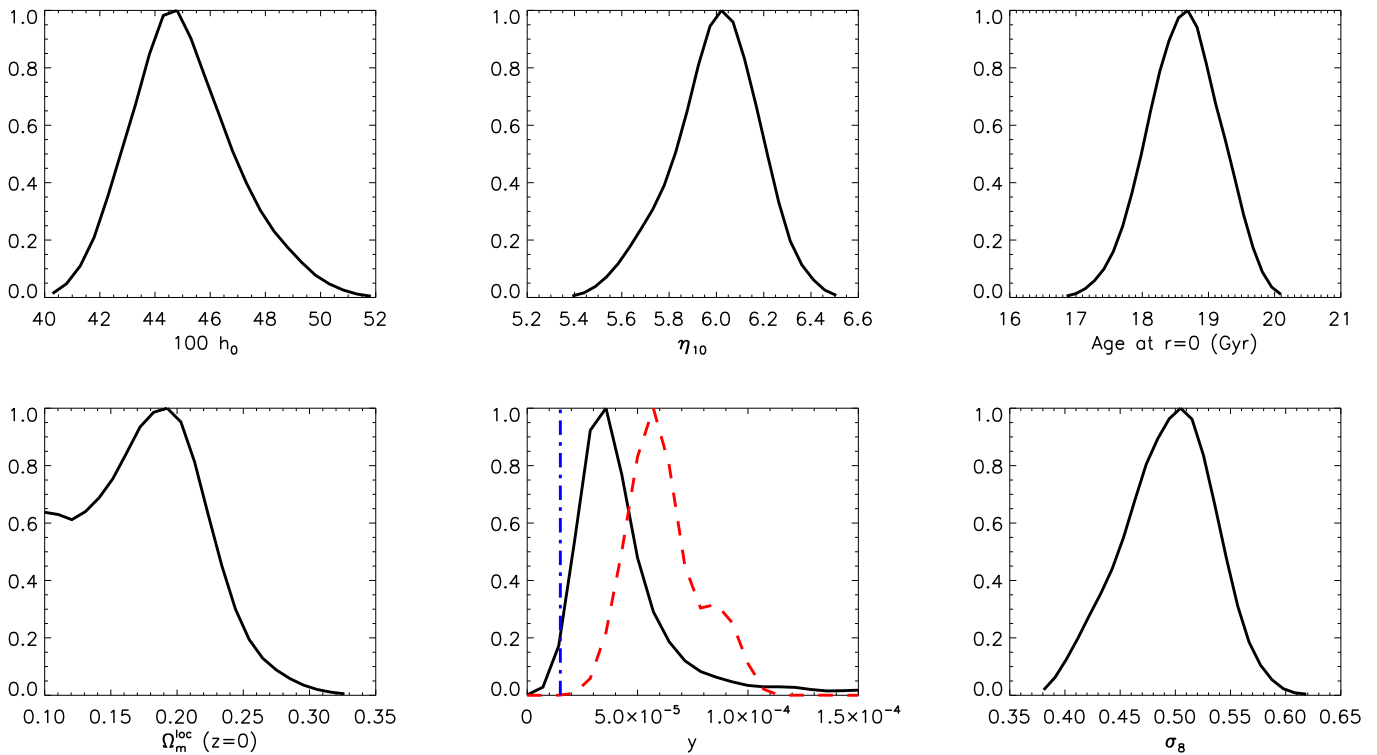}}} \caption{\label{fig:1d}  Selection of marginalized likelihoods for the polynomial + shell profile using the CMB + SN data. For the Compton $y$-distortion the dashed curve shows the addition of radial BAO data, while the vertical line is the COBE $2\sigma$ upper limit~\cite{Fixsen:1996nj}.}
\end{figure*}

There are, fortunately, several other parameters which are good discriminators between void models and \lcdm. We show a selection of marginalized likelihoods for the polynomial + shell profile in Fig.~\ref{fig:1d}, with similar results for the spline and polynomial + cuvature (the exception for curvature being the Compton $y$-distortion, which we discuss later). A general feature of void models is that they require an extremely low local Hubble rate, as first noted in ZMS. We find $h_0 = 0.45  \pm 0.02$ for both the spline and polynomial + shell, and $h_0 = 0.47  \pm 0.02$ for polynomial + curvature. As mentioned in ZMS, {\em local} measurements of $h_0$ (i.e.\ those independent of the cosmological model) are generally much higher (see, e.g., Ref.~\cite{Freedman:2010xv} for a recent review).  For example, recent measurements of Cepheids in SN hosts at $z < 0.1$ give $h_0 = 0.738 \pm 0.024$~\cite{riessetal11}, which rules out the void models at high significance.  These distances are small enough that the dynamics of the void would not significantly affect the local distance ladder.  Even the lower local determinations of $h_0$ in Ref.~\cite{Tammann:2007ge}, who found $h_0 = 0.623 \pm 0.05$, are strongly at odds with the predictions of the void models.  (See~\cite{Riess:2009pu} for a discussion of the reasons for this lower $h_0$ measurement.)

\begin{figure}
\centering \mbox{\resizebox{0.48\textwidth}{!}{\includegraphics[angle=0]{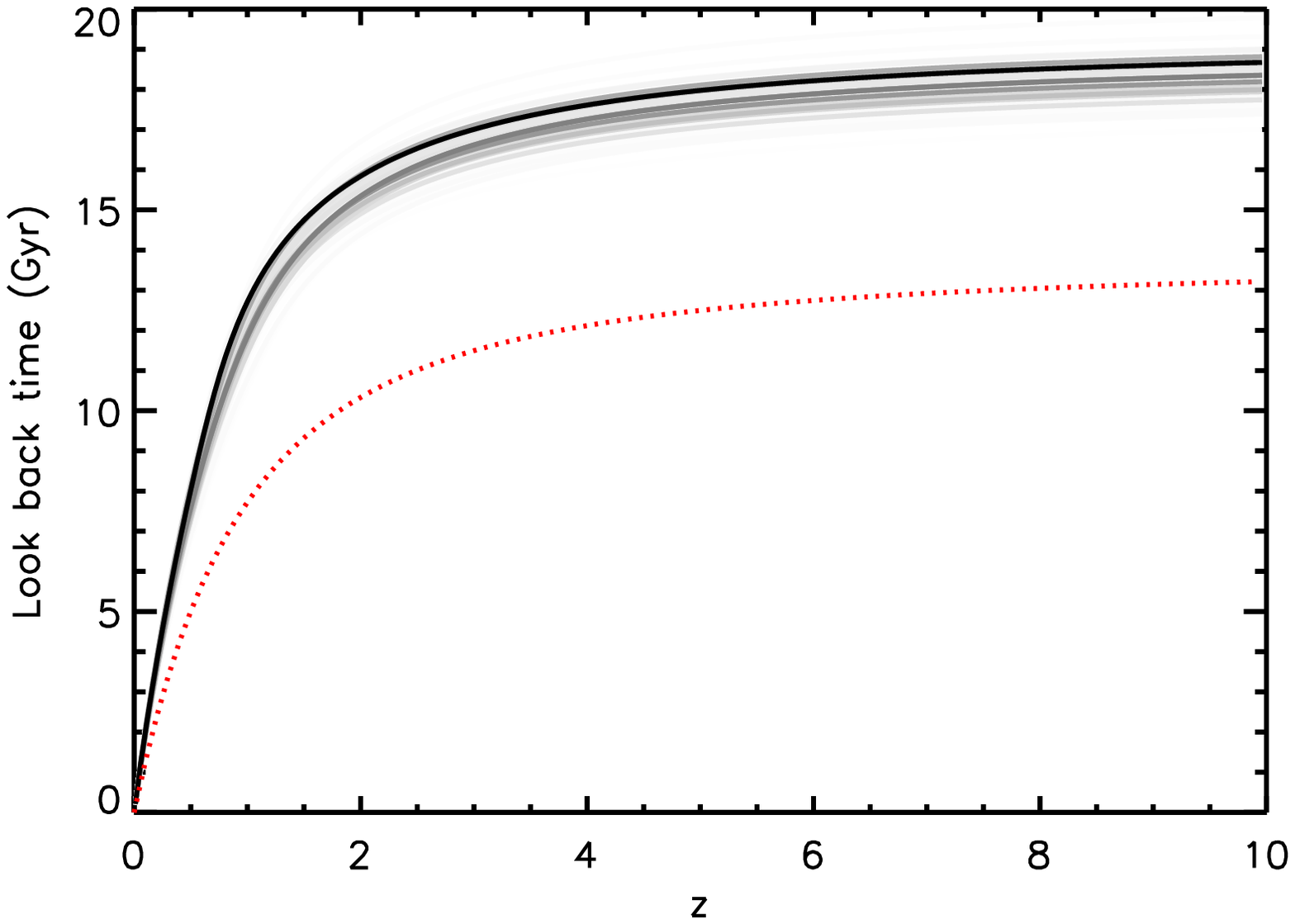}}} \mbox{\resizebox{0.48\textwidth}{!}{\includegraphics[angle=0]{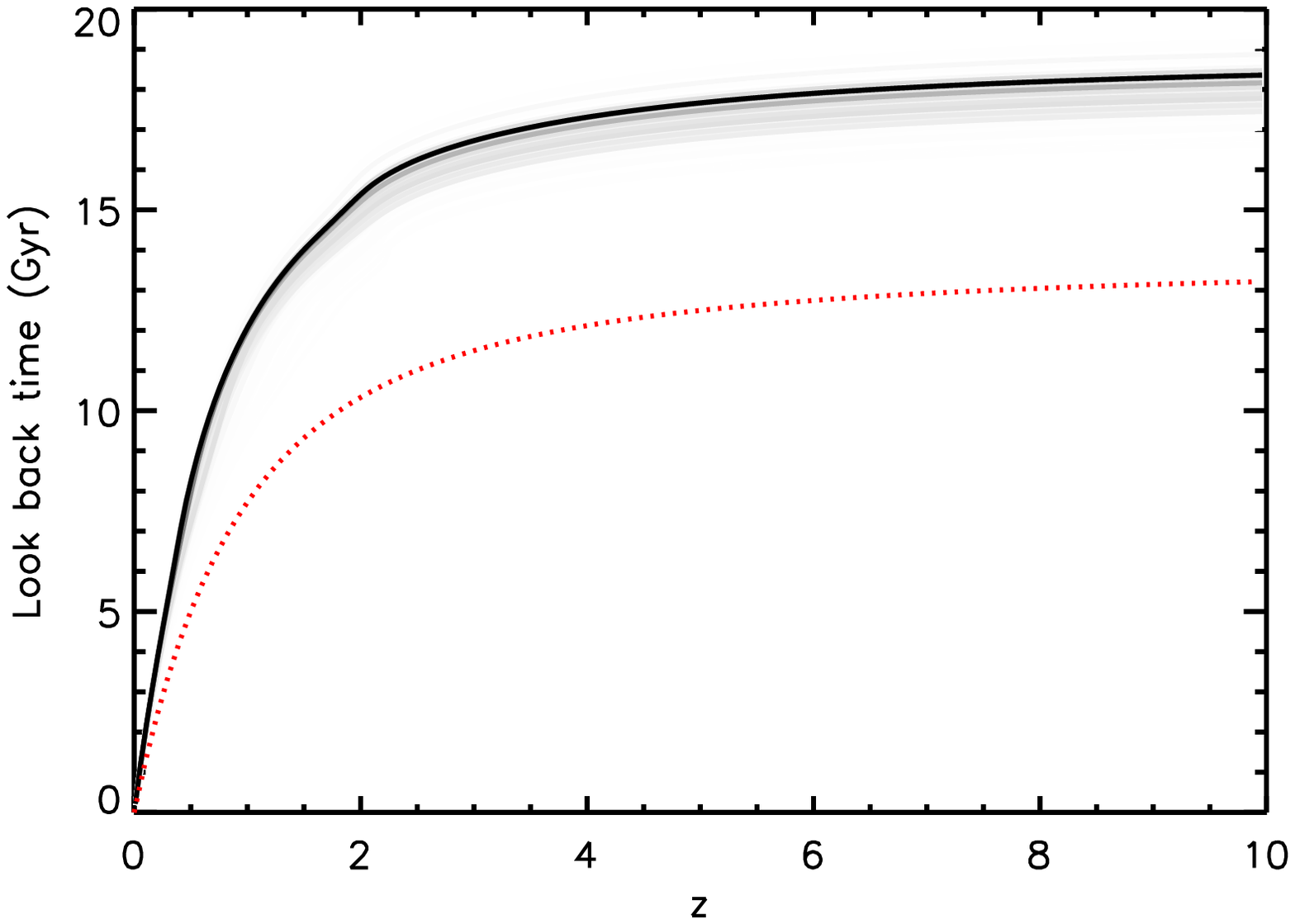}}} \caption{\label{fig:age} Look-back time versus $z$ for the spline (top panel) and polynomial + shell (bottom panel) profiles.  The polynomial + curvature case is very similar. The dotted curve is the look-back time for the WMAP7 \lcdm\ cosmology.}
\end{figure}

The low $h_0$ also contributes to what we refer to as the ``old age'' problem for voids. In Fig.~\ref{fig:age} we plot the look-back time for samples from the MCMC chains. These models are significantly older than \lcdm, whose age is $13.7 \pm 0.1 \,{\rm Gyr}$~\cite{Komatsu:2010fb}. We find ages $18.8 \pm 0.5  \,{\rm Gyr}$ for the spline, $18.6 \pm 0.5 \,{\rm Gyr}$ for the polynomial + shell, and $17.6 \pm 0.4 \,{\rm Gyr}$ for the polynomial + curvature. The look-back time of void models is already equal to the age of \lcdm\ by $z \approx 1$.  This could pose another serious problem for voids, since the existence of observed structures at redshifts $z \gtrsim 5$, which corresponds to look-back times of $17$--$18$ Gyr in void models, could be hard to reconcile with the ages of the oldest known objects in the Universe, i.e.\ globular clusters, which are consistent with \lcdm~\cite{Krauss:2003em}.  Although it is difficult to provide an {\em upper} limit to the age of the Universe from local observations, it is clear that there is a serious problem with void models---in such a model we should be observing the epoch of formation of the oldest known stars at $z \approx 1$!

\subsubsection{Big bang nucleosynthesis}

Observations of elemental abundances provide a constraint on the baryon-photon number density ratio, $\eta \equiv n_{\rm b}/n_{\gamma}$, via our theoretical understanding of the epoch of big bang nucleosynthesis (BBN)~\cite{Steigman:2007xt}.  For FLRW models $\eta$ is constant in time, but this is not generally true for LTB.  (We assume, however, that $\eta$ is spatially constant at early times.)  Therefore we should explicitly calculate $\eta$ for our void model near the LSS (or early enough that curvature is unimportant).  Since the effective EdS model has the same early physics at the LSS as our void model, we can readily show that the early-time value for voids is \beq \eta_{10} \equiv 10^{10} \eta = 273.9 f_{\rm b} \left( \frac{T_0}{T_0^{\rm EdS}} \right)^3 \left( h_0^{\rm EdS} \right)^2. \eeq Recent estimates of the deuterium abundance in metal-poor damped Lyman-$\alpha$ systems, for example, imply that $\eta_{10}=5.8 \pm 0.3$~\cite{BBN}. From our MCMC chains we obtain $ \eta_{10} = 6.1 \pm 0.2 $ for the spline and $ 6.0 \pm 0.2$  for the polynomial + shell and polynomial + curvature. These are entirely consistent with the BBN constraint, which is not surprising since the physics at the LSS in the void model is the same as that of standard \lcdm, and observations of the CMB imply a similarly consistent value of $\eta$ within the standard \lcdm\ framework.


\subsubsection{Dipole}
\label{sec:resultsdipole}

\begin{figure}
\centering \mbox{\resizebox{0.48\textwidth}{!}{\includegraphics[angle=0]{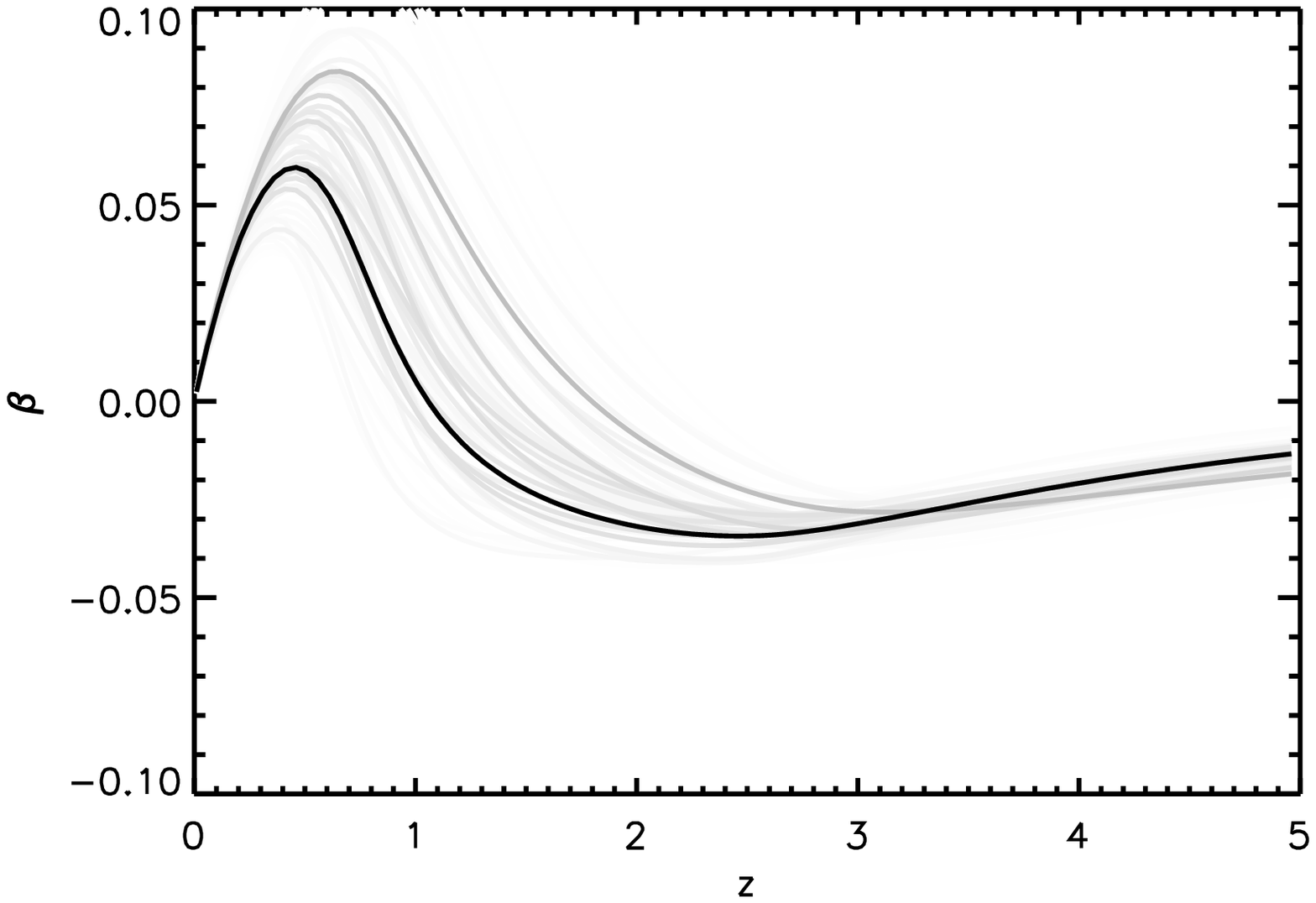}}} \mbox{\resizebox{0.48\textwidth}{!}{\includegraphics[angle=0]{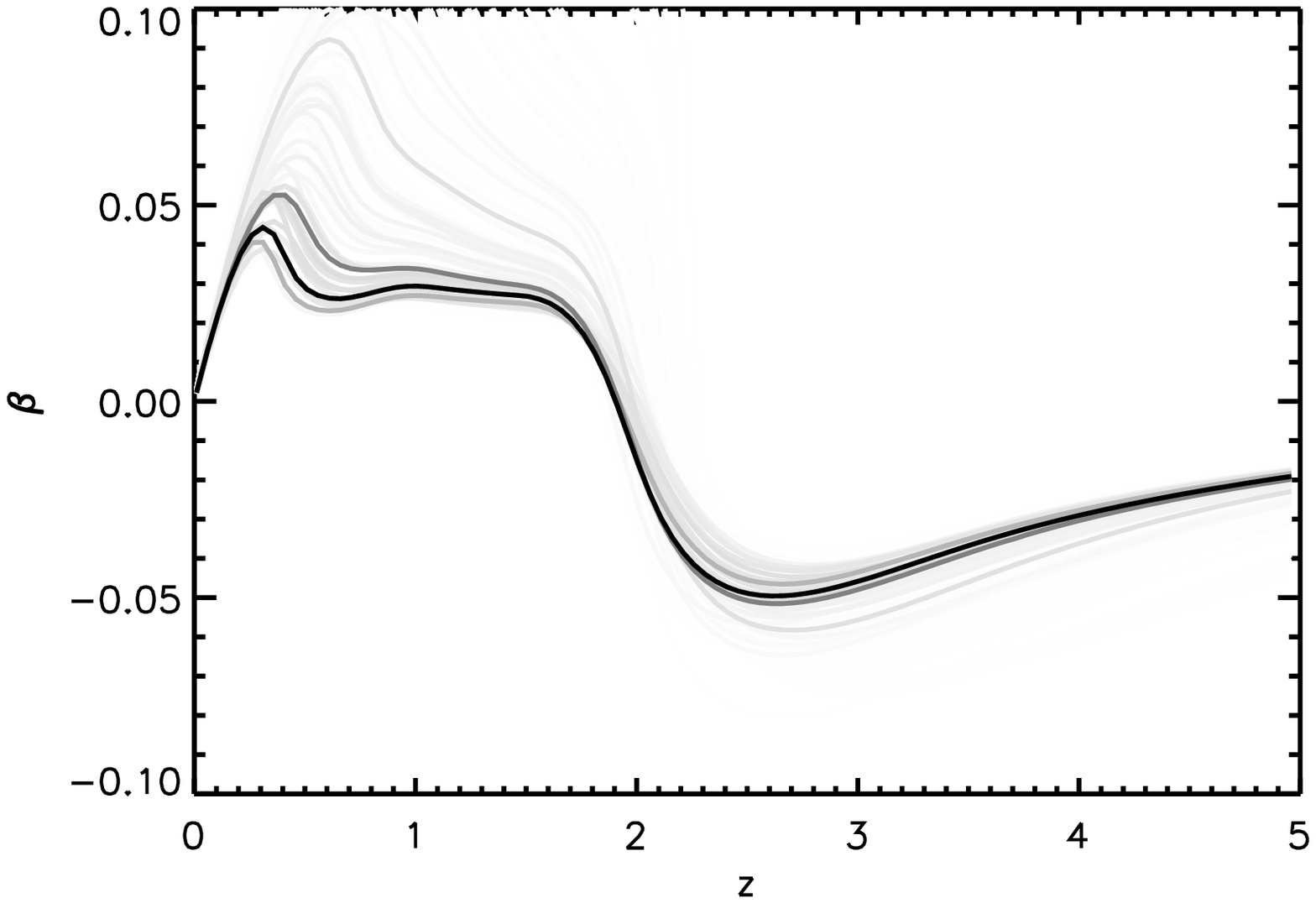}}} \mbox{\resizebox{0.48\textwidth}{!}{\includegraphics[angle=0]{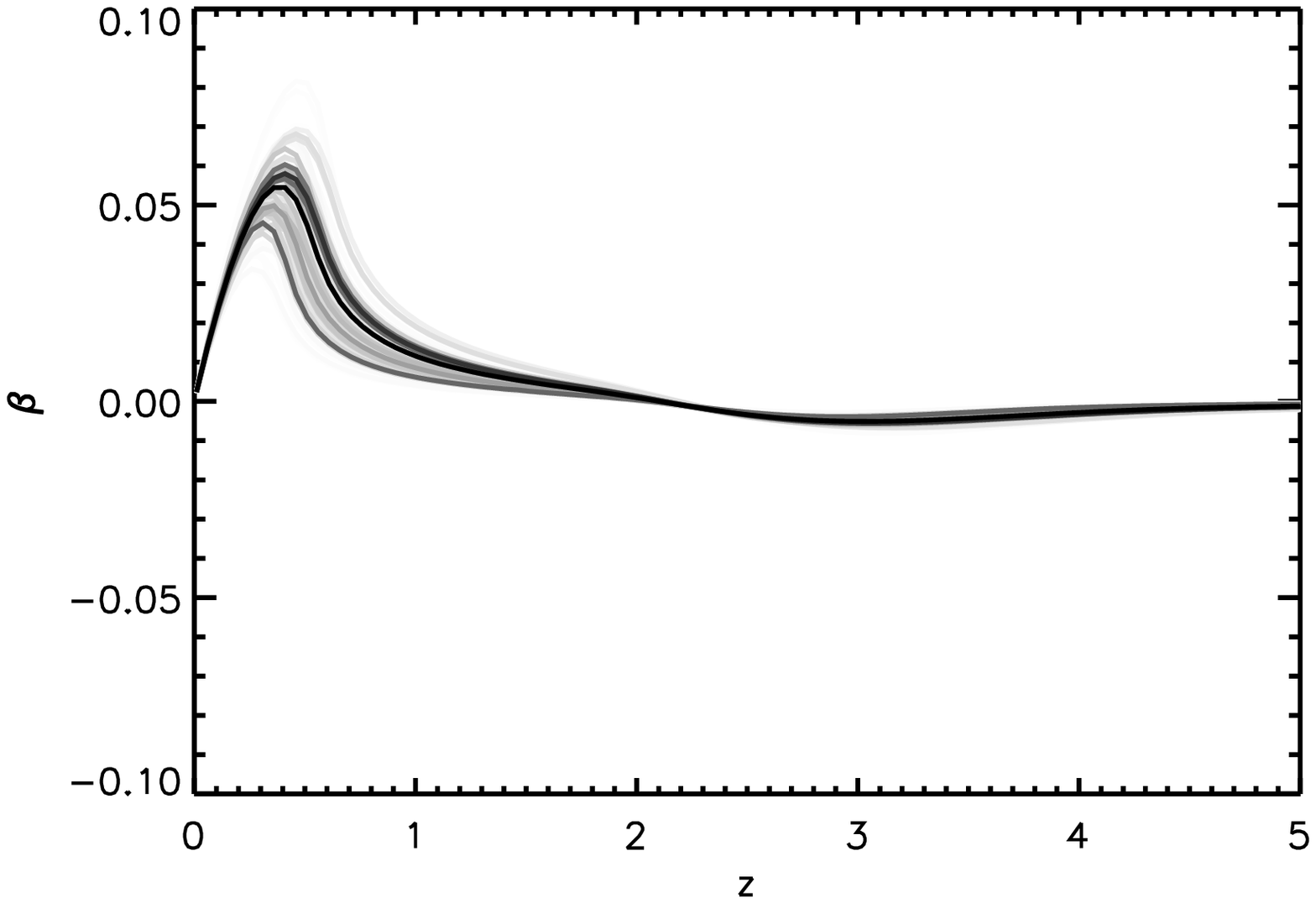}}} \caption{\label{fig:beta} Dipole $\beta$ versus $z$ for the spline (top panel), polynomial + shell (middle panel), and polynomial + curvature (bottom panel) profiles. }
\end{figure}

In Fig.~\ref{fig:beta} we show the dipole $\beta(z)$ generated from our MCMC samples. The peak values  are significantly higher than our local motion with respect to the CMB, for which $\beta \sim 10^{-3}$~\cite{Scott:2010yx}. For small distances from the centre of the void, one can write $\beta$ as a linear function of the proper distance $d$. For the spline and polynomial (shell and curvature) models we find $\beta =  \left(3.8 \pm 0.4 \right) \times 10^{-5} \,d\,{\rm Mpc}^{-1}$, $\left( 3.6 \pm 0.4 \right) \times 10^{-5} \, d\,{\rm Mpc}^{-1}$, and $\left( 3.7 \pm 0.4 \right) \times 10^{-5} \, d\,{\rm Mpc}^{-1}$, respectively. This leads to a nominal constraint on how close we must be to the centre of the void of $\sim$$30 \, {\rm Mpc}$. However, a more detailed analysis including the stochastic (i.e.\ peculiar velocity) and LTB background components leads to weaker constraints~\cite{Foreman:2010uj}.

The dipole plots show several interesting features, which are more transparent in the polynomial + shell case. For $z \lesssim 2$ we have $z_{\rm out} > z_{\rm in}$, so the outgoing photons (those passing though the void centre) receive more redshift than the incoming ones. The initial peak at low $z$ results from the {\em higher} radial expansion rate inside the central void---outgoing photons spend more time in the void so are redshifted more. There is then an extended plateau to $z \sim 2$, even though $\Omega_{\rm m}^{\rm loc} \sim 1$ in this region (see Fig.~\ref{fig:omega}). This plateau is due to the Sach-Wolfe (SW) effect---outgoing photons originating from the LSS are in the region of the overdense shell (on the opposite side of the void), and so redshift out of the gravitational potential. Moving further out in redshift, at $z \gtrsim 2$ we have $z_{\rm in} > z_{\rm out}$. This region results from the {\em lower} expansion rate in the shell, with outgoing photons spending more time in the shell. There is also a SW effect from the central void around $z \sim 3$, but this is subdominant compared to the blueshifting induced by the shell. 

The general features are the same for the spline profile, but the extended SW plateau cannot be resolved.  Importantly, for the polynomial + curvature model there is no dipole induced by the homogenous background curvature. Here the main feature is the peak at low $z$, with the smaller SW signal from the central void also visible around $z \sim 3$.


\subsubsection{Compton $y$-distortion}

The calculation of the $y$-distortion in void models is complicated by several details. First, the integrand in~(\ref{eqn:ydipapprox}) is dependent on $\beta(z)$ and hence (for the case of zero background curvature) the details of the shell. The position of the shell is not constrained by CMB data, since the fit to the CMB is only dependent on the integrated shell density. Second, the integration limit in~(\ref{eqn:ydipapprox}) is the reionization redshift $z_{\rm re}$, which is uncertain in our MCMC analysis. Finally, calculation of $y$ using only the dipole calculated from radial geodesics becomes less accurate in the peripheral region of the void, leading to an error on the value of $y$.

We address these issues by first considering the best-fit polynomial + shell model. This has $y = 3.0 \times 10^{-5}$, which is independent (to the first decimal place) of the choice of $z_{\rm re}$ in the range 8 to 15. The best bound on the $y$-distortion comes from the FIRAS instrument on COBE, which gives $y < 1.5 \times 10^{-5}$ at $2\sigma$ confidence~\cite{Fixsen:1996nj}. It is clear that this limit {\em could} provide important constraints on void models. However, on removal of the outer shell (by manually setting $\delta_1 = 0$) we find $y = 0.1 \times 10^{-5}$, again independent of $z_{\rm re}$, which is a significantly weaker constraint.  This then suggests that by pushing the shell sufficiently far outwards (i.e.\ past $z_{\rm re}$), we may be able to reduce $y$ to satisfy the COBE constraint while still fitting the CMB (by adjusting the shell's amplitude---recall, however, that we must maintain $K(r) < 1$ and avoid shell crossings). Alternatively, one can evade the COBE constraint by considering models with non-zero background curvature. Here, the integrated $\beta (z)$ is much smaller, as shown in Fig.~\ref{fig:beta}, with a best fit of $y = 0.2  \times 10^{-5}$. From our MCMC chains in this case, we find $y < 2.2 \times 10^{-5}$ at $2\sigma$ confidence, i.e.\ the majority of samples are within the COBE limit. 

We investigate the polynomial + shell model further by computing the $y$-distortion for a grid of shell profiles (constrained to have the $\Teds$ and $\Heds_0$ values required to fit the CMB) for the best-fit polynomial central void, but we were unable to find a profile which substantially reduced $y$. Shells close to the LSS must have a small comoving thickness, and these violate $K(r) > 1$. The most distant shell which still fits the CMB is located at $z \sim 20$, but the dipole for this model has an extended SW plateau to $z \sim 10$ and the resulting $y$-distortion is similar to the nominal best fit. The minimal $y$ is found for a shell located at $z \sim 5$ with a thickness of $\delta z \sim 4$. This has $y = 1.0 \times 10^{-5}$ with $z_{\rm re} = 8$, and  $y = 1.5 \times 10^{-5}$ with $z_{\rm re} = 15$. A detailed calculation for this model, using non-radial geodesics~\cite{Zibin_bang}, finds the actual value of $y$ to be around $10$\% higher than the dipole-approximated value.

With these uncertainties in mind, the $y$ values from the MCMC chains of models with shells, which we show in Fig.~\ref{fig:1d}, could be a factor of 2--3 lower if the shell position were carefully tuned. The COBE result then puts models in the tail of the $y$ distribution under severe tension. These models correspond to wider, deeper voids with a larger dipole term, which can be seen in Figs.~\ref{fig:omega} and \ref{fig:beta}.

Finally, note that in Ref.~\cite{Caldwell:2007yu} it was found that very large voids {\em without} overdense shells were already in conflict with the COBE $y$-distortion constraint.  Thus it may seem surprising that our models with large shells, and their consequent dipole contribution (cf.\ Fig.~\ref{fig:beta}), only marginally exceed the COBE limit.  Part of the explanation appears to be that Ref.~\cite{Caldwell:2007yu} imposed $h_0 = 0.73$, while we have found that much lower values of $h_0$ are required to fit the CMB.  Our lower $h_0$ implies a lower density of scatterers and hence lower $y$-distortion. In addition, Ref.~\cite{Caldwell:2007yu} employed a simplified {\em linear} ``Hubble bubble'' model, rather than performing a rigorous LTB treatment.


\subsubsection{Radial BAO}

\begin{figure}
\centering \mbox{\resizebox{0.48\textwidth}{!}{\includegraphics[angle=0]{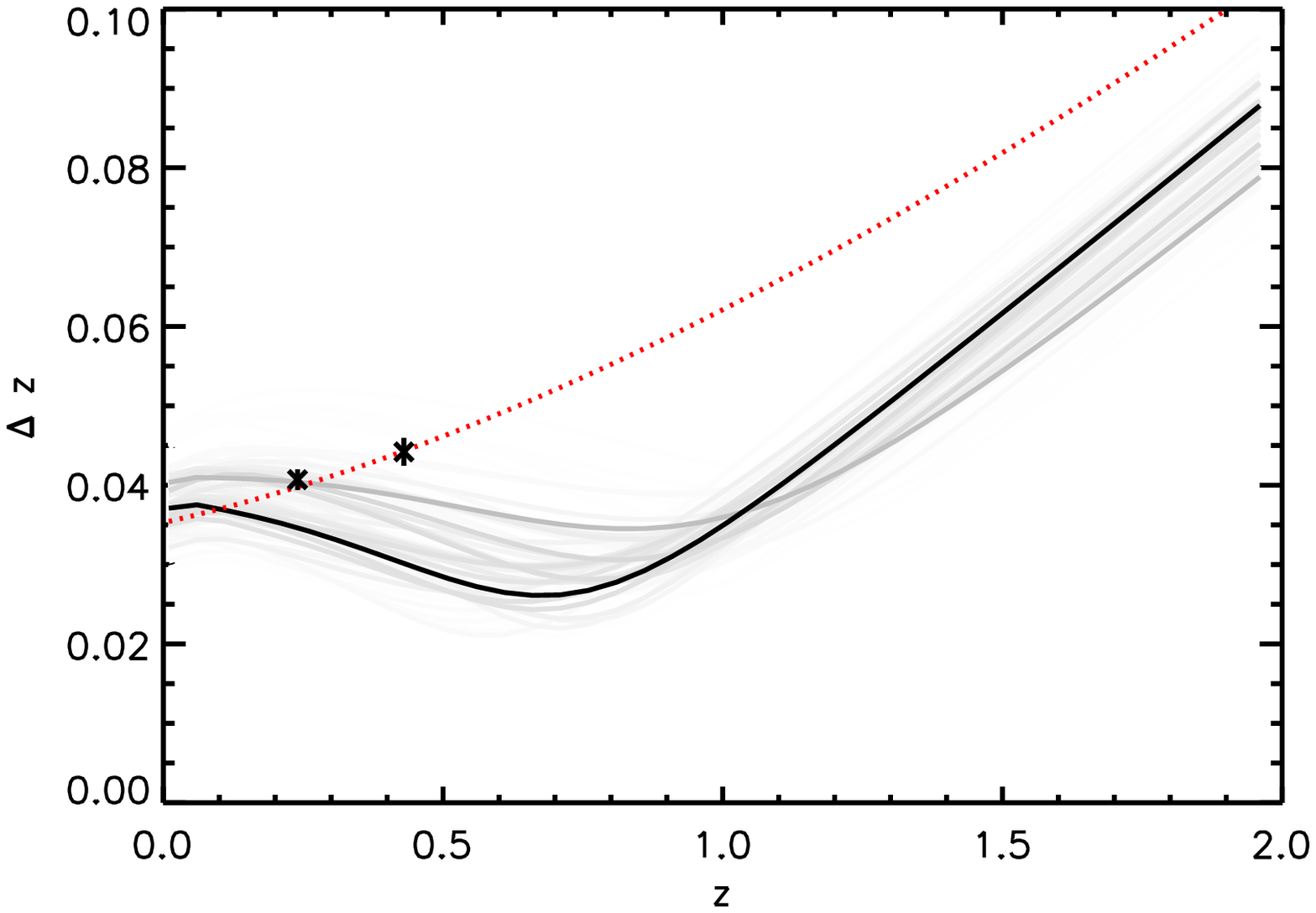}}} \mbox{\resizebox{0.48\textwidth}{!}{\includegraphics[angle=0]{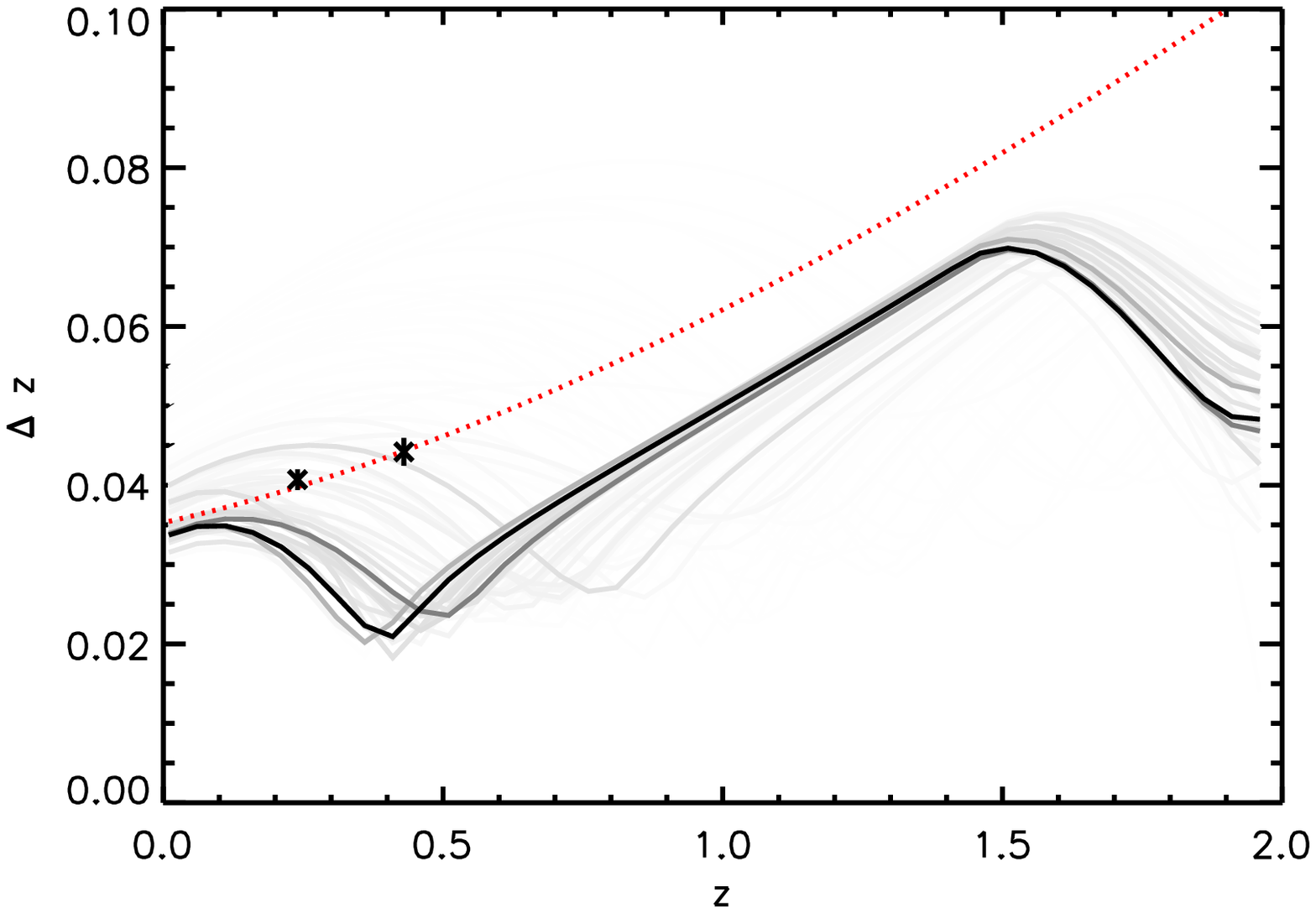}}} \caption{\label{fig:rbao} Radial BAO scale  $\Delta z$ versus $z$ for the spline (top panel) and polynomial + shell (bottom panel) profiles. The polynomial + curvature case is very similar to the polynomial + shell, but without the suppression at  $z \sim 2$. Also shown (dotted curves)  is the best fit \lcdm\ model, along with the measurements from Refs.~\cite{Gaztanaga:2008xz,Gaztanaga:2008de}. }
\end{figure}

Estimates of the radial BAO (RBAO) scale have been made recently in Refs.~\cite{Gaztanaga:2008xz,Gaztanaga:2008de}. These were used in the ZMS analysis (see also~\cite{GarciaBellido:2008yq}), but since that time the systematic errors have been revised and increased by a factor of about a third.  In addition, the statistical significance of the claimed RBAO detection has been questioned in Refs.~\cite{MiraldaEscude:2009uz,Kazin:2010nd} (however, see \cite{cg11} for a counter-argument).  With this caveat in mind, in Fig.~\ref{fig:rbao} we show the RBAO scale $\Delta z$ for our MCMC chains, along with the best fit \lcdm\ model and the measured $\Delta z$ at $z = 0.24$ and 0.43.

One can immediately see that $\Delta z$ is extremely sensitive to the particular profile.  The polynomial + shell models have two regions of strong suppression of $\Delta z$---one on the edge of the central void, where matter piles up after flowing outwards from the void (this can be seen by closely inspecting Fig.~\ref{fig:omega}), the other corresponding to the outer shell.  In the spline models these two regions are indistinguishable.  This suppression of $\Delta z$ is due to the low radial expansion rate in those regions, as explained in Sec.~\ref{sec:formalismbao}. The polynomial + curvature model is very similar to the polynomial + shell, without the suppression at  $z \sim 2$.

Post processing our MCMC chains with the additional RBAO data increases the $\chi^2$ of the best fit by $\Delta \chi^2 = 10.1$, $7.8$, and $7.5$ for the spline, polynomial + shell, and polynomial + curvature cases, respectively. The large increase in $\chi^2$  shows that current RBAO data are {\em already} a strong discriminator from \lcdm. Moreover, RBAO prefers wider void profiles, in opposition to the $y$ constraint. In Fig.~\ref{fig:1d} we show the marginalized $y$ values for the polynomial + shell  model with the addition of RBAO data. Even with factors of 2--3 uncertainties in $y$ (from tuning the position of the shell), voids with shells are strongly disfavoured considering the COBE limit. 

For the polynomial + curvature model we find the posterior is approximately Gaussian, with $y = (1.4 \pm 0.7) \times 10^{-5}$, when including RBAO. Although closer to the COBE limit,  the wider voids favored by RBAO are still compatible with the $y$ constraint when curvature is distributed homogeneously, rather than in the form of an overdense shell.


\subsubsection{Local $\sigma_8$}
\label{sec:resultssigma8}

The marginalized values of the matter power amplitude $\sigma_8$ near the void centre today are  $0.48 \pm 0.04$,  $0.49 \pm 0.04$, and $0.48 \pm 0.04$  for the spline, polynomial + shell, and polynomial + curvature, respectively (see also Fig.~\ref{fig:1d}).  These values are much lower than that in \lcdm~\cite{Komatsu:2010fb}.  The main reason for this is that near the centre the LTB model is well approximated by an open FLRW spacetime, as explained in Sec.~\ref{sec:formalismsigma8}.  For fixed $\Omega_{\rm m}$ today, the growth of structure is much more strongly suppressed in open as opposed to $\Lambda$ models (see, e.g., \cite{ll00}), leading to lower $\sigma_8$ and interesting constraints.  As with the low $H_0$ and old age problems already discussed, such low local $\sigma_8$ values appear to be a generic problem with void models.

Local measurements of $\sigma_8$ include those from weak lensing, the galaxy power spectrum, and cluster abundances. In order to compare our results with observations, we should ensure that the data analysis was not model-dependent, for example by implicitly assuming a \lcdm\ background (as the CMB-based measurements of $\sigma_8$ do).  One such constraint comes from the mass function of local galaxy clusters~\cite{Pierpaoli:2002rh}. There is a strong correlation between the cluster number density and $\sigma_8$, and simulations show a similar functional form in both open and \lcdm\ models~\cite{Jenkins:2000bv}. The complication comes from relating the cluster mass  with an observable property such as temperature  or luminosity. 

Recently, there has been significant progress in understanding systemic errors which affected previous constraints. Ref.~\cite{Henry:2008cg}, for example, provide a detailed discussion of the various sources of error. They use a local sample of clusters (46 of the 48 objects are at $z<0.1$, the other two at $z<0.2$), and find $\sigma_8 \left( \Omega_{\rm m} /0.32 \right)^{0.3} = 0.86 \pm 0.04$ (for $\Omega_{\rm m} < 0.32$) by relating the cluster count to X-ray temperature. The error budget is dominated by uncertainties in cluster physics, and cosmology dependence is negligible at such small $z$. A summary of other recent $\sigma_8$ results can be found in Ref.~\cite{Wen:2010kv}.

For void models, with the $\Omega_{\rm m} = \Omega_{\rm m}^{\rm loc} (z=0)$ likelihood peaked at $\approx 0.2$ (cf.\ Fig.~\ref{fig:1d}), the low value of $\sigma_8$ we have found is incompatible with this limit, at very high significance.


\section{Extensions}
\label{sec:extensions}

\subsection{Modifications to the initial power spectrum} \label{sec:power}

For a simple power-law primordial spectrum and a spatially flat background (as the simplest models of inflation predict), voids {\em without} an outer shell do not fit the CMB at many $\sigma$, since the physics at the LSS is so different from \lcdm.  We found that this result persists even when we allowed for the running of the spectral index, Eq.~(\ref{eq:running}). Intuitively, the redshift from the void is small compared to that from the shell, so without a shell we expect $\Teds \approx T_0$, whereas recall from Sec.~\ref{sec:formalism_CMBspectra} that we need $\Teds \approx 3.4\, {\rm K}$ to fit the CMB. One way to try to compensate for the different physics is to consider non-trivial modifications of the primordial spectrum. A model which has been discussed in the literature (see e.g.~\cite{Blanchard:2003du}) is a broken power law (BPL). This was originally proposed to try to make EdS compatible with observations without the need for dark energy. The four-parameter spectrum has the form \beq \mathcal{P}_{\rm S} (k) =   \left\{ \begin{array}{c}  
A_{\rm S} \left( k/k_0 \right) ^{n_{\rm 1}-1}\,, \quad k<k_1 \\
A_{\rm S} \, k_1^{n_{\rm 1} - n_{\rm 2}} k^{n_{\rm 2}-1} k_0^{1-n_1}\,, \quad k \ge k_1,
\end{array}  \right.  
\eeq where $k_0$ is the normalization scale and a matching condition in $\mathcal{P}_{\rm S} (k) $ is enforced at the break scale $k_1$. The spectral index is $n_1$ for $k < k_1$ and $n_2$ for $k > k_1$. 

\begin{figure} \centering 
\mbox{\resizebox{0.49\textwidth}{!}{\includegraphics[angle=0]{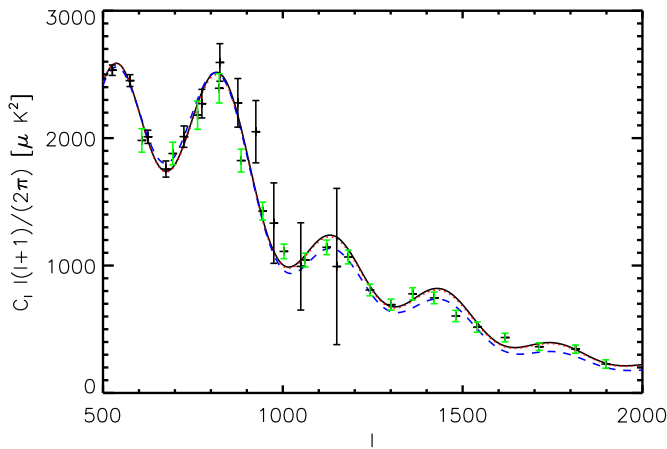}}} \caption{\label{fig:high_ell} High $\ell$ CMB spectrum. The best-fit \lcdm\ model is shown by the solid curve, the effective EdS with power-law spectrum by the dotted curve, and the effective EdS with fixed $\Teds = T_0$ and a broken power law by the dashed curve. Also shown is binned WMAP and ACBAR data.}
\end{figure}

To test whether we could obtain an improved fit with this spectrum, we repeated the effective EdS analysis with the BPL spectrum. We fix  $\Teds = T_0$ since this is a feature of voids without shells. The best-fit likelihood relative to \lcdm\ is $\Delta \chi^2 = -14.2$, and we show the CMB spectrum for each model in Fig.~\ref{fig:high_ell}. The agreement at low $\ell$ is good, but the lack of power at high $\ell$ results in a very poor overall fit with a large $\Delta \chi^2$ (note that the use of the small-scale CMB data was important here). In both cases we do not include lensing of the CMB, but since lensing only smooths out the acoustic peaks we do not expect this to alter our conclusions. This means that current CMB data are already sensitive enough to disfavour EdS with a BPL and $\Teds = T_0$ over \lcdm.  Hence further features in the primordial spectrum would need to be introduced in order to try to improve the fit. The BPL marginalized parameters are $\log_{10}  [k_1 \cdot {\rm Mpc}] = -1.90 \pm 0.04$, with $n_1 = 0.77 \pm 0.01$ and $n_2 = 1.33 \pm 0.08$.  It is worth stressing that this corresponds to a very strongly but oppositely tilted spectrum on either side of the break, i.e.\ to a very large departure from the spectra that the simplest models of inflation predict.  The implication for voids is that shells or background spatial curvature, and their consequent very low $H_0$, are {\em still} required in order to fit the CMB, even with a four-parameter BPL.  Furthermore, any attempt to improve the fit without shells or curvature would entail considerable tuning of the primordial spectrum and large departures from scale invariance.

\subsection{Multi-valued profiles} \label{sec:deep}

\subsubsection{Parameter space}

Multi-valued solutions arise when the radial expansion rate $\dot{Y}'/Y'$, and hence, by Eq.~(\ref{eqn:geo}), $dz/dr$, becomes negative. This can occur in overdense regions, either from matter piling up after flowing outwards from the void, or within the outer shell. In this regime there are three solutions to $r(z)$, and hence three ``branches'' of $d_{\rm A}$ or $\mu$ for a given redshift. 

For the polynomial + shell profile the multi-valued region of parameter space is far from that found using the CMB + SN constraints in Sec.~\ref{sec:results}. The spline allows us to investigate wider voids with a sharper transition from the void to overdense shell. These models have a multi-valued region closer to that allowed by the data, so we focus on these in the following analysis. 

\begin{figure*}
\vspace{0.25cm}\includegraphics[width=\textwidth]{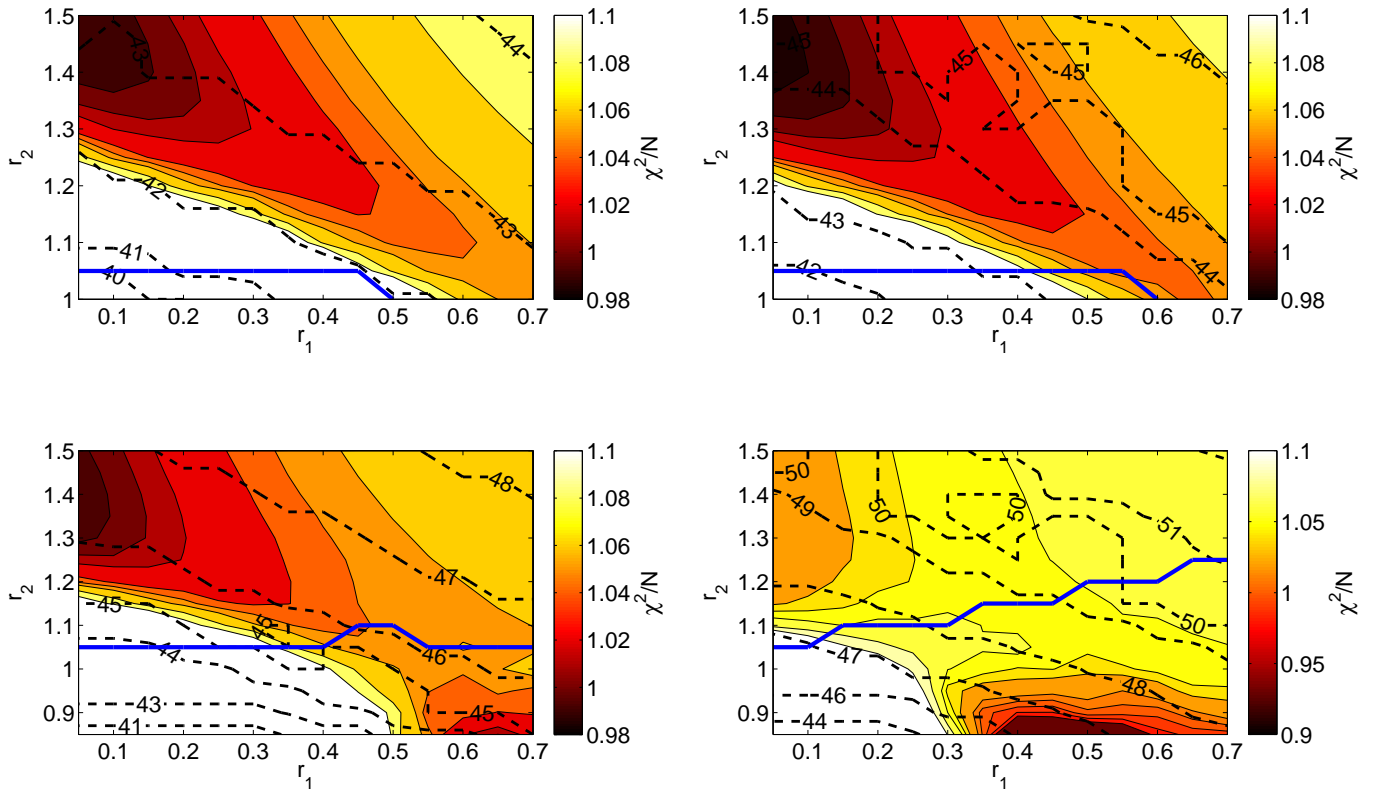} \caption{\label{fig:multi_cont} Multi-valued region of parameter space for the spline profile. The top left panel has $\Omega_{\rm m}^{\rm loc} (z=0) = 0.2$, top right $0.15$, bottom left $0.1$, and bottom right $ 0.05$. Shaded contours show the reduced $\chi^2$ from a fit to SN data (note the different scale in the bottom right panel), and dashed contours the value of $100 \, h_0$. The region at the bottom of each panel below the heavy solid line is the multi-valued regime.}
\end{figure*}

We first perform a survey of the parameter space. To do this we modify our code to use $\{ r_0=0, r_1, r_2, \delta_2 = 0,  \Omega_{\rm m}^{\rm loc} (z=0), \Teds, h_0^{\rm EdS} \}$ as the input parameters rather than $\{ r_0=0, r_1, r_2, \delta_0, \delta_1, \delta_2 = 0,  h_0 \}$ via an iterative search method. We then fix $\Teds$ and $h_0^{\rm EdS}$ to their best-fit values from Sec.~\ref{sec:methodcovarmat}, so the profiles fit the CMB. In Fig.~\ref{fig:multi_cont} we show the reduced $\chi^2$ from a fit to the Union2 SN data for various values of $\Omega_{\rm m}^{\rm loc} (z=0)$. We also indicate the value of $100 \,h_0$ and the region where multi-valued solutions exist. For multi-valued profiles we compute the $\chi^2$ in the ``kindest'' possible way---that is, we compare the observed value with the {\em closest} branch of $\mu$.

\begin{figure}
\centering \mbox{\resizebox{0.48\textwidth}{!}{\includegraphics[angle=0]{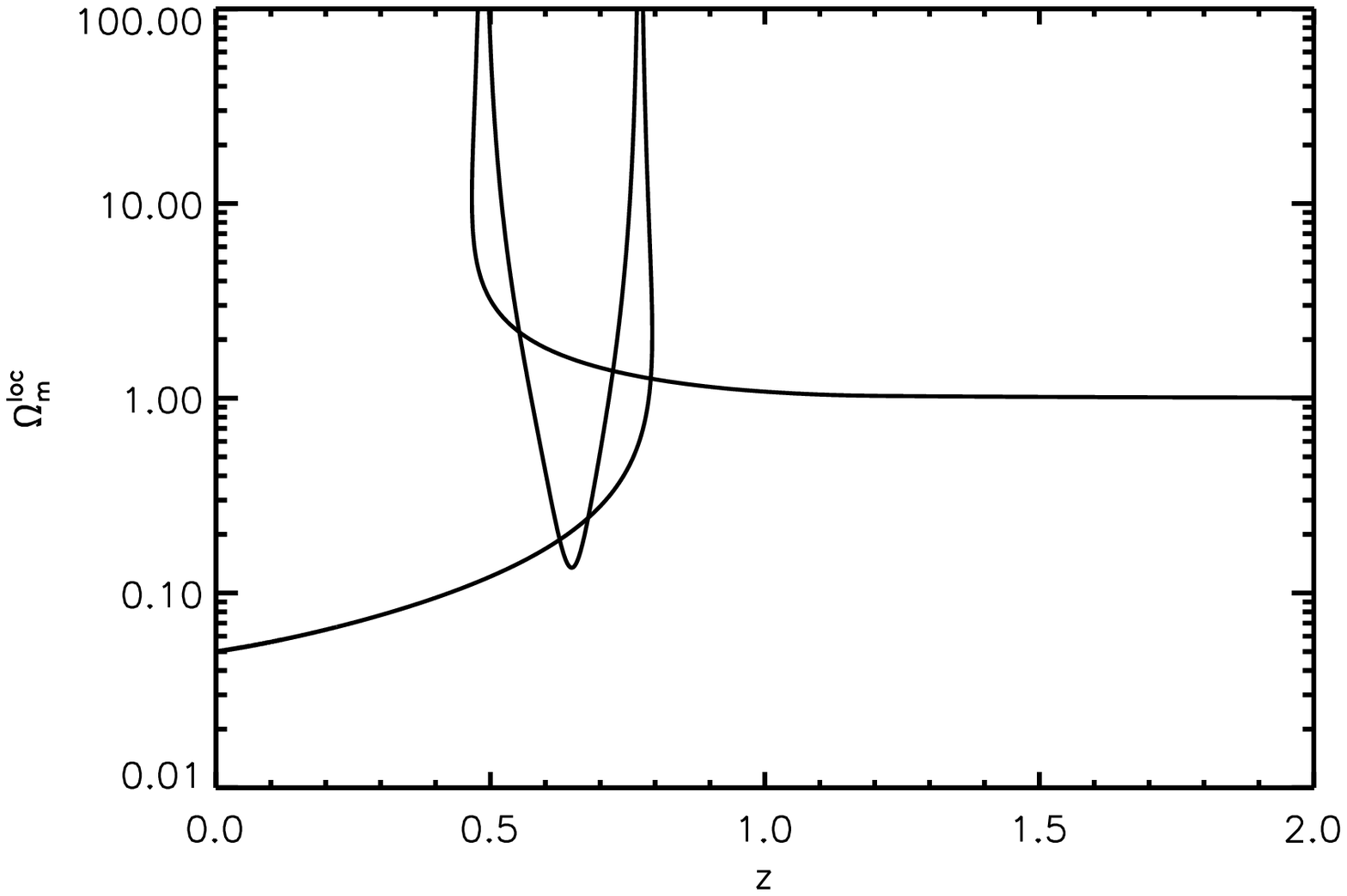}}} \mbox{\resizebox{0.48\textwidth}{!}{\includegraphics[angle=0]{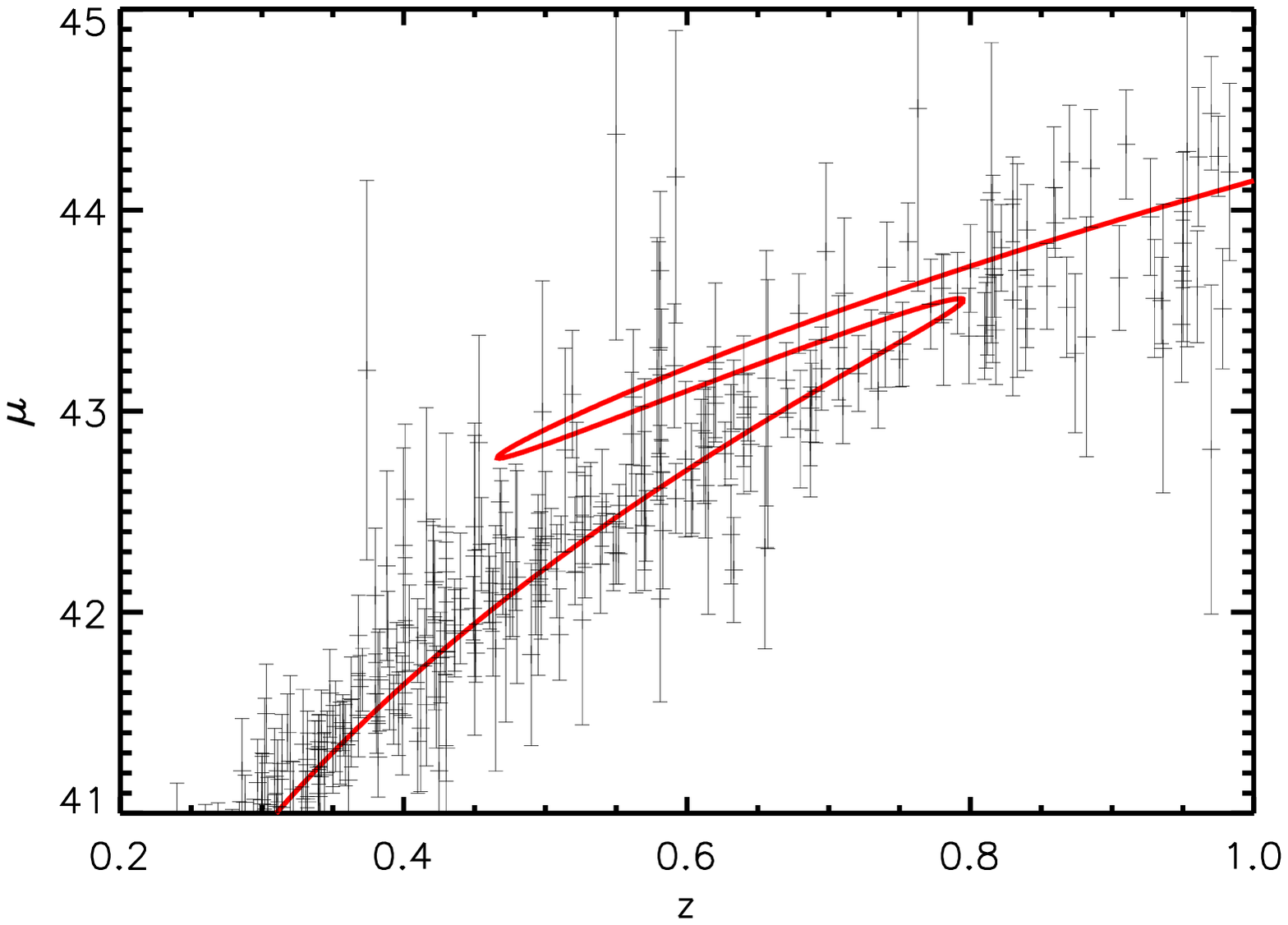}}} \caption{\label{fig:multi} Example of a multi-valued profile for $\Omega_{\rm m}^{\rm loc} (z=0) = 0.05$.  The top panel shows the density parameter and the bottom panel shows the SN magnitude residual, together with the Union2 data. The ``kind'' $\chi^2$ is significantly better than that of the single-valued solutions.}
\end{figure}

For single-valued solutions it is clear that deep voids [i.e.\ models with low  $\Omega_{\rm m}^{\rm loc}(z = 0)$] {\em do not improve the fit to the data}.  However, these deep models have an interesting multi-valued region, where the ``kind'' $\chi^2$ is significantly better than the single-valued solutions. An example of one of these profiles is shown in Fig.~\ref{fig:multi}. The low $\chi^2$ value results from improved fitting of SN outliers around the branches with higher $\mu$.  However, most of the samples are closest to one branch, which has the lowest $\mu$. If this was indeed the true underlying model, we might expect a larger fraction of SNe around the upper branches. An improved $\chi^2$ calculation should therefore take into account the expected distribution of SNe around the branches. 

\subsubsection{Modified supernova fit}

\begin{figure}
\centering \mbox{\resizebox{0.4\textwidth}{!}{\includegraphics{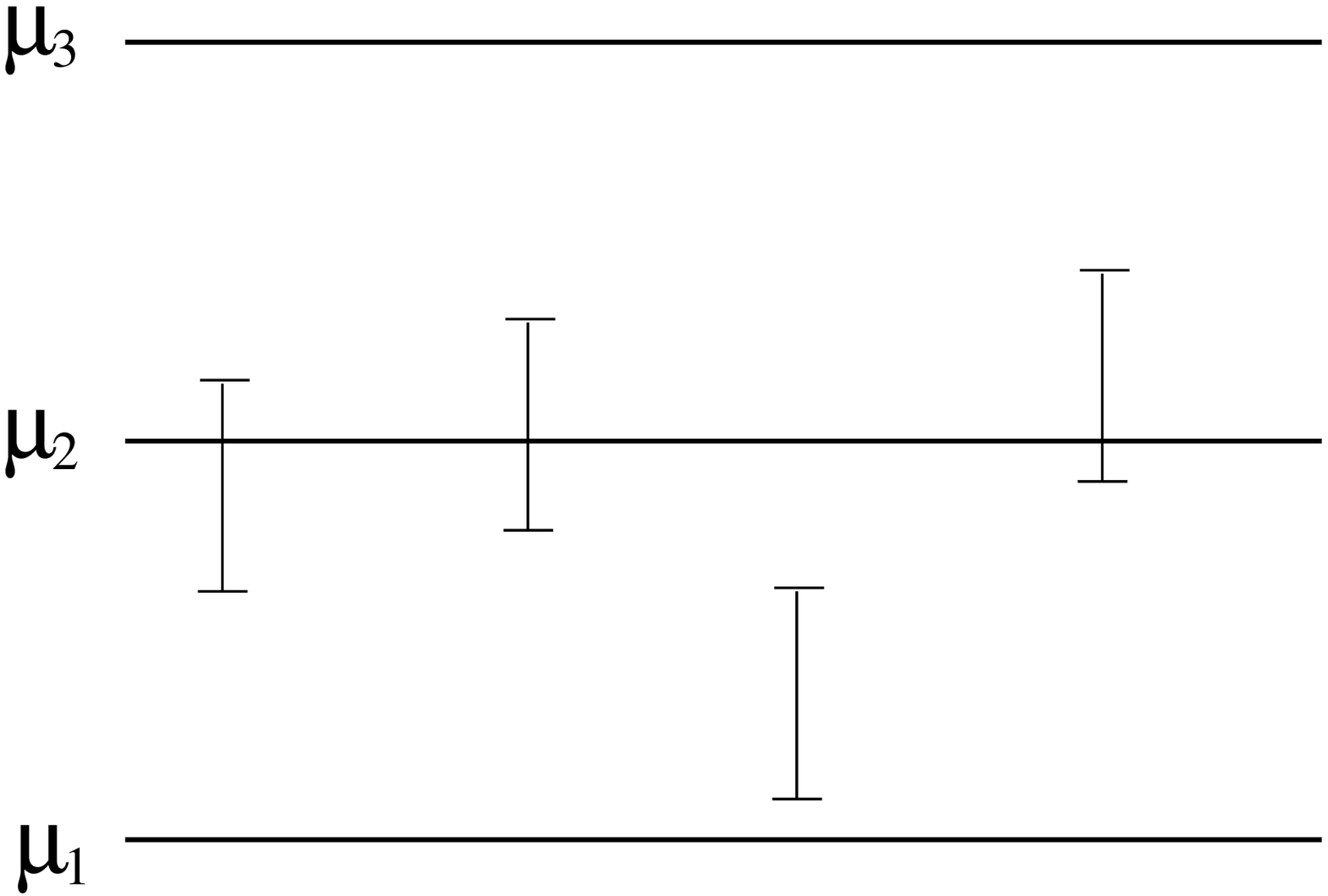}}} \caption{\label{fig:branch} Illustration of SNe distributed in the multi-valued  region of the redshift-luminosity distance relation.  The horizontal axis is a measure of redshift.} 
 \end{figure}

Our modified $\chi^2$ consists of two components: (1) $\chi^2_{\rm kind}$, which we mentioned above; and (2) $\chi^2_{\rm dist}$, from the expected distribution of SNe around each branch. For $\chi^2_{\rm dist}$, suppose there are $N$ supernovae in the multi-valued  region at redshifts $z_i$ ($i = 1,\dots,N$) with three magnitude residuals $\mu_1(z_i) < \mu_2(z_i) < \mu_3(z_i)$, as illustrated in Fig.~\ref{fig:branch}. Each magnitude residual $\mu_i$ has standard deviation $\delta_i$.

Now, suppose all  supernovae  are distributed around branch 2 according to a Gaussian probability function. Then the probability of the $i$'th supernova lying closest to branch 1 or 3, but being associated with branch 2, is, respectively,
\begin{eqnarray}
p_{21, \,i} &=& \frac{1}{2} \left[ 1 - {\rm erf} \left( \frac{ \mu_2-\mu_1}{2 \sqrt{2} \delta_i } \right) \right] \,, \\
p_{23, \,i} &=& \frac{1}{2} \left[ 1 - {\rm erf} \left( \frac{ \mu_3-\mu_2 }{2 \sqrt{2} \delta_i } \right) \right] \,,
\end{eqnarray}
where ${\rm erf}$ is the error function. For branch 2, the probability is $p_{22,\, i} = 1 - p_{21,\, i}  - p_{23,\, i}$. The expected number of supernovae $N_j$ closest to each branch $j$, along with the variance $\sigma_j^2$, can be derived by considering multiple trials from a Bernoulli distribution. The result is \beq N_j = \sum_{i=1}^{N} p_{2j,\, i}\,, \quad \sigma_j^2 =   \sum_{i=1}^{N} p_{2j,\, i} \left( 1 - p_{2j,\, i}\right)\,. \eeq

Next, we allow branch 1 or 3 to be the ``correct''  model. Similar expressions can be derived for the probability of each supernova lying closest to a branch, but being associated with another. Finally, we suppose that supernova $i$ has a {\em  probability} of belonging to each branch $j$. We label this branch probability $b_{j,\, i}$, where $\sum^3_{j=1} b_{j,\, i}=1$. We assume that the branch probability is proportional to the number density of SNe per redshift interval $dz$ at that redshift, such that \beq b_{j,\, i} \propto \frac{dN}{dz d\Omega} = \frac{\rho Y^2 Y'}{\left(1+z\right) \dot{Y}'}. \eeq In doing this we are implicitly assuming that the survey is equally sensitive to each of the branches, and that there is no SN evolution. The expected number of SNe closest to each branch is then
\begin{equation}
N_j = \sum_{i=1}^{N} \sum_{k=1}^3 b_{k, \,i} \, p_{kj,\, i} \,. 
\end{equation}
The variance calculation is more complicated, but we can use the approximation
\begin{equation}
\sigma_j^2 \approx   \sum_{i=1}^{N} \sum_{k=1}^3 p_{kj,\, i} \left( 1 -  p_{kj,\, i}\right)\,.
\end{equation}
This quantity is larger by a factor of approximately 1--2 than the actual variance (we have checked this using numerical realizations), so is a conservative estimate of $\sigma_j^2$. The ``distribution'' component of the $\chi^2$ is then \beq \chi^2_{\rm dist} = \sum_{j=1}^3 \frac{\left( N_j-N_j^{\rm obs} \right)^2}{\sigma_j^2}\,, \eeq where $N_j^{\rm obs}$ is the actual number of SNe closest to each branch.

After modifying the  $\chi^2$ according to this procedure, multi-valued solutions no longer give an improved fit.  In other words, the ``kindest'' approach to $\chi^2$ was too kind.  We investigated the multi-valued parameter space for deep voids over the range $\Omega_{\rm m}^{\rm loc} (z=0) = 0.02$--1 and found that the best fit had a similar $\chi^2/N \approx 1$ to the single-valued models. A more detailed study of the branch probabilities and magnitude errors in these models could be interesting and possibly offer some improvement over the fit to the SNe of ``regular'' voids. 

However, the deep voids still suffer from the same inherent problems in fitting other cosmological data which the single-valued models exhibit.  In particular, they do not fit the CMB without an extremely low $H_0$, as Fig.~\ref{fig:multi_cont} shows.  Note that this contradicts the results of~\cite{Clifton:2009kx}, who found that extremely deep voids could fit the CMB + SN data with reasonable $H_0$, although they stressed that such models were implausible.  The multi-valued regions in deep models are also very close to shell-crossing singularities.  Therefore it is important to treat LTB models with multi-valued regions with great care.  Finally, we note that a dense spherical shell of matter at $z<1$ would likely leave other observable signatures on structure which we have not considered here.

\section{Conclusions}
\label{sec:concl}

We have performed the most complete and up-to-date analysis of the proposal to explain the acceleration of the Universe with a large, nonlinear void. Our main results can be summarized as follows:
\begin{itemize}
\item Models which fit both the CMB and SNe have an extremely low local Hubble rate of $h_0 \approx 0.45 \pm 0.02$ (with the precise constraint being mildly model dependent) and are very old, with an age of around $17.5$--$19 \, {\rm Gyr}$. Both of these are inconsistent with observations.  The use of the {\em full} CMB spectra, rather than just an acoustic angular scale, was essential for obtaining this result.

\item The radial BAO scale is strongly suppressed in the outer void region compared to \lcdm, and the data are a poor fit to the prediction for voids.

\item The Compton $y$-distortion constraint from COBE rules out wider, deeper voids (with shells) even if the overdense shell is tuned to minimize $y$.  Generic shells are inconsistent with the COBE limit and 
{\em all} void models (with shells) are ruled out at high significance in conjunction with radial BAO data. Models with homogenous background curvature can, however, evade the COBE constraint. 

\item The local amplitude of matter fluctuations is extremely low in void models, with $\sigma_8 \approx 0.49 \pm 0.04$. This is due to the suppression of growth compared to \lcdm, and is strongly incompatible with estimates from local galaxy clusters. 

\item The primordial power spectrum would need to be highly tuned and far from scale-invariant in order to attempt to alleviate some of these problems. Running of the spectral index and a four-parameter broken power law were insufficient to change our conclusions.

\item We uncovered a class of models with multi-valued distance-redshift relations. While unusual and interesting to study, they again did not alter our conclusions.

\item All of our results, with the exception of the $y$-distortion constraint, persisted regardless of whether spatial curvature outside the void was distributed in an overdense shell or homogeneously.
\end{itemize}

Thus it appears that void models for acceleration are overwhelmingly at odds with several types of observations.  These conclusions are insensitive to the details of the void profile chosen, as is apparent from the large range of profiles our three parameterizations can capture, or to the presence of background spatial curvature.  Either a large, overdense shell or substantial positive background spatial curvature is needed to fit the CMB, although both move us even farther from the standard inflationary scenario than an isolated void in an EdS background.

   We have assumed that any LTB decaying mode is irrelevant today, although it is likely that a significant decaying mode could help these models evade some of the constraints we have examined here.  However, as discussed in Ref.~\cite{z08}, a substantial decaying mode today requires the early Universe to be extremely inhomogeneous, which is completely at odds with the standard inflationary scenario.  Ideally, though, models with decaying modes should be confronted with observations rather than argued against on such grounds~\cite{Zibin_bang}.

It is worth stressing the crucial role that CMB observations have played in our study.  The situation has a close parallel within the FLRW class of models.  As is well known (see, e.g., \cite{Larson:2010gs}), observations of the CMB alone are consistent with closed models with considerable curvature (and even vanishing $\Lambda$).  However, once local observations of the Hubble rate are included, a flat model is strongly favoured.  Similarly, the CMB alone does not rule out the inhomogeneous-curvature void models, but does in conjunction with local measurements of $H_0$.

Finally, we note that two of the most important assumptions in cosmology are those of the cosmological and Copernican principles.  Therefore, in confronting void models, which blatantly violate both of these principles, with observations, we do more than just examine an unusual approach to the mystery of acceleration.  We put the foundations of modern cosmology themselves to the test.

\section*{Note Added}

   Almost simultaneously with the submission of the first version of this paper, two other works appeared addressing similar questions~\cite{Biswas:2010xm,Clarkson:2010ej}.  The conclusions of these papers appear to be very different from ours.  For example, the abstract of Biswas, Notari, and Valkenburg~\cite{Biswas:2010xm} (hereafter BNV) states {\em ``We find that the inclusion of a nonzero overall curvature drastically improves the goodness of fit of the Void model, bringing it very close to that of a homogeneous universe containing Dark Energy, while by varying the profile one can increase the value of the local Hubble parameter which has been a challenge for these models''}.

   We agree with BNV that a near-compensated void (i.e.\ a void without an overdense shell) in an EdS background cannot fit the CMB regardless of the value of $H_0$, as was already clear in ZMS.  However, one of the main claims of BNV is that the inclusion of non-zero background spatial curvature improves the goodness of fit dramatically.  For this reason we enlarged the parameter space of models in our MCMC analysis to include background curvature (instead of overdense shells).  Our results, as reported above, are essentially unchanged from the overdense shell case.  In particular, it does not matter how the ``positive spatial curvature'' required to fit the CMB is distributed, whether in the form of an overdense shell or homogeneous background curvature: the local Hubble rate $H_0$ is too low in void models.  Importantly, we agree with BNV that the presence of background curvature (or an overdense shell as we found in ZMS) {\em does} dramatically improve the fit to the primary CMB anisotropies.  However, we still find $H_0$ values unrealistically low.

   How do the $H_0$ values presented here compare with those of BNV?  Our results in Sec.~\ref{sec:basicparams} correspond to $2\sigma$ upper limits of $h_0 = 0.50$ for the spline parameterization, $h_0 = 0.49$ for polynomial + shell, and $h_0 = 0.52$ for polynomial + curvature.  The best-fit values of $h_0$ for the CMB + BAO + SN + HST constraints in Table 13 of the published version of BNV range from 0.495 to 0.552 for BNV's profiles A to D.  However, we have noticed a discrepancy with the values in that table.  When the background LTB shear is negligible, the LTB evolution must match that of a spatially curved FLRW model, as we mentioned in Sec.~\ref{sec:formalismsigma8}.  This means that at the origin, as well as asymptotically outside the void, the LTB models must satisfy the FLRW consistency relation
\beq Ht = \oo{\Omega_K^{\rm loc}} - \fr{1 - \Omega_K^{\rm loc}}{2\ld(\Omega_K^{\rm loc}\rd)^{3/2}}
         \cosh^{-1}\ld(\fr{1 + \Omega_K^{\rm loc}}{1 - \Omega_K^{\rm loc}}\rd),
\label{Htexact} \eeq where $\Omega_K^{\rm loc} \equiv 1 - \Omega_{\rm m}^{\rm loc}$, and each quantity is evaluated at the same spacetime event.  We have noticed that the ``in'' values for profiles A to D in the published version of BNV violate the relation~(\ref{Htexact}) at the 5--10\% level.  If the BNV $H_0$ values are adjusted to satisfy Eq.~(\ref{Htexact}), they must be {\em lowered} by 5--10\%, which would bring them in line with our $2\sigma$ upper limits.  Of course, it is not clear in the first place that the $2\sigma$ upper limit of a fit to CMB + SNe (as we have done) is expected to agree well with the best fit to CMB + BAO + SN + HST (as BNV have done).

   In addition, we point out that there exists a relation between the central density parameter and Hubble rate, such that somewhat larger $H_0$ values can be obtained with larger $\Omega_{K,0}^{\rm loc}$ values, as our Fig.~\ref{fig:multi_cont} shows.  BNV's profile D obtains a slightly higher $H_0$ by pushing $\Omega_{K,0}^{\rm loc}$ very high (the original version of BNV contained profiles with extremely large $\Omega_{K,0}^{\rm loc}$).  Nevertheless, we find that even by pushing $\Omega_{K,0}^{\rm loc}$ to unrealistically high values~\cite{Carlberg:1997zp,Fukugita:1997bi}, we cannot significantly increase our upper limits on $H_0$ (see Fig.~\ref{fig:multi_cont}).

   We have also attempted to recreate BNV's profile E.  To calculate the CMB for this model, we must create a (closed) FLRW model that shares the same local density parameter at redshift $z_{\rm m}$ as the (closed) LTB model.  However, we find that the proper circumference of the LTB model at $z_{\rm m}$ exceeds the {\em equatorial} circumference of the FLRW model by tens of percent, so that the matching is not possible.  What this tells us is that profile E does not approach homogeneity sufficiently rapidly outside the void in order to reliably generate an effective model to calculate the CMB.  Hence we cannot quantitatively trust the results for profile E.

   Finally, we note that BNV claimed that their models with the largest $H_0$ were compatible with the local Hubble parameter measurements in Ref.~\cite{Sandage:2006cv}, namely $h_0 = 0.623 \pm 0.063$, which are both lower and have much larger error bars than the more recent result from~\cite{riessetal11}, who find $h_0 = 0.738 \pm 0.024$.  Therefore, as BNV point out themselves, the models they examined exhibit substantially worse fits to the newer Hubble rate data.

   To summarize, we find the largest $h_0$ in void models that provide a good fit to CMB + SNe to be around $0.49$--$0.52$.  It appears that the results of BNV are essentially consistent with ours, so that the Hubble values in void models are strongly at odds with the latest local measurements.

    The second paper that appeared almost simultaneously with ours was Ref.~\cite{Clarkson:2010ej}, whose abstract states {\em ``We allow for the dynamical effects of radiation while analyzing the problem, in contrast to other work which inadvertently fine tunes its spatial profile.  This is a surprisingly important effect and we reach substantially different conclusions.''}  Essentially, the authors of Ref.~\cite{Clarkson:2010ej} claimed that allowing for an $O(1)$ radiation-matter isocurvature mode at early times can provide a loophole to our result that the CMB cannot be fit without a very low $H_0$.  First of all, we point out that the matching procedure we use to generate the effective EdS model does {\em not} ignore radiation at early times; rather, it ignores the shear for $z > z_{\rm m}$ in effectively matching to a FLRW matter + radiation + curvature model at early times.  Our method uses the LTB solution (with test radiation source) at late times when that is a very good approximation, and uses FLRW at early times when {\em that} is a good approximation.

   We maintain that an $O(1)$ radiation-matter isocurvature mode at early times cannot have an important effect at late times.  To see this, recall that observations of the CMB by the central observer specify the physics at the LSS, i.e.\ the matter and radiation densities, via the shape of the $C_\ell$ anisotropy spectrum.  Therefore the presence of any such isocurvature mode would imply matter-radiation ratios different from the standard \lcdm\ values at early times {\em inside} our past light cone.  But any such excess or deficit of radiation would free stream away and could not affect the central observers today, since by construction it does not lie on their past light cone.  At worst, the late time {\em central} effect of an early central radiation over- or underdensity can only be a decaying LTB mode, corresponding at late times to an effectively inhomogeneous bang time, $t_{\rm B}(r)$.  But the amplitude of such a decaying mode could be at greatest $O(t_{\rm eq})$, where $t_{\rm eq}$ is the time of matter-radiation equality.  Therefore, the effect of such a decaying mode today, which would be determined by the ratio $t_{\rm eq}/t_0$, would be negligible.  Nevertheless, we believe that a more thorough examination of the effects of radiation in void models is warranted~\cite{Zibin_bang}.


\begin{acknowledgments}
This research was supported by the Natural Sciences and Engineering Research Council of Canada and the Canadian Space Agency.  The calculations were performed on computing infrastructure purchased with funds from the Canadian Foundation for Innovation and the British Columbia Knowledge Development Fund. 
\end{acknowledgments}

\bibliography{bib}

\begin{thebibliography}{74}
\expandafter\ifx\csname natexlab\endcsname\relax\def\natexlab#1{#1}\fi
\expandafter\ifx\csname bibnamefont\endcsname\relax
  \def\bibnamefont#1{#1}\fi
\expandafter\ifx\csname bibfnamefont\endcsname\relax
  \def\bibfnamefont#1{#1}\fi
\expandafter\ifx\csname citenamefont\endcsname\relax
  \def\citenamefont#1{#1}\fi
\expandafter\ifx\csname url\endcsname\relax
  \def\url#1{\texttt{#1}}\fi
\expandafter\ifx\csname urlprefix\endcsname\relax\def\urlprefix{URL }\fi
\providecommand{\bibinfo}[2]{#2}
\renewcommand{\eprint}[1]{arXiv:\href{http://arxiv.org/abs/#1}{#1}}
\providecommand{\doi}[2]{\href{http://dx.doi.org/#1}{#2}}

\bibitem[{\citenamefont{{Scott}}(2006)}]{scott06}
\bibinfo{author}{\bibfnamefont{D.}~\bibnamefont{{Scott}}},
  \bibinfo{journal}{Can. J. Phys.}
  \textbf{\doi{10.1139/P06-066}{\bibinfo{volume}{84}}},
  \doi{10.1139/P06-066}{\bibinfo{pages}{419}} (\bibinfo{year}{2006}),
  \eprint{astro-ph/0510731}.

\bibitem[{\citenamefont{Caldwell and Stebbins}(2008)}]{Caldwell:2007yu}
\bibinfo{author}{\bibfnamefont{R.~R.} \bibnamefont{Caldwell}} \bibnamefont{and}
  \bibinfo{author}{\bibfnamefont{A.}~\bibnamefont{Stebbins}},
  \bibinfo{journal}{Phys. Rev. Lett.}
  \textbf{\doi{10.1103/PhysRevLett.100.191302}{\bibinfo{volume}{100}}},
  \doi{10.1103/PhysRevLett.100.191302}{\bibinfo{pages}{191302}}
  (\bibinfo{year}{2008}), \eprint{0711.3459} [astro-ph].

\bibitem[{\citenamefont{Garcia-Bellido and
  Haugboelle}(2008{\natexlab{a}})}]{GarciaBellido:2008nz}
\bibinfo{author}{\bibfnamefont{J.}~\bibnamefont{Garcia-Bellido}}
  \bibnamefont{and}
  \bibinfo{author}{\bibfnamefont{T.}~\bibnamefont{Haugboelle}},
  \bibinfo{journal}{JCAP}
  \textbf{\doi{10.1088/1475-7516/2008/04/003}{\bibinfo{volume}{0804}}},
  \doi{10.1088/1475-7516/2008/04/003}{\bibinfo{pages}{003}}
  (\bibinfo{year}{2008}{\natexlab{a}}), \eprint{0802.1523} [astro-ph].

\bibitem[{\citenamefont{Zibin et~al.}(2008)\citenamefont{Zibin, Moss, and
  Scott}}]{Zibin:2008vk}
\bibinfo{author}{\bibfnamefont{J.~P.} \bibnamefont{Zibin}},
  \bibinfo{author}{\bibfnamefont{A.}~\bibnamefont{Moss}}, \bibnamefont{and}
  \bibinfo{author}{\bibfnamefont{D.}~\bibnamefont{Scott}},
  \bibinfo{journal}{Phys. Rev. Lett.}
  \textbf{\doi{10.1103/PhysRevLett.101.251303}{\bibinfo{volume}{101}}},
  \doi{10.1103/PhysRevLett.101.251303}{\bibinfo{pages}{251303}}
  (\bibinfo{year}{2008}), \eprint{0809.3761} [astro-ph].

\bibitem[{\citenamefont{February et~al.}(2010)\citenamefont{February, Larena,
  Smith, and Clarkson}}]{February:2009pv}
\bibinfo{author}{\bibfnamefont{S.}~\bibnamefont{February}},
  \bibinfo{author}{\bibfnamefont{J.}~\bibnamefont{Larena}},
  \bibinfo{author}{\bibfnamefont{M.}~\bibnamefont{Smith}}, \bibnamefont{and}
  \bibinfo{author}{\bibfnamefont{C.}~\bibnamefont{Clarkson}},
  \bibinfo{journal}{Mon. Not. Roy. Astron. Soc.}
  \textbf{\doi{10.1111/j.1365-2966.2010.16627.x}{\bibinfo{volume}{405}}},
  \doi{10.1111/j.1365-2966.2010.16627.x}{\bibinfo{pages}{2231}}
  (\bibinfo{year}{2010}), \eprint{0909.1479} [astro-ph.CO].

\bibitem[{\citenamefont{Tomita}(2009)}]{Tomita:2009ar}
\bibinfo{author}{\bibfnamefont{K.}~\bibnamefont{Tomita}}
  (\bibinfo{year}{2009}), \eprint{0906.1325} [astro-ph.CO].

\bibitem[{\citenamefont{{Hunt} and {Sarkar}}(2010)}]{2010MNRAS.401..547H}
\bibinfo{author}{\bibfnamefont{P.}~\bibnamefont{{Hunt}}} \bibnamefont{and}
  \bibinfo{author}{\bibfnamefont{S.}~\bibnamefont{{Sarkar}}},
  \bibinfo{journal}{Mon. Not. Roy. Astron. Soc.}
  \textbf{\doi{10.1111/j.1365-2966.2009.15670.x}{\bibinfo{volume}{401}}},
  \doi{10.1111/j.1365-2966.2009.15670.x}{\bibinfo{pages}{547}}
  (\bibinfo{year}{2010}), \eprint{0807.4508} [astro-ph].

\bibitem[{\citenamefont{Alnes and Amarzguioui}(2006)}]{Alnes:2006pf}
\bibinfo{author}{\bibfnamefont{H.}~\bibnamefont{Alnes}} \bibnamefont{and}
  \bibinfo{author}{\bibfnamefont{M.}~\bibnamefont{Amarzguioui}},
  \bibinfo{journal}{Phys. Rev.}
  \textbf{\doi{10.1103/PhysRevD.74.103520}{\bibinfo{volume}{D74}}},
  \doi{10.1103/PhysRevD.74.103520}{\bibinfo{pages}{103520}}
  (\bibinfo{year}{2006}), \eprint{astro-ph/0607334}.

\bibitem[{\citenamefont{Blomqvist and Mortsell}(2010)}]{Blomqvist:2009ps}
\bibinfo{author}{\bibfnamefont{M.}~\bibnamefont{Blomqvist}} \bibnamefont{and}
  \bibinfo{author}{\bibfnamefont{E.}~\bibnamefont{Mortsell}},
  \bibinfo{journal}{JCAP}
  \textbf{\doi{10.1088/1475-7516/2010/05/006}{\bibinfo{volume}{1005}}},
  \doi{10.1088/1475-7516/2010/05/006}{\bibinfo{pages}{006}}
  (\bibinfo{year}{2010}), \eprint{0909.4723} [astro-ph.CO].

\bibitem[{\citenamefont{Kodama et~al.}(2010)\citenamefont{Kodama, Saito, and
  Ishibashi}}]{Kodama:2010gr}
\bibinfo{author}{\bibfnamefont{H.}~\bibnamefont{Kodama}},
  \bibinfo{author}{\bibfnamefont{K.}~\bibnamefont{Saito}}, \bibnamefont{and}
  \bibinfo{author}{\bibfnamefont{A.}~\bibnamefont{Ishibashi}},
  \bibinfo{journal}{Prog. Theor. Phys.}
  \textbf{\doi{10.1143/PTP.124.163}{\bibinfo{volume}{124}}},
  \doi{10.1143/PTP.124.163}{\bibinfo{pages}{163}} (\bibinfo{year}{2010}),
  \eprint{1004.3089} [astro-ph.CO].

\bibitem[{\citenamefont{Foreman et~al.}(2010)\citenamefont{Foreman, Moss,
  Zibin, and Scott}}]{Foreman:2010uj}
\bibinfo{author}{\bibfnamefont{S.}~\bibnamefont{Foreman}},
  \bibinfo{author}{\bibfnamefont{A.}~\bibnamefont{Moss}},
  \bibinfo{author}{\bibfnamefont{J.~P.} \bibnamefont{Zibin}}, \bibnamefont{and}
  \bibinfo{author}{\bibfnamefont{D.}~\bibnamefont{Scott}},
  \bibinfo{journal}{Phys. Rev.}
  \textbf{\doi{10.1103/PhysRevD.82.103532}{\bibinfo{volume}{D82}}},
  \doi{10.1103/PhysRevD.82.103532}{\bibinfo{pages}{103532}}
  (\bibinfo{year}{2010}), \eprint{1009.0273} [astro-ph.CO].

\bibitem[{\citenamefont{Tomita}(2000)}]{Tomita:1999qn}
\bibinfo{author}{\bibfnamefont{K.}~\bibnamefont{Tomita}},
  \bibinfo{journal}{Astrophys. J.}
  \textbf{\doi{10.1086/308277}{\bibinfo{volume}{529}}},
  \doi{10.1086/308277}{\bibinfo{pages}{38}} (\bibinfo{year}{2000}),
  \eprint{astro-ph/9906027}.

\bibitem[{\citenamefont{Goodwin et~al.}(1999)\citenamefont{Goodwin, Thomas,
  Barber, Gribbin, and Onuora}}]{Goodwin:1999ej}
\bibinfo{author}{\bibfnamefont{S.~P.} \bibnamefont{Goodwin}},
  \bibinfo{author}{\bibfnamefont{P.~A.} \bibnamefont{Thomas}},
  \bibinfo{author}{\bibfnamefont{A.~J.} \bibnamefont{Barber}},
  \bibinfo{author}{\bibfnamefont{J.}~\bibnamefont{Gribbin}}, \bibnamefont{and}
  \bibinfo{author}{\bibfnamefont{L.~I.} \bibnamefont{Onuora}}
  (\bibinfo{year}{1999}), \eprint{astro-ph/9906187}.

\bibitem[{\citenamefont{Celerier}(2000)}]{Celerier:1999hp}
\bibinfo{author}{\bibfnamefont{M.-N.} \bibnamefont{Celerier}},
  \bibinfo{journal}{Astron. Astrophys.} \textbf{\bibinfo{volume}{353}},
  \bibinfo{pages}{63} (\bibinfo{year}{2000}), \eprint{astro-ph/9907206}.

\bibitem[{\citenamefont{{Lema{\^i}tre}}(1933)}]{lemaitre33}
\bibinfo{author}{\bibfnamefont{G.}~\bibnamefont{{Lema{\^i}tre}}},
  \bibinfo{journal}{Ann. Soc. Sci. Bruxelles} \textbf{\bibinfo{volume}{53}},
  \bibinfo{pages}{51} (\bibinfo{year}{1933}).

\bibitem[{\citenamefont{Tolman}(1934)}]{tolman34}
\bibinfo{author}{\bibfnamefont{R.~C.} \bibnamefont{Tolman}},
  \bibinfo{journal}{Proc. Nat. Acad. Sci.} \textbf{\bibinfo{volume}{20}},
  \bibinfo{pages}{169} (\bibinfo{year}{1934}).

\bibitem[{\citenamefont{Bondi}(1947)}]{bondi47}
\bibinfo{author}{\bibfnamefont{H.}~\bibnamefont{Bondi}}, \bibinfo{journal}{Mon.
  Not. Roy. Astron. Soc.} \textbf{\bibinfo{volume}{107}}, \bibinfo{pages}{410}
  (\bibinfo{year}{1947}).

\bibitem[{\citenamefont{Yoo et~al.}(2008)\citenamefont{Yoo, Kai, and
  Nakao}}]{Yoo:2008su}
\bibinfo{author}{\bibfnamefont{C.-M.} \bibnamefont{Yoo}},
  \bibinfo{author}{\bibfnamefont{T.}~\bibnamefont{Kai}}, \bibnamefont{and}
  \bibinfo{author}{\bibfnamefont{K.-i.} \bibnamefont{Nakao}},
  \bibinfo{journal}{Prog. Theor. Phys.}
  \textbf{\doi{10.1143/PTP.120.937}{\bibinfo{volume}{120}}},
  \doi{10.1143/PTP.120.937}{\bibinfo{pages}{937}} (\bibinfo{year}{2008}),
  \eprint{0807.0932} [astro-ph].

\bibitem[{\citenamefont{Garcia-Bellido and
  Haugboelle}(2008{\natexlab{b}})}]{GarciaBellido:2008gd}
\bibinfo{author}{\bibfnamefont{J.}~\bibnamefont{Garcia-Bellido}}
  \bibnamefont{and}
  \bibinfo{author}{\bibfnamefont{T.}~\bibnamefont{Haugboelle}},
  \bibinfo{journal}{JCAP}
  \textbf{\doi{10.1088/1475-7516/2008/09/016}{\bibinfo{volume}{0809}}},
  \doi{10.1088/1475-7516/2008/09/016}{\bibinfo{pages}{016}}
  (\bibinfo{year}{2008}{\natexlab{b}}), \eprint{0807.1326} [astro-ph].

\bibitem[{\citenamefont{Alnes et~al.}(2006)\citenamefont{Alnes, Amarzguioui,
  and Gron}}]{Alnes:2005rw}
\bibinfo{author}{\bibfnamefont{H.}~\bibnamefont{Alnes}},
  \bibinfo{author}{\bibfnamefont{M.}~\bibnamefont{Amarzguioui}},
  \bibnamefont{and} \bibinfo{author}{\bibfnamefont{O.}~\bibnamefont{Gron}},
  \bibinfo{journal}{Phys. Rev.}
  \textbf{\doi{10.1103/PhysRevD.73.083519}{\bibinfo{volume}{D73}}},
  \doi{10.1103/PhysRevD.73.083519}{\bibinfo{pages}{083519}}
  (\bibinfo{year}{2006}), \eprint{astro-ph/0512006}.

\bibitem[{\citenamefont{Bolejko and Wyithe}(2009)}]{Bolejko:2008cm}
\bibinfo{author}{\bibfnamefont{K.}~\bibnamefont{Bolejko}} \bibnamefont{and}
  \bibinfo{author}{\bibfnamefont{J.~S.~B.} \bibnamefont{Wyithe}},
  \bibinfo{journal}{JCAP}
  \textbf{\doi{10.1088/1475-7516/2009/02/020}{\bibinfo{volume}{0902}}},
  \doi{10.1088/1475-7516/2009/02/020}{\bibinfo{pages}{020}}
  (\bibinfo{year}{2009}), \eprint{0807.2891} [astro-ph].

\bibitem[{\citenamefont{Garcia-Bellido and
  Haugboelle}(2009)}]{GarciaBellido:2008yq}
\bibinfo{author}{\bibfnamefont{J.}~\bibnamefont{Garcia-Bellido}}
  \bibnamefont{and}
  \bibinfo{author}{\bibfnamefont{T.}~\bibnamefont{Haugboelle}},
  \bibinfo{journal}{JCAP}
  \textbf{\doi{10.1088/1475-7516/2009/09/028}{\bibinfo{volume}{0909}}},
  \doi{10.1088/1475-7516/2009/09/028}{\bibinfo{pages}{028}}
  (\bibinfo{year}{2009}), \eprint{0810.4939} [astro-ph].

\bibitem[{\citenamefont{Yoo et~al.}(2010)\citenamefont{Yoo, Nakao, and
  Sasaki}}]{Yoo:2010qy}
\bibinfo{author}{\bibfnamefont{C.-M.} \bibnamefont{Yoo}},
  \bibinfo{author}{\bibfnamefont{K.-i.} \bibnamefont{Nakao}}, \bibnamefont{and}
  \bibinfo{author}{\bibfnamefont{M.}~\bibnamefont{Sasaki}},
  \bibinfo{journal}{JCAP}
  \textbf{\doi{10.1088/1475-7516/2010/07/012}{\bibinfo{volume}{1007}}},
  \doi{10.1088/1475-7516/2010/07/012}{\bibinfo{pages}{012}}
  (\bibinfo{year}{2010}), \eprint{1005.0048} [astro-ph.CO].

\bibitem[{\citenamefont{Alexander et~al.}(2009)\citenamefont{Alexander, Biswas,
  Notari, and Vaid}}]{Alexander:2007xx}
\bibinfo{author}{\bibfnamefont{S.}~\bibnamefont{Alexander}},
  \bibinfo{author}{\bibfnamefont{T.}~\bibnamefont{Biswas}},
  \bibinfo{author}{\bibfnamefont{A.}~\bibnamefont{Notari}}, \bibnamefont{and}
  \bibinfo{author}{\bibfnamefont{D.}~\bibnamefont{Vaid}},
  \bibinfo{journal}{JCAP}
  \textbf{\doi{10.1088/1475-7516/2009/09/025}{\bibinfo{volume}{0909}}},
  \doi{10.1088/1475-7516/2009/09/025}{\bibinfo{pages}{025}}
  (\bibinfo{year}{2009}), \eprint{0712.0370} [astro-ph].

\bibitem[{\citenamefont{Vonlanthen et~al.}(2010)\citenamefont{Vonlanthen,
  Rasanen, and Durrer}}]{Vonlanthen:2010cd}
\bibinfo{author}{\bibfnamefont{M.}~\bibnamefont{Vonlanthen}},
  \bibinfo{author}{\bibfnamefont{S.}~\bibnamefont{Rasanen}}, \bibnamefont{and}
  \bibinfo{author}{\bibfnamefont{R.}~\bibnamefont{Durrer}},
  \bibinfo{journal}{JCAP}
  \textbf{\doi{10.1088/1475-7516/2010/08/023}{\bibinfo{volume}{1008}}},
  \doi{10.1088/1475-7516/2010/08/023}{\bibinfo{pages}{023}}
  (\bibinfo{year}{2010}), \eprint{1003.0810} [astro-ph.CO].

\bibitem[{\citenamefont{Clarkson et~al.}(2009)\citenamefont{Clarkson, Clifton,
  and February}}]{ccf09}
\bibinfo{author}{\bibfnamefont{C.}~\bibnamefont{Clarkson}},
  \bibinfo{author}{\bibfnamefont{T.}~\bibnamefont{Clifton}}, \bibnamefont{and}
  \bibinfo{author}{\bibfnamefont{S.}~\bibnamefont{February}},
  \bibinfo{journal}{JCAP}
  \textbf{\doi{10.1088/1475-7516/2009/06/025}{\bibinfo{volume}{0906}}},
  \doi{10.1088/1475-7516/2009/06/025}{\bibinfo{pages}{025}}
  (\bibinfo{year}{2009}), \eprint{0903.5040} [astro-ph.CO].

\bibitem[{\citenamefont{Zibin}(2008)}]{z08}
\bibinfo{author}{\bibfnamefont{J.~P.} \bibnamefont{Zibin}},
  \bibinfo{journal}{Phys. Rev.}
  \textbf{\doi{10.1103/PhysRevD.78.043504}{\bibinfo{volume}{D78}}},
  \doi{10.1103/PhysRevD.78.043504}{\bibinfo{pages}{043504}}
  (\bibinfo{year}{2008}), \eprint{0804.1787} [astro-ph].

\bibitem[{\citenamefont{Dunsby et~al.}(2010)\citenamefont{Dunsby, Goheer,
  Osano, and Uzan}}]{Dunsby:2010ts}
\bibinfo{author}{\bibfnamefont{P.}~\bibnamefont{Dunsby}},
  \bibinfo{author}{\bibfnamefont{N.}~\bibnamefont{Goheer}},
  \bibinfo{author}{\bibfnamefont{B.}~\bibnamefont{Osano}}, \bibnamefont{and}
  \bibinfo{author}{\bibfnamefont{J.-P.} \bibnamefont{Uzan}},
  \bibinfo{journal}{JCAP}
  \textbf{\doi{10.1088/1475-7516/2010/06/017}{\bibinfo{volume}{1006}}},
  \doi{10.1088/1475-7516/2010/06/017}{\bibinfo{pages}{017}}
  (\bibinfo{year}{2010}), \eprint{1002.2397} [astro-ph.CO].

\bibitem[{\citenamefont{{Silk}}(1977)}]{silk77}
\bibinfo{author}{\bibfnamefont{J.}~\bibnamefont{{Silk}}},
  \bibinfo{journal}{Astron. Astrophys.} \textbf{\bibinfo{volume}{59}},
  \bibinfo{pages}{53} (\bibinfo{year}{1977}).

\bibitem[{\citenamefont{Lewis et~al.}(2000)\citenamefont{Lewis, Challinor, and
  Lasenby}}]{Lewis:1999bs}
\bibinfo{author}{\bibfnamefont{A.}~\bibnamefont{Lewis}},
  \bibinfo{author}{\bibfnamefont{A.}~\bibnamefont{Challinor}},
  \bibnamefont{and} \bibinfo{author}{\bibfnamefont{A.}~\bibnamefont{Lasenby}},
  \bibinfo{journal}{Astrophys. J.}
  \textbf{\doi{10.1086/309179}{\bibinfo{volume}{538}}},
  \doi{10.1086/309179}{\bibinfo{pages}{473}} (\bibinfo{year}{2000}),
  \eprint{astro-ph/9911177}.

\bibitem[{\citenamefont{Mustapha et~al.}(1998)\citenamefont{Mustapha, Bassett,
  Hellaby, and Ellis}}]{Mustapha:1997xb}
\bibinfo{author}{\bibfnamefont{N.}~\bibnamefont{Mustapha}},
  \bibinfo{author}{\bibfnamefont{B.~A.} \bibnamefont{Bassett}},
  \bibinfo{author}{\bibfnamefont{C.}~\bibnamefont{Hellaby}}, \bibnamefont{and}
  \bibinfo{author}{\bibfnamefont{G.~F.~R.} \bibnamefont{Ellis}},
  \bibinfo{journal}{Class. Quant. Grav.}
  \textbf{\doi{10.1088/0264-9381/15/8/016}{\bibinfo{volume}{15}}},
  \doi{10.1088/0264-9381/15/8/016}{\bibinfo{pages}{2363}}
  (\bibinfo{year}{1998}), \eprint{gr-qc/9708043}.

\bibitem[{\citenamefont{Biswas et~al.}(2007)\citenamefont{Biswas, Mansouri, and
  Notari}}]{Biswas:2006ub}
\bibinfo{author}{\bibfnamefont{T.}~\bibnamefont{Biswas}},
  \bibinfo{author}{\bibfnamefont{R.}~\bibnamefont{Mansouri}}, \bibnamefont{and}
  \bibinfo{author}{\bibfnamefont{A.}~\bibnamefont{Notari}},
  \bibinfo{journal}{JCAP}
  \textbf{\doi{10.1088/1475-7516/2007/12/017}{\bibinfo{volume}{0712}}},
  \doi{10.1088/1475-7516/2007/12/017}{\bibinfo{pages}{017}}
  (\bibinfo{year}{2007}), \eprint{astro-ph/0606703}.

\bibitem[{\citenamefont{Jarosik et~al.}(2011)}]{Jarosik:2010iu}
\bibinfo{author}{\bibfnamefont{N.}~\bibnamefont{Jarosik}} \bibnamefont{et~al.},
  \bibinfo{journal}{Astrophys. J. Supp.}
  \textbf{\doi{10.1088/0067-0049/192/2/14}{\bibinfo{volume}{192}}},
  \doi{10.1088/0067-0049/192/2/14}{\bibinfo{pages}{14}} (\bibinfo{year}{2011}),
  \eprint{1001.4744} [astro-ph.CO].

\bibitem[{\citenamefont{Komatsu et~al.}(2011)}]{Komatsu:2010fb}
\bibinfo{author}{\bibfnamefont{E.}~\bibnamefont{Komatsu}} \bibnamefont{et~al.}
  (\bibinfo{collaboration}{WMAP}), \bibinfo{journal}{Astrophys. J. Supp.}
  \textbf{\doi{10.1088/0067-0049/192/2/18}{\bibinfo{volume}{192}}},
  \doi{10.1088/0067-0049/192/2/18}{\bibinfo{pages}{18}} (\bibinfo{year}{2011}),
  \eprint{1001.4538} [astro-ph.CO].

\bibitem[{\citenamefont{Amanullah et~al.}(2010)}]{Amanullah:2010vv}
\bibinfo{author}{\bibfnamefont{R.}~\bibnamefont{Amanullah}}
  \bibnamefont{et~al.}, \bibinfo{journal}{Astrophys. J.}
  \textbf{\doi{10.1088/0004-637X/716/1/712}{\bibinfo{volume}{716}}},
  \doi{10.1088/0004-637X/716/1/712}{\bibinfo{pages}{712}}
  (\bibinfo{year}{2010}), \eprint{1004.1711} [astro-ph.CO].

\bibitem[{\citenamefont{Fixsen}(2009)}]{Fixsen:2009ug}
\bibinfo{author}{\bibfnamefont{D.~J.} \bibnamefont{Fixsen}},
  \bibinfo{journal}{Astrophys. J.}
  \textbf{\doi{10.1088/0004-637X/707/2/916}{\bibinfo{volume}{707}}},
  \doi{10.1088/0004-637X/707/2/916}{\bibinfo{pages}{916}}
  (\bibinfo{year}{2009}), \eprint{0911.1955} [astro-ph.CO].

\bibitem[{\citenamefont{Zibin}(2011)}]{Zibin_bang}
\bibinfo{author}{\bibfnamefont{J.~P.} \bibnamefont{Zibin}}
  (\bibinfo{year}{2011}), \bibinfo{note}{in preparation}.

\bibitem[{\citenamefont{Goodman}(1995)}]{Goodman:1995dt}
\bibinfo{author}{\bibfnamefont{J.}~\bibnamefont{Goodman}},
  \bibinfo{journal}{Phys. Rev.}
  \textbf{\doi{10.1103/PhysRevD.52.1821}{\bibinfo{volume}{D52}}},
  \doi{10.1103/PhysRevD.52.1821}{\bibinfo{pages}{1821}} (\bibinfo{year}{1995}),
  \eprint{astro-ph/9506068}.

\bibitem[{\citenamefont{Stebbins}(2007)}]{Stebbins}
\bibinfo{author}{\bibfnamefont{A.}~\bibnamefont{Stebbins}}
  (\bibinfo{year}{2007}), \eprint{astro-ph/0703541}.

\bibitem[{\citenamefont{Eisenstein and Hu}(1998)}]{eisenstein}
\bibinfo{author}{\bibfnamefont{D.~J.} \bibnamefont{Eisenstein}}
  \bibnamefont{and} \bibinfo{author}{\bibfnamefont{W.}~\bibnamefont{Hu}},
  \bibinfo{journal}{Astrophys. J.}
  \textbf{\doi{10.1086/305424}{\bibinfo{volume}{496}}},
  \doi{10.1086/305424}{\bibinfo{pages}{605}} (\bibinfo{year}{1998}),
  \eprint{astro-ph/9709112}.

\bibitem[{\citenamefont{Eisenstein et~al.}(2005)}]{Eisenstein:2005su}
\bibinfo{author}{\bibfnamefont{D.~J.} \bibnamefont{Eisenstein}}
  \bibnamefont{et~al.} (\bibinfo{collaboration}{SDSS}),
  \bibinfo{journal}{Astrophys. J.}
  \textbf{\doi{10.1086/466512}{\bibinfo{volume}{633}}},
  \doi{10.1086/466512}{\bibinfo{pages}{560}} (\bibinfo{year}{2005}),
  \eprint{astro-ph/0501171}.

\bibitem[{\citenamefont{Lewis and Bridle}(2002)}]{cosmomc}
\bibinfo{author}{\bibfnamefont{A.}~\bibnamefont{Lewis}} \bibnamefont{and}
  \bibinfo{author}{\bibfnamefont{S.}~\bibnamefont{Bridle}},
  \bibinfo{journal}{Phys. Rev.}
  \textbf{\doi{10.1103/PhysRevD.66.103511}{\bibinfo{volume}{D66}}},
  \doi{10.1103/PhysRevD.66.103511}{\bibinfo{pages}{103511}}
  (\bibinfo{year}{2002}), \eprint{astro-ph/0205436}.

\bibitem[{\citenamefont{Reichardt et~al.}(2009)}]{acbar}
\bibinfo{author}{\bibfnamefont{C.~L.} \bibnamefont{Reichardt}}
  \bibnamefont{et~al.}, \bibinfo{journal}{Astrophys. J.}
  \textbf{\doi{10.1088/0004-637X/694/2/1200}{\bibinfo{volume}{694}}},
  \doi{10.1088/0004-637X/694/2/1200}{\bibinfo{pages}{1200}}
  (\bibinfo{year}{2009}), \eprint{0801.1491} [astro-ph].

\bibitem[{\citenamefont{Jones et~al.}(2006)}]{boom}
\bibinfo{author}{\bibfnamefont{W.~C.} \bibnamefont{Jones}}
  \bibnamefont{et~al.}, \bibinfo{journal}{Astrophys. J.}
  \textbf{\doi{10.1086/505559}{\bibinfo{volume}{647}}},
  \doi{10.1086/505559}{\bibinfo{pages}{823}} (\bibinfo{year}{2006}),
  \eprint{astro-ph/0507494}.

\bibitem[{\citenamefont{Sievers et~al.}(2009)}]{cbi}
\bibinfo{author}{\bibfnamefont{J.~L.} \bibnamefont{Sievers}}
  \bibnamefont{et~al.} (\bibinfo{year}{2009}), \eprint{0901.4540}
  [astro-ph.CO].

\bibitem[{\citenamefont{Pryke et~al.}(2009)}]{quad}
\bibinfo{author}{\bibfnamefont{C.}~\bibnamefont{Pryke}} \bibnamefont{et~al.}
  (\bibinfo{collaboration}{QUaD}), \bibinfo{journal}{Astrophys. J.}
  \textbf{\doi{10.1088/0004-637X/692/2/1247}{\bibinfo{volume}{692}}},
  \doi{10.1088/0004-637X/692/2/1247}{\bibinfo{pages}{1247}}
  (\bibinfo{year}{2009}), \eprint{0805.1944} [astro-ph].

\bibitem[{\citenamefont{Komatsu and Seljak}(2002)}]{sztemplate}
\bibinfo{author}{\bibfnamefont{E.}~\bibnamefont{Komatsu}} \bibnamefont{and}
  \bibinfo{author}{\bibfnamefont{U.}~\bibnamefont{Seljak}},
  \bibinfo{journal}{Mon. Not. Roy. Astron. Soc.}
  \textbf{\doi{10.1046/j.1365-8711.2002.05889.x}{\bibinfo{volume}{336}}},
  \doi{10.1046/j.1365-8711.2002.05889.x}{\bibinfo{pages}{1256}}
  (\bibinfo{year}{2002}), \eprint{astro-ph/0205468}.

\bibitem[{\citenamefont{Carlberg et~al.}(1997)}]{Carlberg:1997zp}
\bibinfo{author}{\bibfnamefont{R.~G.} \bibnamefont{Carlberg}}
  \bibnamefont{et~al.} (\bibinfo{year}{1997}), \eprint{astro-ph/9711272}.

\bibitem[{\citenamefont{Fukugita et~al.}(1998)\citenamefont{Fukugita, Hogan,
  and Peebles}}]{Fukugita:1997bi}
\bibinfo{author}{\bibfnamefont{M.}~\bibnamefont{Fukugita}},
  \bibinfo{author}{\bibfnamefont{C.~J.} \bibnamefont{Hogan}}, \bibnamefont{and}
  \bibinfo{author}{\bibfnamefont{P.~J.~E.} \bibnamefont{Peebles}},
  \bibinfo{journal}{Astrophys. J.}
  \textbf{\doi{10.1086/306025}{\bibinfo{volume}{503}}},
  \doi{10.1086/306025}{\bibinfo{pages}{518}} (\bibinfo{year}{1998}),
  \eprint{astro-ph/9712020}.

\bibitem[{\citenamefont{{Freedman} and {Madore}}(2010)}]{Freedman:2010xv}
\bibinfo{author}{\bibfnamefont{W.~L.} \bibnamefont{{Freedman}}}
  \bibnamefont{and} \bibinfo{author}{\bibfnamefont{B.~F.}
  \bibnamefont{{Madore}}}, \bibinfo{journal}{Annu. Rev. Astron. Astrophys.}
  \textbf{\doi{10.1146/annurev-astro-082708-101829}{\bibinfo{volume}{48}}},
  \doi{10.1146/annurev-astro-082708-101829}{\bibinfo{pages}{673}}
  (\bibinfo{year}{2010}), \eprint{1004.1856} [astro-ph.CO].

\bibitem[{\citenamefont{{Riess} et~al.}(2011)}]{riessetal11}
\bibinfo{author}{\bibfnamefont{A.~G.} \bibnamefont{{Riess}}}
  \bibnamefont{et~al.}, \bibinfo{journal}{Astrophys. J.}
  \textbf{\doi{10.1088/0004-637X/730/2/119}{\bibinfo{volume}{730}}},
  \doi{10.1088/0004-637X/730/2/119}{\bibinfo{pages}{119}}
  (\bibinfo{year}{2011}), \eprint{1103.2976} [astro-ph.CO].

\bibitem[{\citenamefont{Tammann et~al.}(2008)\citenamefont{Tammann, Sandage,
  and Reindl}}]{Tammann:2007ge}
\bibinfo{author}{\bibfnamefont{G.~A.} \bibnamefont{Tammann}},
  \bibinfo{author}{\bibfnamefont{A.}~\bibnamefont{Sandage}}, \bibnamefont{and}
  \bibinfo{author}{\bibfnamefont{B.}~\bibnamefont{Reindl}},
  \bibinfo{journal}{Astrophys. J.}
  \textbf{\doi{10.1086/529508}{\bibinfo{volume}{679}}},
  \doi{10.1086/529508}{\bibinfo{pages}{52}} (\bibinfo{year}{2008}),
  \eprint{0712.2346} [astro-ph].

\bibitem[{\citenamefont{Riess et~al.}(2009)}]{Riess:2009pu}
\bibinfo{author}{\bibfnamefont{A.~G.} \bibnamefont{Riess}}
  \bibnamefont{et~al.}, \bibinfo{journal}{Astrophys. J.}
  \textbf{\doi{10.1088/0004-637X/699/1/539}{\bibinfo{volume}{699}}},
  \doi{10.1088/0004-637X/699/1/539}{\bibinfo{pages}{539}}
  (\bibinfo{year}{2009}), \eprint{0905.0695} [astro-ph.CO].

\bibitem[{\citenamefont{Fixsen et~al.}(1996)}]{Fixsen:1996nj}
\bibinfo{author}{\bibfnamefont{D.~J.} \bibnamefont{Fixsen}}
  \bibnamefont{et~al.}, \bibinfo{journal}{Astrophys. J.}
  \textbf{\doi{10.1086/178173}{\bibinfo{volume}{473}}},
  \doi{10.1086/178173}{\bibinfo{pages}{576}} (\bibinfo{year}{1996}),
  \eprint{astro-ph/9605054}.

\bibitem[{\citenamefont{Krauss and Chaboyer}(2003)}]{Krauss:2003em}
\bibinfo{author}{\bibfnamefont{L.~M.} \bibnamefont{Krauss}} \bibnamefont{and}
  \bibinfo{author}{\bibfnamefont{B.}~\bibnamefont{Chaboyer}},
  \bibinfo{journal}{Science}
  \textbf{\doi{10.1126/science.1075631}{\bibinfo{volume}{299}}},
  \doi{10.1126/science.1075631}{\bibinfo{pages}{65}} (\bibinfo{year}{2003}).

\bibitem[{\citenamefont{Steigman}(2007)}]{Steigman:2007xt}
\bibinfo{author}{\bibfnamefont{G.}~\bibnamefont{Steigman}},
  \bibinfo{journal}{Ann. Rev. Nucl. Part. Sci.}
  \textbf{\doi{10.1146/annurev.nucl.56.080805.140437}{\bibinfo{volume}{57}}},
  \doi{10.1146/annurev.nucl.56.080805.140437}{\bibinfo{pages}{463}}
  (\bibinfo{year}{2007}), \eprint{0712.1100} [astro-ph].

\bibitem[{\citenamefont{{Pettini} et~al.}(2008)}]{BBN}
\bibinfo{author}{\bibfnamefont{M.}~\bibnamefont{{Pettini}}}
  \bibnamefont{et~al.}, \bibinfo{journal}{Mon. Not. Roy. Astron. Soc.}
  \textbf{\doi{10.1111/j.1365-2966.2008.13921.x}{\bibinfo{volume}{391}}},
  \doi{10.1111/j.1365-2966.2008.13921.x}{\bibinfo{pages}{1499}}
  (\bibinfo{year}{2008}), \eprint{0805.0594} [astro-ph].

\bibitem[{\citenamefont{Scott and Smoot}(2010)}]{Scott:2010yx}
\bibinfo{author}{\bibfnamefont{D.}~\bibnamefont{Scott}} \bibnamefont{and}
  \bibinfo{author}{\bibfnamefont{G.~F.} \bibnamefont{Smoot}}
  (\bibinfo{year}{2010}), \eprint{1005.0555} [astro-ph.CO].

\bibitem[{\citenamefont{Gaztanaga
  et~al.}(2009{\natexlab{a}})\citenamefont{Gaztanaga, Cabre, and
  Hui}}]{Gaztanaga:2008xz}
\bibinfo{author}{\bibfnamefont{E.}~\bibnamefont{Gaztanaga}},
  \bibinfo{author}{\bibfnamefont{A.}~\bibnamefont{Cabre}}, \bibnamefont{and}
  \bibinfo{author}{\bibfnamefont{L.}~\bibnamefont{Hui}}, \bibinfo{journal}{Mon.
  Not. Roy. Astron. Soc.}
  \textbf{\doi{10.1111/j.1365-2966.2009.15405.x}{\bibinfo{volume}{399}}},
  \doi{10.1111/j.1365-2966.2009.15405.x}{\bibinfo{pages}{1663}}
  (\bibinfo{year}{2009}{\natexlab{a}}), \eprint{0807.3551} [astro-ph].

\bibitem[{\citenamefont{Gaztanaga
  et~al.}(2009{\natexlab{b}})\citenamefont{Gaztanaga, Miquel, and
  Sanchez}}]{Gaztanaga:2008de}
\bibinfo{author}{\bibfnamefont{E.}~\bibnamefont{Gaztanaga}},
  \bibinfo{author}{\bibfnamefont{R.}~\bibnamefont{Miquel}}, \bibnamefont{and}
  \bibinfo{author}{\bibfnamefont{E.}~\bibnamefont{Sanchez}},
  \bibinfo{journal}{Phys. Rev. Lett.}
  \textbf{\doi{10.1103/PhysRevLett.103.091302}{\bibinfo{volume}{103}}},
  \doi{10.1103/PhysRevLett.103.091302}{\bibinfo{pages}{091302}}
  (\bibinfo{year}{2009}{\natexlab{b}}), \eprint{0808.1921} [astro-ph].

\bibitem[{\citenamefont{Miralda-Escude}(2009)}]{MiraldaEscude:2009uz}
\bibinfo{author}{\bibfnamefont{J.}~\bibnamefont{Miralda-Escude}}
  (\bibinfo{year}{2009}), \eprint{0901.1219} [astro-ph].

\bibitem[{\citenamefont{Kazin et~al.}(2010)}]{Kazin:2010nd}
\bibinfo{author}{\bibfnamefont{E.~A.} \bibnamefont{Kazin}}
  \bibnamefont{et~al.}, \bibinfo{journal}{Astrophys. J.}
  \textbf{\doi{10.1088/0004-637X/719/2/1032}{\bibinfo{volume}{719}}},
  \doi{10.1088/0004-637X/719/2/1032}{\bibinfo{pages}{1032}}
  (\bibinfo{year}{2010}), \eprint{1004.2244} [astro-ph.CO].

\bibitem[{\citenamefont{{Cabr{\'e}} and {Gazta{\~n}aga}}(2011)}]{cg11}
\bibinfo{author}{\bibfnamefont{A.}~\bibnamefont{{Cabr{\'e}}}} \bibnamefont{and}
  \bibinfo{author}{\bibfnamefont{E.}~\bibnamefont{{Gazta{\~n}aga}}},
  \bibinfo{journal}{Mon. Not. Roy. Astron. Soc.} p. \bibinfo{pages}{L211}
  (\bibinfo{year}{2011}), \eprint{1011.2729} [astro-ph.CO].

\bibitem[{\citenamefont{Liddle and Lyth}(2000)}]{ll00}
\bibinfo{author}{\bibfnamefont{A.~R.} \bibnamefont{Liddle}} \bibnamefont{and}
  \bibinfo{author}{\bibfnamefont{D.~H.} \bibnamefont{Lyth}},
  \emph{\bibinfo{title}{Cosmological Inflation and Large-Scale Structure}}
  (\bibinfo{publisher}{Cambridge University Press},
  \bibinfo{address}{Cambridge}, \bibinfo{year}{2000}).

\bibitem[{\citenamefont{Pierpaoli et~al.}(2003)\citenamefont{Pierpaoli,
  Borgani, Scott, and White}}]{Pierpaoli:2002rh}
\bibinfo{author}{\bibfnamefont{E.}~\bibnamefont{Pierpaoli}},
  \bibinfo{author}{\bibfnamefont{S.}~\bibnamefont{Borgani}},
  \bibinfo{author}{\bibfnamefont{D.}~\bibnamefont{Scott}}, \bibnamefont{and}
  \bibinfo{author}{\bibfnamefont{M.~J.} \bibnamefont{White}},
  \bibinfo{journal}{Mon. Not. Roy. Astron. Soc.}
  \textbf{\doi{10.1046/j.1365-8711.2003.06525.x}{\bibinfo{volume}{342}}},
  \doi{10.1046/j.1365-8711.2003.06525.x}{\bibinfo{pages}{163}}
  (\bibinfo{year}{2003}), \eprint{astro-ph/0210567}.

\bibitem[{\citenamefont{Jenkins et~al.}(2001)}]{Jenkins:2000bv}
\bibinfo{author}{\bibfnamefont{A.}~\bibnamefont{Jenkins}} \bibnamefont{et~al.},
  \bibinfo{journal}{Mon. Not. Roy. Astron. Soc.}
  \textbf{\doi{10.1046/j.1365-8711.2001.04029.x}{\bibinfo{volume}{321}}},
  \doi{10.1046/j.1365-8711.2001.04029.x}{\bibinfo{pages}{372}}
  (\bibinfo{year}{2001}), \eprint{astro-ph/0005260}.

\bibitem[{\citenamefont{Henry et~al.}(2009)}]{Henry:2008cg}
\bibinfo{author}{\bibfnamefont{J.~P.} \bibnamefont{Henry}}
  \bibnamefont{et~al.}, \bibinfo{journal}{Astrophys. J.}
  \textbf{\doi{10.1088/0004-637X/691/2/1307}{\bibinfo{volume}{691}}},
  \doi{10.1088/0004-637X/691/2/1307}{\bibinfo{pages}{1307}}
  (\bibinfo{year}{2009}), \eprint{0809.3832} [astro-ph].

\bibitem[{\citenamefont{{Wen} et~al.}(2010)\citenamefont{{Wen}, {Han}, and
  {Liu}}}]{Wen:2010kv}
\bibinfo{author}{\bibfnamefont{Z.~L.} \bibnamefont{{Wen}}},
  \bibinfo{author}{\bibfnamefont{J.~L.} \bibnamefont{{Han}}}, \bibnamefont{and}
  \bibinfo{author}{\bibfnamefont{F.~S.} \bibnamefont{{Liu}}},
  \bibinfo{journal}{Mon. Not. Roy. Astron. Soc.}
  \textbf{\doi{10.1111/j.1365-2966.2010.16930.x}{\bibinfo{volume}{407}}},
  \doi{10.1111/j.1365-2966.2010.16930.x}{\bibinfo{pages}{533}}
  (\bibinfo{year}{2010}), \eprint{1004.3337} [astro-ph.CO].

\bibitem[{\citenamefont{Blanchard et~al.}(2003)\citenamefont{Blanchard,
  Douspis, Rowan-Robinson, and Sarkar}}]{Blanchard:2003du}
\bibinfo{author}{\bibfnamefont{A.}~\bibnamefont{Blanchard}},
  \bibinfo{author}{\bibfnamefont{M.}~\bibnamefont{Douspis}},
  \bibinfo{author}{\bibfnamefont{M.}~\bibnamefont{Rowan-Robinson}},
  \bibnamefont{and} \bibinfo{author}{\bibfnamefont{S.}~\bibnamefont{Sarkar}},
  \bibinfo{journal}{Astron. Astrophys.}
  \textbf{\doi{10.1051/0004-6361:20031425}{\bibinfo{volume}{412}}},
  \doi{10.1051/0004-6361:20031425}{\bibinfo{pages}{35}} (\bibinfo{year}{2003}),
  \eprint{astro-ph/0304237}.

\bibitem[{\citenamefont{Clifton et~al.}(2009)\citenamefont{Clifton, Ferreira,
  and Zuntz}}]{Clifton:2009kx}
\bibinfo{author}{\bibfnamefont{T.}~\bibnamefont{Clifton}},
  \bibinfo{author}{\bibfnamefont{P.~G.} \bibnamefont{Ferreira}},
  \bibnamefont{and} \bibinfo{author}{\bibfnamefont{J.}~\bibnamefont{Zuntz}},
  \bibinfo{journal}{JCAP}
  \textbf{\doi{10.1088/1475-7516/2009/07/029}{\bibinfo{volume}{0907}}},
  \doi{10.1088/1475-7516/2009/07/029}{\bibinfo{pages}{029}}
  (\bibinfo{year}{2009}), \eprint{0902.1313} [astro-ph.CO].

\bibitem[{\citenamefont{Larson et~al.}(2011)}]{Larson:2010gs}
\bibinfo{author}{\bibfnamefont{D.}~\bibnamefont{Larson}} \bibnamefont{et~al.},
  \bibinfo{journal}{Astrophys. J. Supp.}
  \textbf{\doi{10.1088/0067-0049/192/2/16}{\bibinfo{volume}{192}}},
  \doi{10.1088/0067-0049/192/2/16}{\bibinfo{pages}{16}} (\bibinfo{year}{2011}),
  \eprint{1001.4635} [astro-ph.CO].

\bibitem[{\citenamefont{Biswas et~al.}(2010)\citenamefont{Biswas, Notari, and
  Valkenburg}}]{Biswas:2010xm}
\bibinfo{author}{\bibfnamefont{T.}~\bibnamefont{Biswas}},
  \bibinfo{author}{\bibfnamefont{A.}~\bibnamefont{Notari}}, \bibnamefont{and}
  \bibinfo{author}{\bibfnamefont{W.}~\bibnamefont{Valkenburg}},
  \bibinfo{journal}{JCAP}
  \textbf{\doi{10.1088/1475-7516/2010/11/030}{\bibinfo{volume}{1011}}},
  \doi{10.1088/1475-7516/2010/11/030}{\bibinfo{pages}{030}}
  (\bibinfo{year}{2010}), \eprint{1007.3065} [astro-ph.CO].

\bibitem[{\citenamefont{{Clarkson} and {Regis}}(2011)}]{Clarkson:2010ej}
\bibinfo{author}{\bibfnamefont{C.}~\bibnamefont{{Clarkson}}} \bibnamefont{and}
  \bibinfo{author}{\bibfnamefont{M.}~\bibnamefont{{Regis}}},
  \bibinfo{journal}{JCAP}
  \textbf{\doi{10.1088/1475-7516/2011/02/013}{\bibinfo{volume}{2}}},
  \doi{10.1088/1475-7516/2011/02/013}{\bibinfo{pages}{13}}
  (\bibinfo{year}{2011}), \eprint{1007.3443} [astro-ph.CO].

\bibitem[{\citenamefont{Sandage et~al.}(2006)}]{Sandage:2006cv}
\bibinfo{author}{\bibfnamefont{A.}~\bibnamefont{Sandage}} \bibnamefont{et~al.},
  \bibinfo{journal}{Astrophys. J.}
  \textbf{\doi{10.1086/508853}{\bibinfo{volume}{653}}},
  \doi{10.1086/508853}{\bibinfo{pages}{843}} (\bibinfo{year}{2006}),
  \eprint{astro-ph/0603647}.

\end{thebibliography}

\end{document}